\documentclass[12pt]{iopart}
\usepackage{graphicx}
\def\lesssim{\mathbin{\lower 3pt\hbox 
      {$\rlap{\raise 5pt\hbox{$\char'074$}}\mathchar"7218$}}} 
\def\gtrsim{\mathbin{\lower 3pt\hbox
      {$\rlap{\raise 5pt\hbox{$\char'076$}}\mathchar"7218$}}} 

\begin{document}

\title[Neutron Star Surface Emission]
{Surface Emission from Neutron Stars and Implications for the Physics of their Interiors}

\author{Feryal \"Ozel}
\address{Department of Astronomy,
University of Arizona, 933 N. Cherry Ave, Tucson, AZ 85721}
\ead{fozel@email.arizona.edu}
\begin{abstract}
Neutron stars are associated with diverse physical phenomena that take
place in conditions characterized by ultrahigh densities as well as
intense gravitational, magnetic, and radiation fields. Understanding
the properties and interactions of matter in these regimes remains one
of the challenges in compact object astrophysics. Photons emitted from
the surfaces of neutron stars provide direct probes of their
structure, composition, and magnetic fields. In this review, I discuss
in detail the physics that governs the properties of emission from the
surfaces of neutron stars and their various observational
manifestations. I present the constraints on neutron star radii, core
and crust composition, and magnetic field strength and topology
obtained from studies of their broadband spectra, evolution of thermal
luminosity, and the profiles of pulsations that originate on their
surfaces.
\end{abstract}

\tableofcontents
\maketitle

%\twocolumn

\section{Introduction}

Neutron stars are some of the most fascinating astrophysical objects.
During their lifetimes, they are connected to a great variety of
phenomena, from the supernova explosions that accompany their births
to the bursts of gravitational waves and gamma-rays that are emitted
during their inspiral into other compact objects. More than 2000
neutron stars have been discovered to date in the Galaxy with a large
diversity of observational appearance from pulsating sources in the
radio to bright persistent sources in the X-rays.

Despite their ubiquity, a large number of outstanding problems remain
in neutron star physics, owing primarily to the extreme physical
conditions present in their interiors and surfaces. Indeed, neutron
stars are the stellar objects that possess the most extreme densities
and the strongest magnetic fields found in the universe. They also
harbor the strongest gravitational fields among all astrophysical
objects that still have a surface. Modeling their properties and
evolution requires detailed knowledge of physics in regimes that are
inaccessible to terrestrial experiments. Conversely, observations of
neutron stars offer unique probes of the behavior of matter and
radiation in these extraordinary conditions.

Neutron star studies make use of different observables related to the
dynamics of binary systems, the structure and energetics of radio
pulses, and the spectra and variability properties of their high
energy radiation. Among these various probes, direct observations of
the emission from a neutron star surface provide the most powerful
diagnostic not only of the conditions of its surface but also of its
interior composition and physics. 

Neutron star surfaces, however, are often hidden from view, as they
are overpowered by non-thermal magnetospheric emission in isolated
sources or by accretion luminosity in contact binaries. Nevertheless,
a remarkable and growing number of sources in different classes show
unequivocal evidence for surface emission. During the last 15 years,
high energy observatories with superb angular, spectral, and timing
resolutions have revealed surface emission from isolated cooling
neutron stars (see \"Ogelman 1995; Page et al.\ 2009), accreting
sources during periods of quiescence (e.g., Degenaar et al.\ 2011b),
thermonuclear bursters (Strohmayer \& Bildsten 2006), and isolated
neutron stars powered by magnetic field decay (Woods \& Thompson
2004). The emission in these different sources probes a wide range of
conditions such as temperatures and magnetic field strengths, as well
as of spin periods, ages, and evolutionary histories.

Studying the observed surface emission and using it as a tool to
understand the properties of neutron stars themselves requires
detailed models of their atmospheres, which shape the spectrum and
pattern of radiation originating in the stellar crust or core. The
great progress in the last decade in detecting surface emission from a
variety of neutron stars has been accompanied by the development of
atmospheric models that span a wide range of magnetic field strengths,
compositions, temperatures, and ionization states. Theoretical models
are now routinely compared to high quality observations, with an
emphasis in deriving the fundamental properties of neutron stars from
their surface emission.

Observations of neutron stars in quiescence and of neutron stars
during thermonuclear bursts have led to the first constraining
measurements of neutron star radii. These studies indicate most likely
values of 9~km$< R <$12~km and have been used to constrain the
equation of state of ultradense matter; equations of state that
predict radii as large as 15~km are disfavored by observations
(\"Ozel, Baym, \& G\"uver 2010; Steiner, Lattimer, \& Brown 2010).

Observations of cooling isolated neutron stars with high sensitivity
instruments, which have been modeled with magnetic atmosphere models
and used to study the evolution of the thermal properties of neutron
stars at different ages, led to significant constraints on neutron
star crusts as well as on their interior composition. Comparison of
these results with theoretical cooling curves revealed that it is
highly unlikely for neutron star cores to have the compositions and
the high densities necessary for direct Urca processes to play a role
in neutron star cooling (Yakovlev \& Pethick 2004). Similar
observations of the young neutron star in the supernova remnant Cas~A
provide tantalizing evidence for the transition to superfluidity in
its core (Page et al.\ 2011; Shternin et al.\ 2011).

Our understanding of how magnetic fields can power and shape the
emission from neutron star surfaces has dramatically advanced thanks
to the studies of the highly magnetic Anomalous X-ray Pulsars and Soft
Gamma-ray Repeaters (Woods \& Thompson 2004). Identifiable effects of
ultrastrong magnetic fields both on their thermal spectra and in
post-burst phenomena have been observed from these sources. The
polarization of the vacuum in strong magnetic fields has been
incorporated into neutron star atmosphere calculations that are
actively being compared to high energy resolution observations of
magnetars in order to definitively identify the signatures of this QED
phenomenon.

There has also been significant progress in the past decade on the
development of other probes of neutron star gravity and spacetimes,
such as the comparison of the observed pulse profiles in accretion
powered millisecond pulsars to model lightcurves computed for hotspots
on rotating neutron stars. Because the amplitudes and shapes of
pulsations are sensitive to the surface gravity, these first studies
have been able to place promising constraints on the neutron star
masses and radii. These constraints will improve in the near future
with the availability of higher signal-to-noise data with new X-ray
satellites. It is also eagerly anticipated that the existing probes of
neutron star spacetimes will be augmented by observations of
gravitational waves and neutrinos emitted from neutron stars to reveal
the detailed structure of these compact objects.

In this review, I start by discussing the relevant physical processes
that power the surface emission observed from isolated and accreting
neutron stars. I then present in Section 2.2-2.4 the various factors
that enter models of neutron star atmospheres and determine the flux,
the spectrum, and the anisotropy of their surface emission. In
Section~3, I summarize the observational properties of numerous types
of neutron stars from which surface emission has been detected, as
well as the prevailing models that aim to explain their properties.
Sections 4-6 are devoted to discussing in detail how the results on
neutron star radii, compositions, as well as magnetic field geometry
and evolution are obtained from comparisons of the theoretical models
to the observations presented in Section~3. In Section~7, I explore a
number of approaches to probing neutron star physics that complement
the studies of their surface emission. Finally, I conclude in
Section~8 with the prospects for furthering our understanding of
neutron star physics through progress in theoretical models as well as
with future observations.

\section{Physics of the Neutron Star Surface Emission}

\subsection{Energy Sources}

The surface emission observed from neutron stars can be powered by a
variety of physical mechanisms, including the release of the internal
heat of the star, accretion, nuclear reactions on the neutron star
surface, magnetic field decay, and rotational energy. \\

\noindent {\bf Radiation of residual heat in young neutron stars} \\
Following a supernova explosion, a hot proto-neutron star, with a core
temperature of $\sim 10^{11}$~K, cools through the combination of
neutrino emission from its interior and photon cooling from its
surface.

Calculations of the thermal evolution of a neutron star after the
supernova explosion traditionally divide the object into a stellar
interior and an outer heat blanketing envelope that are separated at a
boundary radius $R_b$ that corresponds to a density of $\sim
10^{10}$~g~cm$^{-3}$ (see Yakovlev \& Pethick 2004 and Page \& Reddy
2006 for reviews of neutron star cooling). The interior becomes
isothermal within a few years after the birth of the neutron star,
whereas the envelope sustains strong temperature gradients.

The envelope is $\approx 100$~m deep, has a very short thermal
relaxation time and negligible neutrino emissivity. For these reasons,
it is modeled as a stationary, plane-parallel envelope in hydrostatic
equilibrium. Solving the heat transport equation through this boundary
leads to a relation between the temperature $T_b$ at the bottom of the
layer and the effective temperature $T_{\rm eff}$ that specifies the
radiative flux emerging from the neutron star surface (see below).

\begin{figure}
\centerline{\includegraphics[scale=0.55]{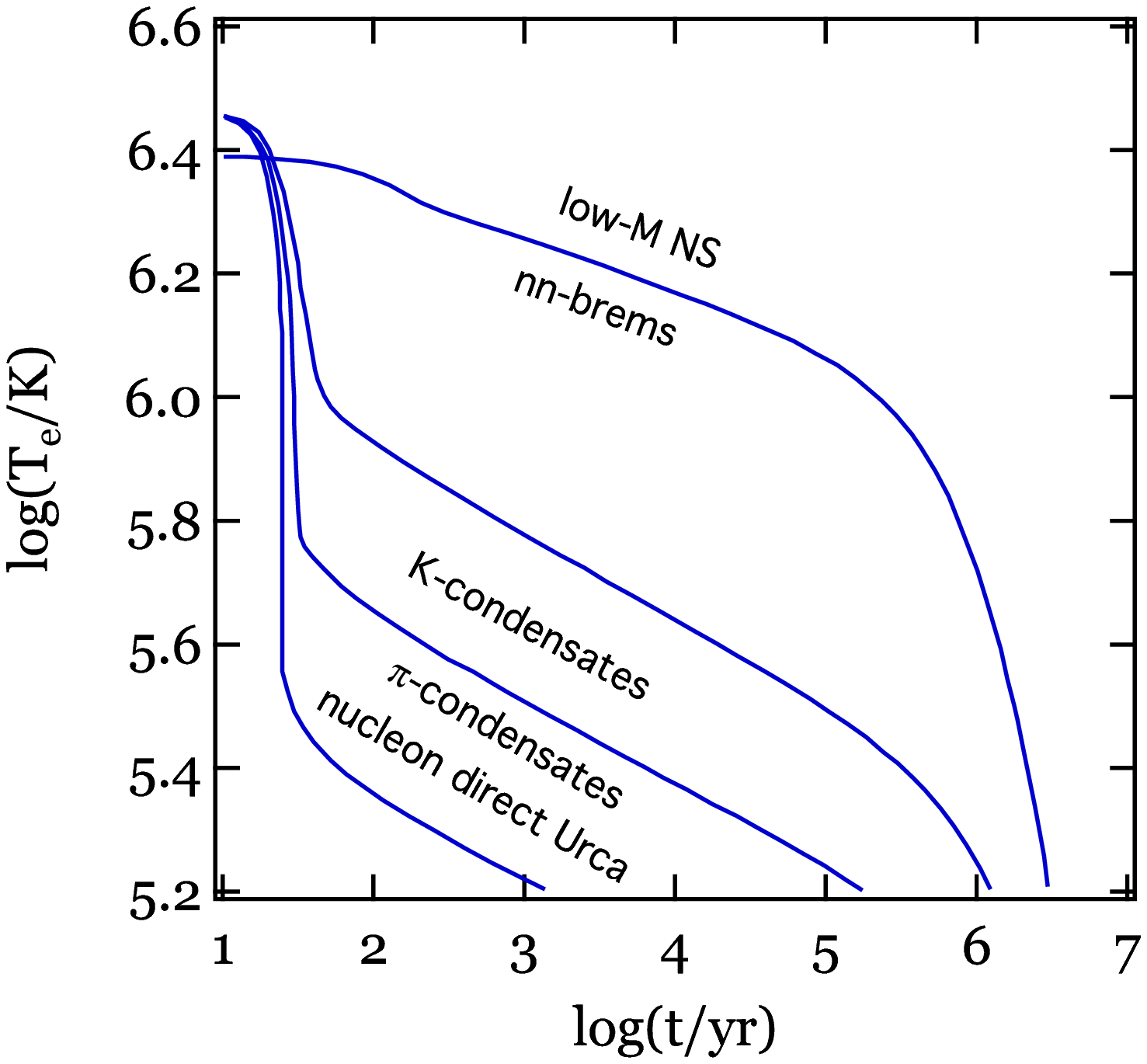}
   \includegraphics[scale=0.55]{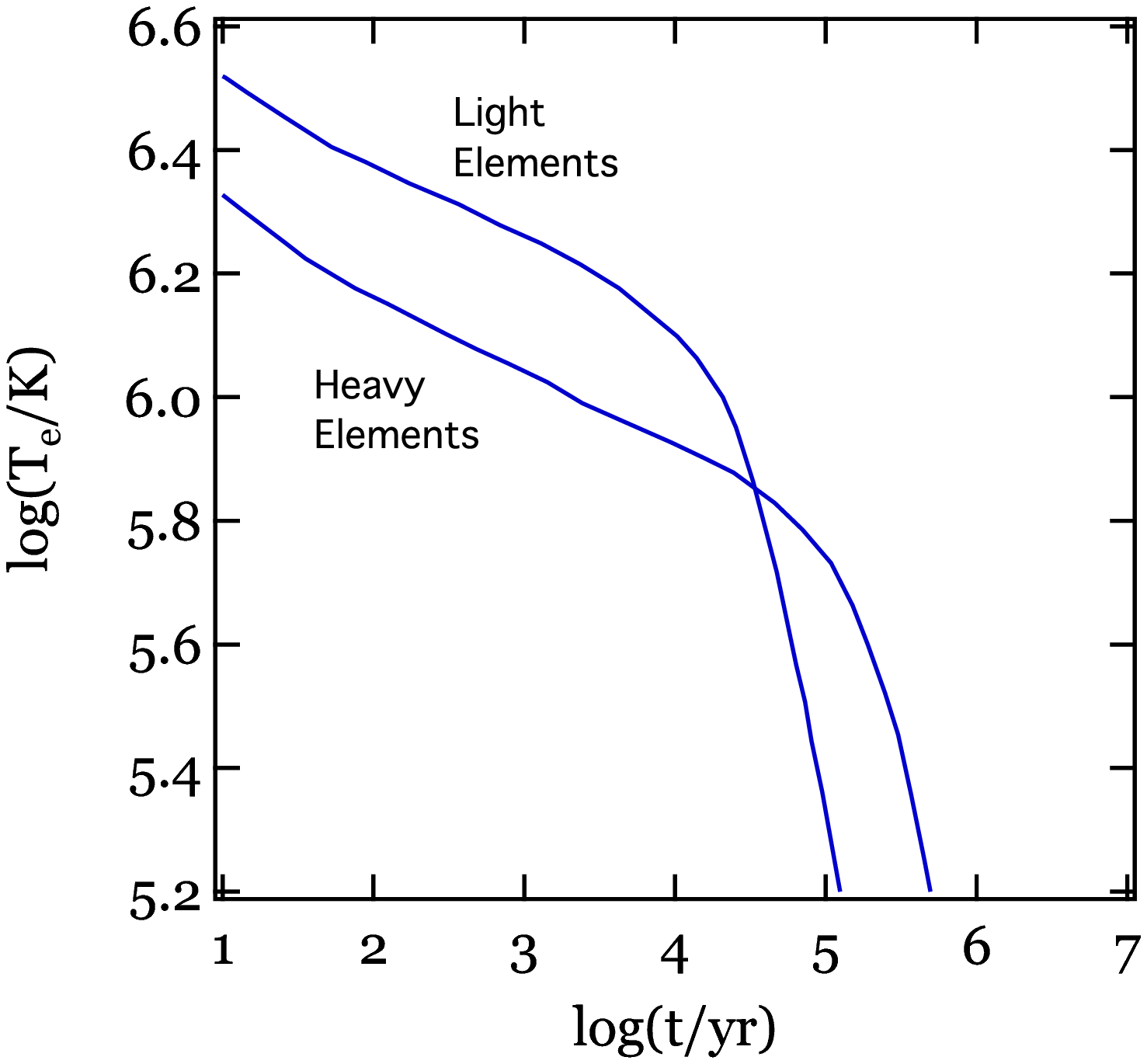}}
\caption{\footnotesize {\em (Left)\/} Cooling curves showing the 
fast and slow cooling as a function of neutron star composition and
mass, adapted from Yakovlev \& Pethick (2004). {\em (Right)\/} Cooling
curves showing the sensitivity of the cooling rate to the composition
of the envelope, adapted from Page et al.\ (2004). By varying the
amount of light elements in the envelope, any trajectory between these
two extremes is possible.}
\label{fig:cooling}
\end{figure}

Gudmundsson et al.\ (1982, 1983) modeled the $T_b-T_{\rm eff}$
relation and found that it most sensitively depends on the properties
of a layer within the envelope where the ions are in liquid phase and
the conductivity is determined by electron conduction. Most
importantly, if this layer contains an appreciable amount of light
elements (up to oxygen), which is allowed given the density and the
temperature of the layer, the effective temperature is higher for a
given base temperature $T_b$ (Potekhin et al.\ 1997). As a result,
neutron stars with light elements in this layer are both brighter and
have shorter cooling timescales than neutron stars that have only
heavy elements in this layer (see Figure~\ref{fig:cooling}). The
magnetic field in the envelope also has a modest effect on the heat
transport, enhancing it along the field lines and suppressing it in
the perpendicular direction. This leads to a non-uniform surface
temperature distribution (Greenstein \& Hartke 1983; Heyl \& Hernquist
1998; Geppert, K\"uker, \& Page 2004). Moreover, for $B >
10^{11-12}$~G, the effective temperature is slightly enhanced for a
given boundary temperature (see Yakovlev \& Pethick 2004).

The temperature of the boundary layer is determined by the thermal
evolution of the core, which is described in general relativity by the
following equations for a spherically symmetric star (Thorne 1977)
\begin{eqnarray}
&&
\frac{e^{-\lambda-2\Phi}}{4 \pi r^2}\frac{\partial}{\partial r}
(e^{2\Phi L_r}) = -Q - \frac{c_T}{e^\Phi}\frac{\partial T}{\partial t} 
\nonumber \\
&&\frac{L_r}{4 \pi \kappa r^2} = e^{-\lambda-\Phi} \frac{\partial (T e^\Phi)}{\partial r}.
\end{eqnarray}
Here $r$ is the Schwarzschild radial coordinate, $\Phi$ is the
gravitational potential such that $\exp(\Phi/c^2)$ specifies the
gravitational redshift, $\exp(\lambda) = [1-2Gm(r)/rc^2]^{-1/2}$ is
the volume correction factor, $m(r)$ is the mass enclosed within a
sphere of radius $r$, and $c$ is the speed of light. In the same
equation, $T$ is the local temperature in the stellar interior, $Q$ is
the neutrino emissivity, $c_T$ is the heat capacity per unit volume,
$\kappa$ is the thermal conductivity, and $L_r$ is the local
luminosity transported through a sphere of radius $r$ excluding
neutrinos. These equations show that the thermal evolution of the core
depends mostly on the heat capacity of the core and the neutrino
emissivity. The major contribution to the heat capacity comes from the
various degenerate constituents of the core, primarily from
non-superfluid neutrons. The heat capacity in a superfluid
neutron/proton core is significantly reduced.

The neutrino emissivity depends on the composition and the properties
of the stellar interior and is characterized as fast or slow depending
on the dominant emission channel. The direct Urca process from
nucleons, hyperons, meson condensates, and quark matter leads to fast
cooling.  Because it has a density threshold, direct Urca operates
only in the inner core. It also requires an appreciable proton
fraction, which introduces a strong dependence on the equation of
state (see Shapiro \& Teukolsky 1986; Yakovlev \& Pethick 2004 and
references therein for a more detailed discussion). Slow neutrino
emission, on the other hand, takes place throughout the core of the
neutron star via modified Urca and nucleon-nucleon bremsstrahlung
processes. The significant effect of baryon superfluidity on neutrino
emissivity is discussed in Yakovlev et al.\ (2001; see references
therein). While the presence of a gap in the baryon energy spectrum
suppresses the rate of neutrino emission, the formation and breaking
of Cooper pairs enhances the neutrino cooling rate.

Models of neutron star cooling fall under three broad categories
depending on the relative importance of the neutrino emission
processes discussed above. Standard cooling models include only the
effects of modified Urca and nucleon-nucleon bremsstrahlung processes
(see Pethick 1992), while the minimal cooling paradigm extends
standard cooling to include the additional neutrino emission from
Cooper pairs in a superfluid core (Page et al.\ 2004, 2009). Fast
cooling models incorporate direct Urca processes and generally result
in lower surface temperatures and shorter timescales over which
thermal emission from cooling neutron stars is detectable. Examples of
cooling curves in Figure~\ref{fig:cooling} show the different cooling
rates via these processes for a variety of interior compositions.

The thermal evolution of a neutron star in all models is characterized
by three stages. For the first 10$-$100 years after the birth of the
neutron star, the stellar envelope thermally relaxes by radiating its
thermal energy and its temperature remains decoupled from that of the
core. In the following $10^5$ years, the temperature of the crust is
set by the temperature of the isothermal core, which cools by neutrino
emission. During this stage, the neutrino luminosity $L_\nu$ remains
much larger than the photon luminosity $L_\gamma$ down to a core
temperature of ${\rm few} \times 10^8$~K. The heat that diffuses
outward from the core to the surface maintains it at a temperature of
$T \sim 10^{5.5-6.5}$~K and gives rise to the thermal emission
observed from young neutron stars. Beyond $10^5$ years, the thermal
evolution of the neutron star enters the photon cooling stage during
which the neutrino luminosity falls below the photon luminosity and
the cooling of the core proceeds via the escape of radiation from the
neutron star surface.

The equation of state of the neutron star interior, its superfluidity,
and the central density affect the time period over which the thermal
emission from the neutron star is observable (see Page et al.\
2004). Nevertheless, a large variety of models predict that a young,
cooling neutron star is detectable for $\gtrsim 10^5$~yr in the soft
X-rays down to a luminosity of $\sim 10^{32}$~erg~s$^{-1}$. I will
discuss the observations of thermal emission from young isolated
neutron stars in Section 3 and the constraints obtained for the
neutron star cooling and interior in Section 5. \\

\noindent {\bf  Reradiation of deposited heat between accretion episodes} \\
Electron capture and pycnonuclear reactions occurring primarily at
densities $\sim 10^{12}$~g~cm$^{-3}$ in the deep crust of a
transiently accreting neutron star release an amount of energy $Q_{\rm
nuc}\simeq 1.5-2$~MeV for each accreted baryon (Haensel \& Zdunik
1990, 2008).  Brown, Bildsten, \& Rutledge (1998) showed that if a
fraction $f$ of this energy is not lost through neutrino emission but
is instead deposited as heat in the stellar interior, then these
reactions maintain the core at a temperature $T \sim (5-10) \times
10^7$~K.  When accretion halts, the envelope relaxes to a thermal
equilibrium set by the flux from the hot core and reradiates the
deposited energy.  Brown et al.\ (1998) calculated the amount of
deposited heat that is thermally reradiated from the core as a
function of the time-averaged accretion rate and found
\begin{equation}
L_q \simeq f Q_{\rm nuc} \frac{<\dot{M}>}{m_u}\simeq 9 \times
10^{32}\; f \; \frac{Q_{\rm nuc}}{1.5~MeV} \frac{<\dot{M}>}{10^{-11}
M_\odot~yr^{-1}} ~{\rm erg~s}^{-1}
\label{eq:lq}
\end{equation}
for typical time-averaged accretion rates of $\sim 10^{-11} \;
M_\odot$~yr$^{-1}$ inferred for low-mass X-ray transients; here, $m_u$
is the mass per nucleon.  For a storage efficiency $f=0.01-1$, this
quiescent luminosity is of the same order as the luminosity observed
from transient neutron stars during the quiescent phases between high
mass accretion episodes, or outbursts, and is, therefore, sufficient
to power the quiescent emission (see Section~5.2 and Page \& Reddy
2006). (Note that the possible contribution from continued accretion
to the luminosity observed during the quiescent phases and the
variability in the quiescent luminosity will be discussed in the next
section). \\

\noindent {\bf Thermonuclear bursts} \\
In weakly magnetic ($B < 10^{10}$~G) accreting neutron stars,
recurrent X-rays flashes with a rise time of $\sim 1$~s and a duration
of $\sim 10$~s are frequently observed. During these so-called Type~I
X-ray bursts, the luminosity emitted from the neutron star surface can
dwarf the accretion luminosity by up to two orders of magnitude. This
phenomenon has been detected from nearly one hundred neutron stars
that accrete hydrogen and helium rich matter from a binary companion
(see Galloway et al.\ 2008 for a recent compilation and Lewin, van
Paradijs, \& Taam 1993 for an earlier review).

The energy source of these flashes has been identified as the unstable
burning of helium via the $3\alpha$ reaction (sometimes in the
presence of hydrogen) in the newly accreted layers on the neutron star
surface. Hansen \& van Horn (1975) were the first to point out that
nuclear burning in shells of fuel on neutron star surfaces are
thermally unstable for a wide range of mass accretion rates. The
association of Type-I bursts with runaway thermonuclear processes is
supported by the agreement between the recurrence time between the
flashes ($\sim$ hours) and the energy generated in each flash (see
Joss 1977; Lamb \& Lamb 1978).

Thermonuclear flashes are triggered when continued accretion
compresses the material accumulating on the neutron star surface,
causing a rise in the density and temperature throughout the freshly
accreted layer.  The recurrence time is set by the ignition conditions
at the bottom of this layer, when the rate of energy generation by the
$3\alpha$ reaction,
\begin{equation}
\epsilon_{3\alpha} = 5.3 \times 10^{21}\; Y^3 
\left(\frac{\rho}{10^5 {\rm g}~{\rm cm}^{-3}}\right)^2
 \left(\frac{T}{10^8~{\rm K}}\right)^{-3} 
\exp\left(\frac{-44 \times 10^8~{\rm K}}{T}\right)
{\rm erg}~{\rm s}^{-1}~{\rm g}^{-1}, 
\end{equation}
which has a very steep dependence on temperature, rapidly exceeds
the cooling rate. Here, $Y$ denotes the helium mass fraction and
$\rho$ and $T$ denote the density and temperature of the material,
respectively. 

The consistency between the total energy released per burst and the
yield expected from the freshly accreted helium layer is established
through an observational comparison between the integrated accretion
flux $F_{\rm accr}$ and the burst fluence $\int F_{\rm burst} dt$
through the parameter $\alpha$:
\begin{equation}
\alpha \equiv \frac{\int F_{\rm accr} dt}{\int F_{\rm burst} dt}. 
\end{equation}
Theoretically, this is expected to be approximately equal to the ratio
of the gravitational potential energy released per nucleon during
accretion to the energy released per nucleon, $E_{\rm nuc}$, during
thermonuclear burning, i.e.,
\begin{equation}
\alpha \approx \frac{GM m_u}{R E_{\rm nuc}} = 
44 \left(\frac{M}{1.4~M_\odot}\right) \left(\frac{R}{10~{\rm
km}}\right)^{-1}
\left(\frac{E_{\rm nuc}}{4.4~{\rm MeV~nucleon}^{-1}}\right)^{-1}.
\end{equation} 
The distribution of $\alpha$ values observed from Type I bursts indeed
peaks at $\sim 40$ (see Galloway et al.\ 2008 for a more detailed
discussion).

There are several mass accretion rate regimes that lead to
thermonuclear flashes with different characteristics (Fujimoto, Hanawa, 
\& Miyaji 1981; Fushiki \& Lamb 1987; Narayan \& Heyl 2003). These 
regimes are typically expressed in units of the Eddington mass
accretion rate
\begin{equation}
\dot{M}_{\rm E} = \frac{8 \pi m_p c R}{(1+X) \sigma_{\rm T}} = 
1.8 \times 10^{-8} \left(\frac{R}{10~{\rm km}}\right)
\left(\frac{1+X}{1.7}\right)^{-1} M_\odot~{\rm yr}^{-1}, 
\end{equation}
where $X$ is the hydrogen mass fraction of the accreted material. Note
that the relevant quantity in determining the stability of nuclear
burning is the local mass accretion rate per unit area, which has been
converted to a total mass accretion rate here by assuming that
accretion is uniform over the neutron star surface. At the lowest mass
accretion rates, $\dot{m} = \dot{M}/\dot{M}_{\rm E} \lesssim 0.01$,
hydrogen burning is unstable and in turn triggers unstable helium
burning. At intermediate mass accretion rates, $0.01 \lesssim \dot{m} 
\lesssim 0.1$, hydrogen burns stably into helium between bursts, forming 
a helium layer at the base of the accreted material. The temperature
of the fuel layer rises until the point of helium ignition is
reached. At high accretion rates, $0.1 \lesssim \dot{m} \lesssim 0.9$,
helium ignites unstably in a hydrogen-rich environment because steady
burning of hydrogen does not proceed fast enough to convert all of the
hydrogen into helium. At even higher mass accretion rates, helium
burning is also stable and thermonuclear flashes are not expected and
rarely observed. \\

\noindent {\bf Magnetic field decay} \\ 
The decay of a very strong magnetic field ($B > 10^{14}$~G) as well as
the release of heat from the stellar crust fracturing under the stress
from such a strong field has been suggested as a source of energy for
neutron star surface emission (e.g., Arras, Cumming, \& Thompson 2004;
Aguilera, Pons, \& Miralles 2008). This is especially relevant for a
group of isolated sources called the Anomalous X-ray Pulsars and the
Soft Gamma-ray Repeaters, from which surface emission is observed both
as persistent emission in the X-rays and in the aftermath of energetic
bursts in the X-ray and $\gamma$-ray bands (see Section 3.6).

Several lines of arguments lead to the magnetic fields as the power
source. First, if the observed rate of spindown is due to magnetic 
braking, then the dipolar magnetic field strength at the stellar
surface is 
\begin{equation}
B = 2 \times 10^{14} \left(\frac{P}{6~{\rm s}}\right)^{1/2} \left(\frac{\dot{P}}
{10^{-11}~{\rm s}~{\rm s}^{-1}} \right)^{1/2}~{\rm G}
\label{eq:dip_field}
\end{equation}
for these neutron stars, given that their spin periods cluster in the
$5-12$~s range while their period derivatives are measured to be
$\dot{P} \sim 10^{-11}$~s~s$^{-1}$ (Kouveliotou et al.\ 1998; see also
Section 3, and the review by Woods \& Thompson 2004). Note that in
individual sources, significant variations in the period derivatives
are measured and, therefore, these dipole magnetic field strengths are
only approximate (see also the discussion in Section 6 and
Figure~\ref{fig:mag_field}).

Second, the observed persistent luminosity, $L_X \approx
10^{34-36}$~erg~s$^{-1}$, significantly exceeds the spindown power
\begin{equation}
\dot{E}_{\rm rot} = I \Omega \dot{\Omega} = 1.8 \times 10^{33} 
\left(\frac{I}{10^{45}~{\rm g}~{\rm cm}^2}\right)
\left(\frac{\dot{P}}{10^{-11}~{\rm s}~{\rm s}^{-1}}\right)
\left(\frac{P}{6~{\rm s}}\right)^{-3}~{\rm erg~s}^{-1},  
\end{equation}
where $I$ is the moment of inertia of the neutron star. Deep searches
for companions or accretion disks around these neutron stars (e.g.,
Hulleman et al.\ 2001; Wang, Kaspi, \& Higdon 2007) rule out accretion
as the source of energy for the persistent emission or the bursts (but
see the arguments for the contribution from accretion in Ertan et al.\
2007, 2009).

Finally, the luminosity during bursts is much higher than the
persistent luminosity, with $E \sim 10^{44}$~erg released over several
hundred seconds, with a pronounced peak within roughly the first
second. The radiative fluxes during the bursts are highly
super-Eddington and require either that the radiating plasma is
confined, most likely magnetically, and/or that the photon-electron
scattering cross sections are radically reduced by the presence of a
strong magnetic field, up to 4 orders of magnitude below the
non-magnetic Thomson cross section (Paczynski 1992; Ulmer 1994; Miller
1995).

The mechanisms for magnetic field decay (e.g., Goldreich \&
Reisenegger 1992), the resulting heating rate, as well as the
properties of neutron star surface emission and thermal evolution have
been investigated in a variety of settings (e.g., Arras et al.\ 2004;
Aguilera et al.\ 2008). These studies have found that energy injection
due to field decay is consistent with the luminosities and the
durations of thermal emission observed from these strongly magnetic
neutron stars. \\

\noindent {\bf Particle bombardment onto polar caps} \\
Relativistic electron-positron pairs produced in the magnetospheres of
pulsars bombard the polar caps on the neutron star surface.  The polar
caps, heated as a result of the particle bombardment, reradiate this
energy as thermal emission peaked in the UV to X-ray wavelengths
(Ruderman \& Sutherland 1975; Arons 1981; Zhang \& Harding 2000;
Harding \& Muslimov 2001). 

An order-of-magnitude estimate for the bombardment rate can be
obtained by considering a density of charges in the magnetosphere
given by the Goldreich-Julian model (1969), moving at the speed of
light, incident on the polar cap that has a radius $R_p$ (Zhang \&
Harding 2000),
\begin{equation}
\dot{N} = c\;N_{\rm GJ}\; \pi R^2_P \approx 10^{30} \left(\frac{R}{10^6~{\rm cm}}\right)^3
\left(\frac{B}{10^{12}~{\rm G}} \right) \left(\frac{P}{1~{\rm s}}\right)^{-2} {\rm s}^{-1}.
\end{equation}
If the positrons have an average Lorentz factor of $\gamma$, then the
rate of energy deposition in the polar cap is proportional to the
bombardment rate and this average Lorentz factor. The detailed
calculations of the Lorentz factors as well as of the rate of energy
deposition depend on the structure of the pulsar magnetosphere and in
the mechanism of particle acceleration, both of which require
numerical modeling (see, e.g., Zhang \& Harding 2000). Under
reasonable assumptions, these models can account for the polar cap
emission observed in the X-rays from a small number of
rotation-powered pulsars.

\subsection{Factors determining the flux, spectrum, and the anisotropy 
of surface emission}

In nearly all situations discussed above, energy generation occurs in
the neutron star crust or in its core, at depths larger than the scale
height of its atmosphere. The only exception is the case of
bombardment of the surface by magnetospheric particles, in which
energy is deposited throughout the neutron star atmosphere.

The observational appearance of a neutron star is determined, however,
by the physical processes that take place in the outermost layer of
its surface, which we call the photosphere. The photosphere
reprocesses the radiation that originates from the deeper layers and
determines the spectrum, anisotropy, and polarization of the radiation
that reaches a distant observer. 

Modeling this surface emission is usually carried out under the
assumption of hydrostatic equilibrium in the atmosphere. This
assumption is valid as long as the flux of radiation emitted from the
neutron star does not lead to a force that exceeds the gravitational
force on the surface. Defining the local Eddington limit as the
luminosity at which the radiation and gravitational forces are in
balance,
\begin{equation}
L_{\rm Edd} = \frac{8 \pi G M m_p c}{(1+X) \sigma_{\rm T}}, 
\end{equation}
provides a useful condition on the validity of the hydrostatic
equilibrium assumption: as long as the radiative flux remains below
the corresponding Eddington flux, the hydrostatic equilibrium
condition is satisfied.

The very large gravitational acceleration on the neutron star surface, 
\begin{equation}
g \approx G M/R^2 \approx 1.9 \times 10^{14} 
\left(\frac{M}{1.4~M_\odot}\right)
\left(\frac{R}{10~{\rm km}}\right)^{-2}~{\rm cm~s}^{-2}, 
\end{equation}
leads to a very small scale height for its atmosphere,
\begin{equation}
h \approx 2 \frac{k_B T}{m_p g} \simeq 8.8
\left(\frac{T}{10^7~{\rm K}}\right)
\left(\frac{R}{10~{\rm km}}\right)^2
\left(\frac{M}{1.4~M_\odot}\right)^{-1}~{\rm cm}, 
\end{equation}
where $T$ is the temperature in the atmosphere, $k_{\rm B}$ is the
Boltzmann constant, $m_p$ is the mass of the proton, and the material
is taken to be fully ionized. This scale height is clearly much
smaller than the radius of the neutron star and justifies the
approximation of a plane parallel atmosphere that is usually employed
in calculations.

The timescale over which photons exchange energy with matter in the
photosphere can be estimated as 
\begin{equation}
t_e = \frac{\frac{\rho}{m_p} k_{\rm B}T}{\chi F} 
\end{equation}
where the numerator is the thermal energy per unit volume in the
atmosphere, $\chi$ is the extinction coefficient, and 
$F=L/4\pi R^2$ is the flux of radiation
through the atmosphere such that the
denominator measures the rate of energy exchange. If we take the 
Thompson cross section to be a characteristic scale for the extinction 
coefficient, this timescale becomes  
\begin{equation}
t_e = 
4 \pi R^2 \frac{k_{\rm B}T}{\sigma_{\rm T} L} \simeq 
2.6 \times 10^{-8} \left(\frac{R}{10~{\rm km}}\right)^2
\left(\frac{T}{10^7~{\rm K}}\right)
\left(\frac{L}{10^{36}~{\rm erg~s}^{-1}}\right)~{\rm s}.
\end{equation}
which is much shorter than the characteristic timescale in which the
interior cools. This is true even in the case of thermonuclear bursts
which occur closest to the surface layers and have cooling times of
the order of a few seconds. Because of this, the outermost layers of
the star are in radiative equilibrium, characterized by constant flux
$F$. Frequently, the constant flux is given in terms of an equivalent
effective temperature defined as $\sigma_{\rm B} T^4_{\rm eff} = F$,
where $\sigma_{\rm B}$ is the Stefan-Boltzmann constant.

In radiative equilibrium, the constant flux of radiation that diffuses
outward from the deep layers determines the temperature gradient in
the atmosphere via the relation
\begin{equation}
F_\nu = -\frac{4 \pi}{3}\left(\frac{1}{\chi_\nu}\frac{\partial B_\nu}{\partial T}
\right) \left(\frac{dT}{dz}\right),
\end{equation}
which is formally valid in the high optical depth limit. Here,
$\chi_\nu$ is the energy dependent extinction (absorption plus
scattering) coefficient, $B_\nu$ is the blackbody function, and all
quantities have been evaluated in the local frame. The extinction
coefficient, in turn, depends on the composition of the neutron star
surface, its density profile, and its magnetic field strength. As a
result, the radiation emerging from the surface in total depends on
four parameters: the effective temperature, the gravitational
acceleration at the surface, the composition, and the magnetic field
strength. Note that any of these parameters may vary across the
neutron star surface, including the gravitational acceleration when
the neutron star is oblate due to fast spin. \\

\noindent{\bf Surface Composition} \\
The composition of the neutron star surface is determined by three
different processes in the formation and the subsequent life of the
star: the abundance and amount of fallback material during the
supernova explosion at birth, the abundance and amount of material
accreted from a binary companion or from the interstellar medium, and
the gravitational settling of the heavy elements. 

Assuming that there is no continuing accretion, the time it takes for
an element of atomic number $A=2Z$ to sediment through an atmosphere
of atomic hydrogen on the neutron star surface is (Brown, Bildsten, \&
Chang 2002 and references therein)
\begin{equation}
t_{\rm sed} \approx 10^5 \left(\frac{g}{10^{14}~{\rm cm~s}^{-2}}\right)^{-2}
\left(\frac{\rho}{10^5~{\rm g~cm}^{-3}}\right)^{1.3}
\left(\frac{T}{10^7~{\rm K}}\right)^{0.3}\;Z^{-0.7}~{\rm s}.
\end{equation}
The density at an electron scattering optical depth of $\tau_{\rm T}$ 
is given by hydrostatic equilibrium and scales as 
\begin{equation}
\rho = \frac{g m_p^2 \tau_T}{2 \sigma_T k_B T}= 1.5 
\left(\frac{g}{10^{14}~{\rm cm~s}^{-2}}\right)
\left(\frac{T}{10^7~{\rm K}}\right)^{-1}
\left(\frac{\tau_T}{10}\right)~{\rm g~cm}^{-3}.
\end{equation}
Inserting this density into the expression for the sedimentation
timescale, we find 
\begin{equation}
t_{\rm sed} = 0.013
\left(\frac{g}{10^{14}~{\rm cm~s}^{-2}}\right)^{-0.7}
\left(\frac{T}{10^7~{\rm K}}\right)^{-1.6}
\left(\frac{\tau_T}{10}\right)^{1.3}
\left(\frac{Z}{8}\right)^{-0.7}~{\rm s}.
\end{equation}
As a result, even an element as light as oxygen can remain in the
photosphere and give rise to atomic lines in the neutron star spectrum
only if there are either no lighter elements present or if it is being
continuously replenished by accretion or by convection, e.g., from the
deeper layer where thermonuclear bursts take place.

Light elements are expected to be present on the neutron star surface
either by fallback or by accretion from a binary companion or from the
interstellar medium. In the case of accretion, light elements can also
be produced by spallation even if the accreted material consists
entirely of heavy elements (Bildsten et al.\ 1992). The amount of
hydrogen that is necessary to cover the surface of a neutron star down
to its photosphere can be estimated, for a non-magnetic neutron star,
by
\begin{equation}
m_{\rm H} = 4 \pi R^2 h N_{\rm p} m_p 
\end{equation}
where $h$ is the scale height of the hydrogen layer above the
photosphere and the gas is assumed to be ionized so that the hydrogen
density is equal to the proton density, $N_H = N_p$. For simplicity,
if we take the electron density $N_e$ to be constant down to an
electron scattering optical depth of unity, i.e., $\tau = N_e \sigma_T
h$, then the mass of hydrogen needed becomes
\begin{equation}
m_{\rm H} = 1.6 \times 10^{-20} \left(\frac{\tau}{1}\right)
\left(\frac{R}{10~km}\right)^2 M_\odot, 
\end{equation}
which is indeed a very small amount of hydrogen. Even for magnetic
neutron stars where the scattering cross section is significantly
reduced (see below), so that the depth of the layer down to an optical
depth of unity in the X-ray band can be 5-6 orders of magnitude
larger, less than $10^{-10} M_\odot$ of hydrogen accreted from the
interstellar medium, a companion, or during the supernova explosion is
sufficient to provide a light element atmosphere that determines the
formation of the surface spectrum.

The typical mass accretion rates for neutron stars in contact binaries
ranges from $10^{-11}~M_\odot~{\rm yr}^{-1}$ to $10^{-8}~M_\odot~{\rm
yr}^{-1}$. Clearly, it takes a trivial amount of time for the required
amount of hydrogen to be accumulated on the neutron star surface in
this situation. We can also estimate the amount of time it takes for a
neutron star to accrete this amount of hydrogen from the interstellar
medium. Using the Bondi-Hoyle formalism, the mass accretion rate is
given by
\begin{equation}
\dot{M}_{\rm BH} = \frac{4 \pi (GM)^2 \rho_{\rm ISM}}{v^3},
\end{equation}
where $\rho_{\rm ISM}$ is the density of the interstellar medium and
$v$ denotes the spatial velocity of the neutron star. (Note that
accretion rates in realistic situations, as predicted by detailed
numerical models, often have additional scalings; see Ruffert 1996 and
the discussion in Perna et al.\ 2003). For typical values of the
parameters of the interstellar medium and of the neutron star
velocities (Arzoumanian et al.\ 2002), the mass accretion rate becomes
\begin{equation}
\dot{M}_{\rm BH} = 10^{-17} 
\left(\frac{M}{1.4~M_\odot}\right)^2
\left(\frac{\rho_{\rm ISM}}{2 \times 10^{-24}~{\rm g~cm}^{-3}}\right)
\left(\frac{v}{100~{\rm km~s}^{-1}}\right)~M_\odot~{\rm yr}^{-1}. 
\end{equation}
The timescale for the accretion of sufficient hydrogen to cover the
stellar surface, therefore, ranges between $<< 1$~yr for a
non-magnetic neutron star to $\sim 10^3$~yr for a highly magnetic one
(in which the photosphere occurs at larger column densities, see
below), unless there is a physical reason, such as the propeller
mechanism (llarionov \& Sunyaev 1975; Romanova et al.\ 2004), that
inhibits accretion onto the neutron star surface. Coupled with the
short timescale for the sedimentation of heavy elements, this result
justifies the dominant use of hydrogen and helium compositions when
modeling the emission from neutron star surfaces. \\

\noindent {\bf Magnetic Field Strength}\\
The magnetic field of a neutron star has profound effects on the
properties of its surface and, therefore, on the emission originating
from it. Magnetic field strengths can vary between $<10^{8-9}$~G in
steadily accreting or recycled neutron stars to $\sim 10^{15}$~G in
magnetars (see the census of sources in the next section). The lower
end of this range is considered ``unmagnetized'' due to the negligible
effect of the magnetic field on the properties of the plasma, the
hydrodynamics, the structure of the atoms, or the photon-electron
interaction cross sections. In the upper end of this range, on the
other hand, the magnetic force dominates over the other forces at play
in the surface layers (with the exception of gravity) and becomes the
most important parameter that determines the emission properties.

The electron cyclotron energy 
\begin{equation}
E_{cycl, e} = \frac{\hbar e B}{m_e c} = \hbar {\omega_{c,e}} =
11.6\;\left(\frac{B}{10^{12}~{\rm G}}\right)~{\rm keV}
\label{eq:cycl}
\end{equation}
is a natural unit to use when quantifying the strength and the effects
of the magnetic field because it can easily be compared to the
electron rest energy, the Fermi energy, the Coulomb energy, and the
thermal energy of the matter in the surface. For example, even at the
fairly modest field strength of $10^{12}$~G, the cyclotron energy
exceeds by an order of magnitude the typical surface thermal energies
of $\lesssim 1$~keV.  The field strength at which the cyclotron energy
equals the electron rest energy defines the quantum critical field
\begin{equation}
B_Q \equiv \frac{m_e^2 c^3}{e \hbar} = 4.4 \times 10^{13}~{\rm G}, 
\end{equation}
beyond which we find a regime in which new physical processes start to
take place.  Finally, comparison of the electron cyclotron energy to
the characteristic energy scale in the atom $E_{\rm atom} = 1~{\rm
Ryd} = e^2 /(2 a_0) = 13.6$~eV, where $a_0$ is the Bohr radius, helps
define the regime in which the cyclotron energy significantly exceeds
the typical Coulomb energy. This happens at a field strength $B \simeq 
1.2 \times 10^9$~G, above which magnetic field effects considerably 
alter the properties of atoms and molecules.

At $E_{\rm cycl} \leq E_{\rm atom}$, a perturbative treatment of the
effects of the magnetic field gives rise to the well-known Zeeman
splitting of atomic energy levels.  In the opposite regime, Coulomb
forces act as a perturbation to magnetic forces. The electrons are
confined to the ground Landau level and the Coulomb force binds the
electrons along the magnetic field direction. This results in two
important changes in the atomic structure: the atoms become
cylindrical and the spacing between the energy levels becomes
larger. A simplified treatment for the hydrogen atom gives 160~eV for
the ground state energy at $10^{12}$~G and 540~eV at $10^{14}$~G,
compared to 13.6~eV for the non-magnetic case (see the review of
Harding \& Lai 2006 and references therein for the approximations and
a detailed discussion).

The cylindrical atoms can form covalent bonds along the magnetic field
direction to make up linear molecular chains, which in turn can form
three dimensional condensates. The corresponding dissociation energies
are smaller than the ionization energies and scale as $E_{\rm dis} =
[\ln(E_{\rm cycl} /E_{\rm atom})]^2$.

\begin{figure}
\centering
   \includegraphics[width=7.6cm,height=7cm,clip=true]{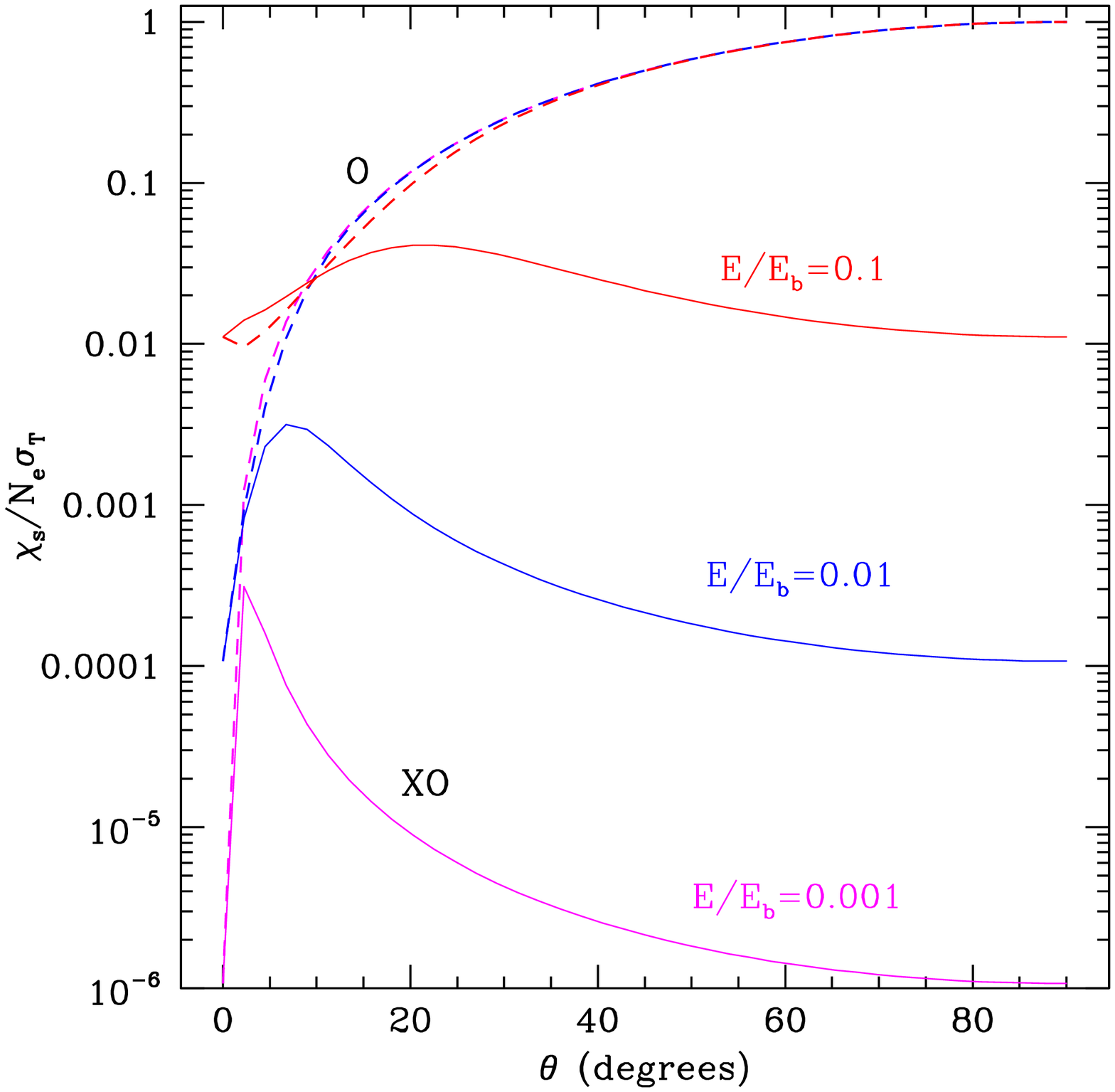}
   \includegraphics[width=7.6cm,height=7cm]{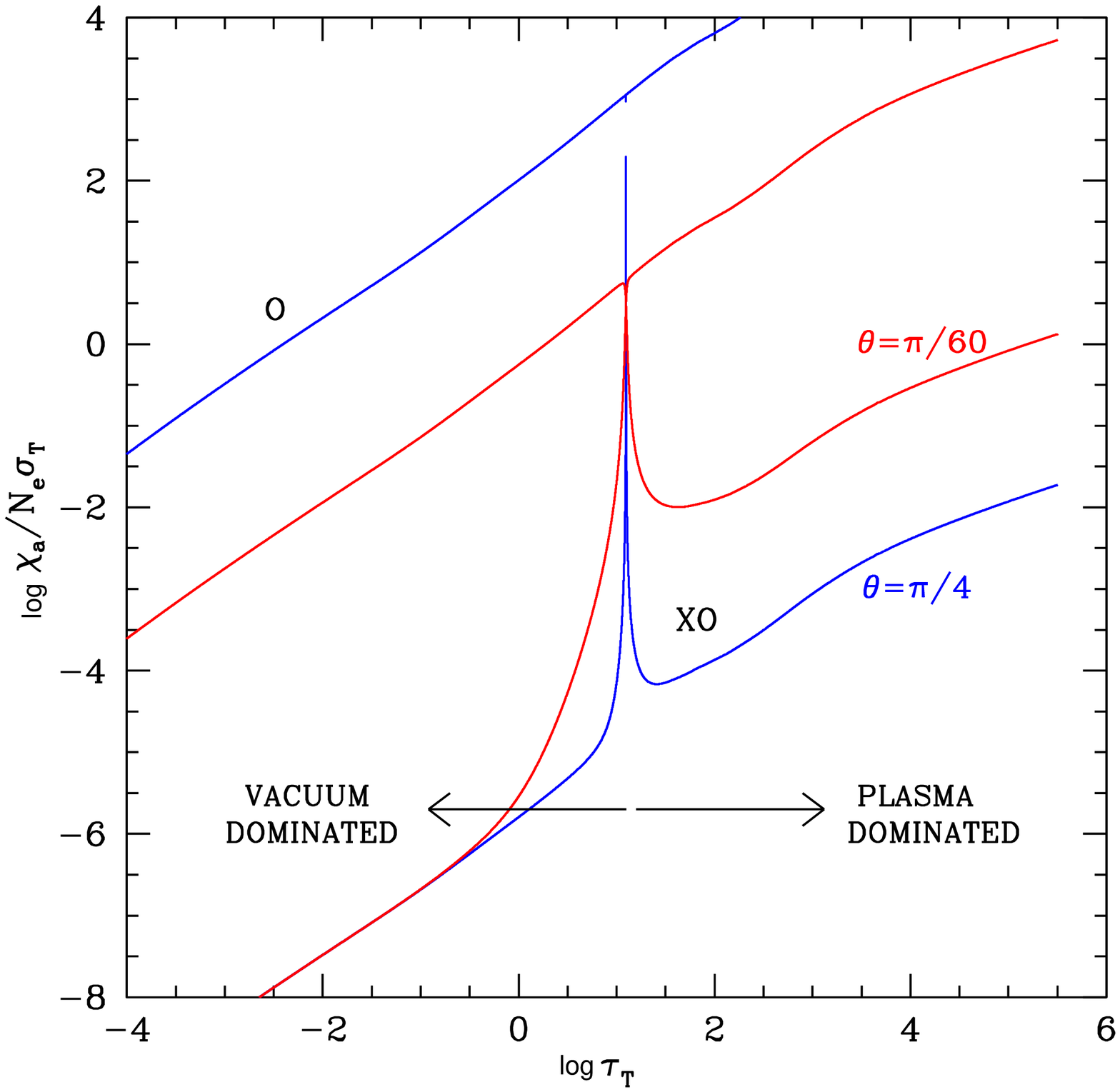}
\caption{\footnotesize {\em (Left)\/} The extinction coefficient for 
scattering for the extraordinary (solid line) and the ordinary (dashed
line) modes as a function of the direction of propagation $\theta$
with respect to the magnetic field orientation, for different values
of the photon energy $E$ in units of the cyclotron energy, which is
denoted here as $E_b$. {\em (Right)\/} The absorption coefficients of
the extraordinary (XO) and ordinary (O) modes as a function of Thomson
optical depth in the atmosphere of a neutron star for two directions
of propagation, for a photon energy of 0.5 keV in a $10^{15}$~G
magnetic field (adapted from \"Ozel 2001). For each mode, the two
curves show different directions of photon propagation with respect to
the magnetic field. The sharp feature at $\log \tau_{\rm T} \simeq 1$
occurs at the transition from plasma-dominated to vacuum-dominated
interaction cross sections.}
\mbox{}
\label{fig:mag_op} 
\end{figure}

Last but not least, the magnetic field strength drastically affects
the propagation of photons through the neutron star atmosphere and
hence the photon-electron interaction cross sections. Because the
electrons are confined to the ground Landau level in the direction
perpendicular to the magnetic field, the interaction cross sections
exhibit a very strong dependence on polarization and the direction of
propagation. In most circumstances, photon propagation can be
described in terms of two orthogonal polarization modes. The cross
section for the so-called extraordinary (or perpendicular) mode, which
describes the case where the electric field vector of radiation is
perpendicular to the magnetic field, is suppressed by a factor
$(E/E_b)^2$ with respect to the non-magnetic case; see
Figure~\ref{fig:mag_op}. The cross section for the ordinary (parallel)
mode remains comparable to the non-magnetic case for direction of
propagation perpendicular to the magnetic field orientation but is
also suppressed when the photons are propagating along the field
(Gnedin \& Pavlov 1974; Pavlov \& Shibanov 1979; \"Ozel 2001; Ho \&
Lai 2001; van Adelsberg \& Lai 2006).

The interaction of the photons with the electrons and protons in the
magnetic atmosphere is characterized by strong resonances at the
cyclotron energy, which leads to potentially detectable absorption
features in the surface spectra. The electron cyclotron energy is
given in equation~(\ref{eq:cycl}) and falls within the 0.5-10~keV
X-ray band, where neutron star surface emission is often detected, if
the magnetic field strength is between $5 \times 10^{10}$~G to
$10^{12}$~G (see section 2.4 for a discussion of gravitational
redshifts). The cyclotron energy for the protons is a factor $m_p/m_e$
higher than the electron cyclotron energy. Therefore, for a proton
cyclotron feature to fall in the soft X-ray band, the surface magnetic
field needs to the between $10^{14}$~G to $2 \times 10^{15}$~G.

One other process that influences the propagation of photons in the
outermost layers of a neutron star atmosphere in a strong magnetic
field is the polarization of the magnetic vacuum itself. When the
plasma density is low, the photons interact primarily with virtual
electron-positron pairs in the vacuum. The polarization of the virtual
pairs in the presence of the strong magnetic field leads to a
significant change in the permeability and dielectric tensor of the
vacuum with a magnitude that scales as $(B/B_Q)^2$ (Adler
1971). Because of this dependence, vacuum polarization effects are
significant at field strengths $B \gtrsim B_Q$. In the presence of a
plasma with a density gradient, which is characteristic of neutron
star atmospheres, vacuum polarization gives rise to a resonance when
the normal modes of photon propagation change from being mostly
circularly polarized at high electron densities (deeper in the
atmosphere) to being mostly linearly polarized at low densities. Thus,
at a critical density that depends on photon energy, the conversion of
photons between the two polarizarion modes is highly enhanced,
accompanied by a change in the opacities of the normal modes (Adler
1971; Tsai \& Erber 1975; Mezsaros \& Ventura 1979; Kaminker et al.\
1982; see Meszaros 1992 for a review). These resonances modify the
emerging spectrum and give rise to broad absorption-like features
(Bulik \& Miller 1997; \"Ozel 2001, 2003; Lai \& Ho 2002).

\subsection{Atmosphere Models}

The properties of the radiation emerging from a neutron star surface
have been investigated in different settings, taking into account the
various considerations of flux, composition, and magnetic field
strength discussed above. Emission models, such as those discussed
below, typically employ as parameters the effective temperature
$T_{\rm eff}$, magnetic field strength $B$, and surface gravity $g$,
in addition to specifying the composition of the material making up
the atmosphere.

The justification of the frequently employed assumptions of
plane-parallel geometry, radiative equilibrium, and hydrostatic
equilibrium conditions were discussed in the previous section.  The
ionization state of the atmosphere is determined by solving the
(magnetic) Saha equation. We can broadly classify the results
depending on the effective temperature of the atmosphere and the
magnetic field of the neutron star. 

\begin{figure}
\centering
   \includegraphics[width=7.6cm,height=7cm,clip=true]{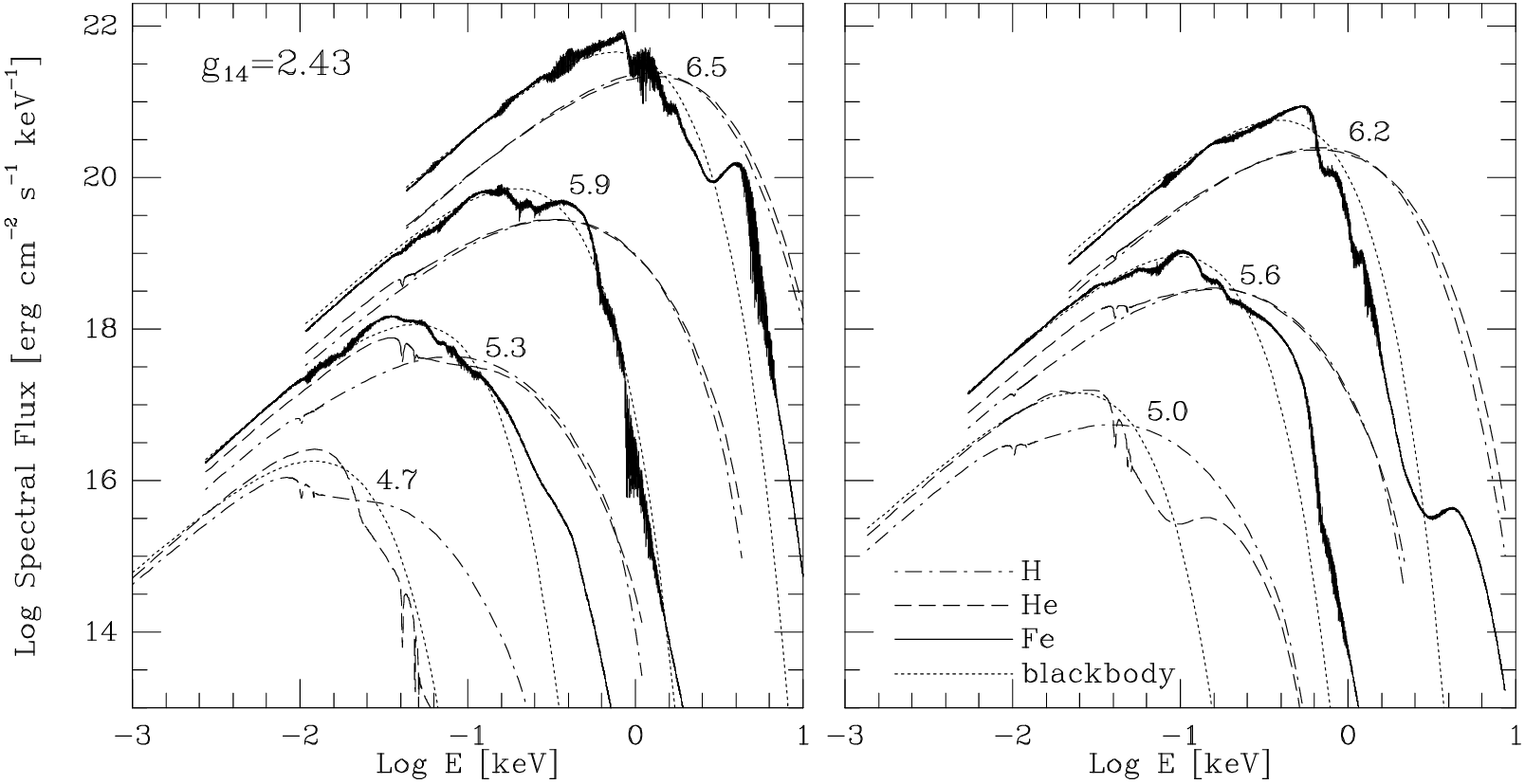}
   \includegraphics[width=7.6cm,height=7cm]{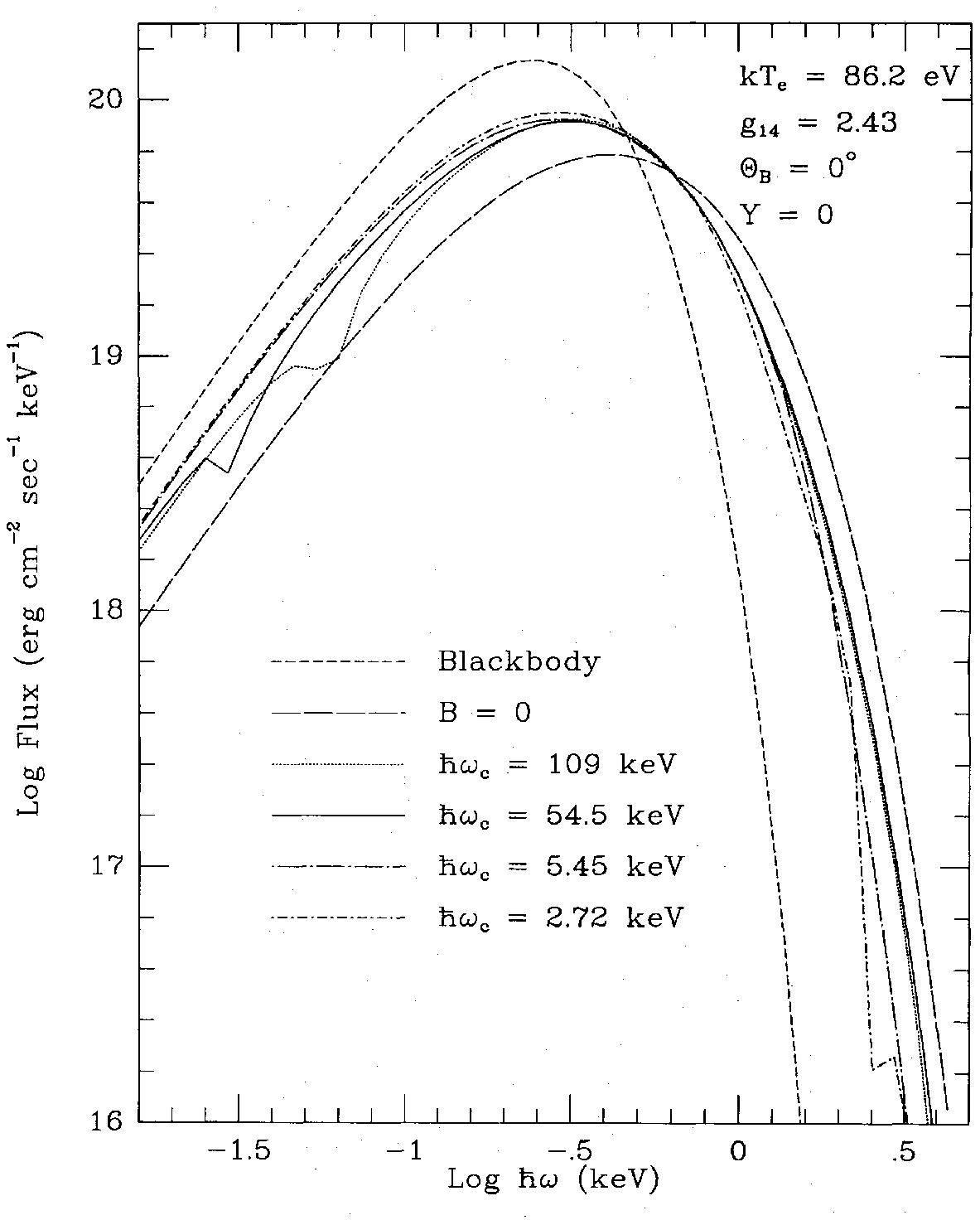}
   \includegraphics[width=7.9cm,height=6.9cm,clip=true]{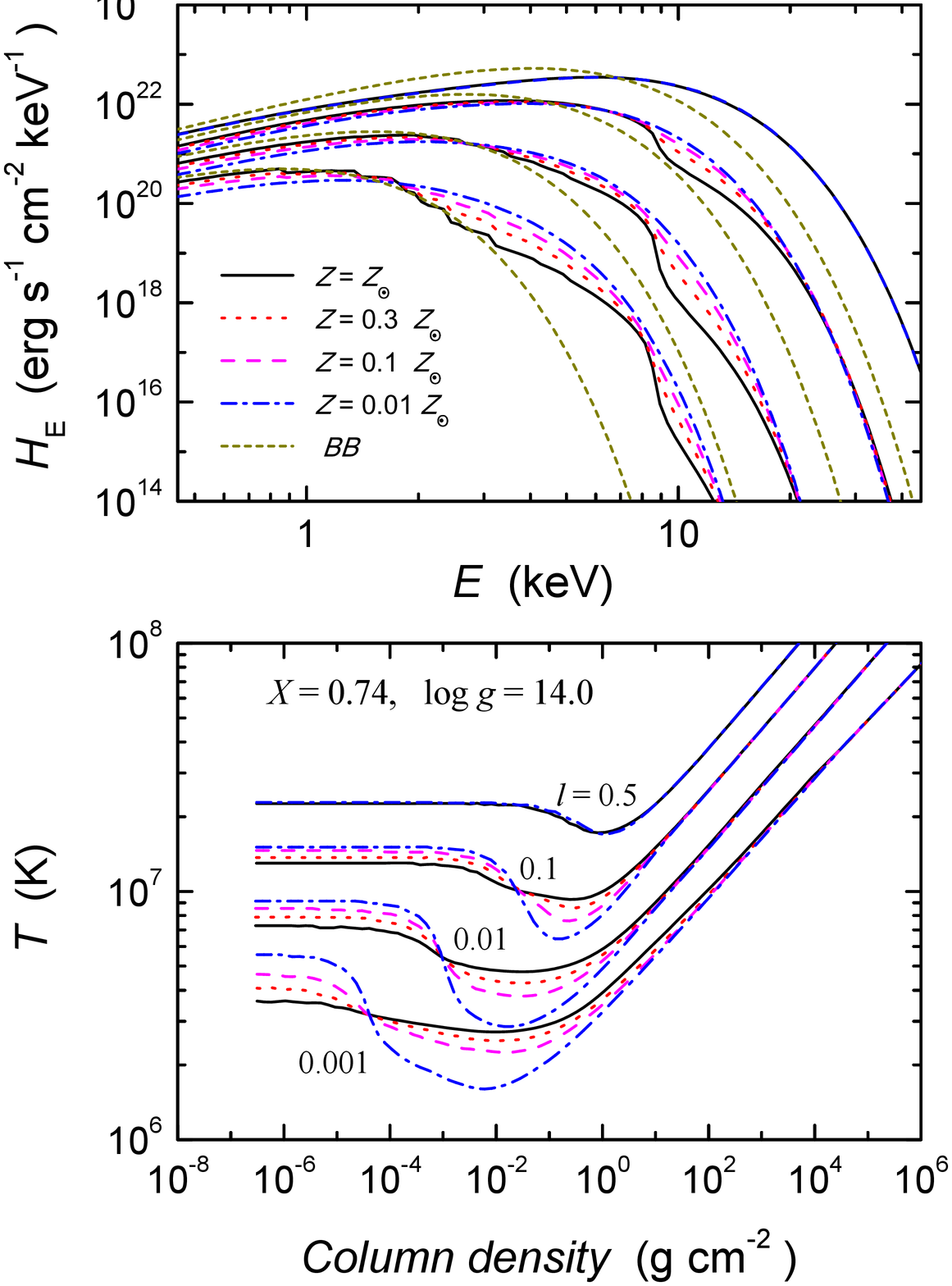}
   \includegraphics[width=7.6cm,height=6.9cm,clip=true]{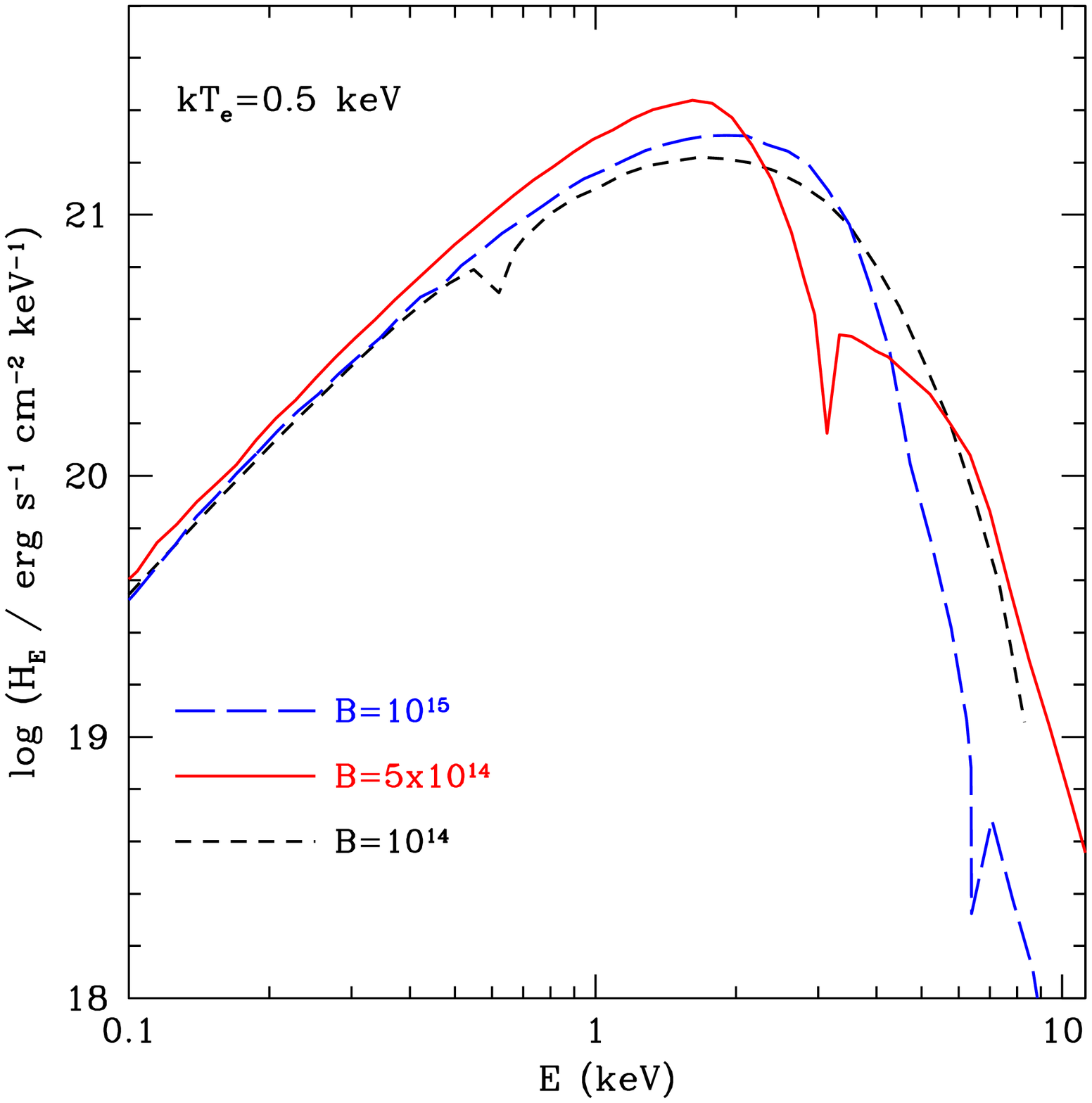}
\caption{\footnotesize Model spectra for four different types of neutron 
stars. In all panels, the emerging flux is shown against photon energy
in keV.  The Eddington flux $H_E$ is related to the spectral flux
$F_E$ by $F_E = 4 \pi H_E$. A range of results are shown for different
compositions, magnetic field strengths, and effective temperatures
(labeled in K or $\rm{eV}=1.16 \times 10^4$~K). {\em (Top Left)\/}
non-magnetic, cool neutron stars; each group of curves correspond to a
different effective temperature ($T_{\rm eff} = 10^{4.7-6.5}$~K) with
a range of compositions: dash-dotted lines represent H, dashed lines
represent He, and solid lines represent Fe. Dotted lines show the
blackbody at the effective temperature (Zavlin et al.\ 1996) {\em (Top
Right)\/} cool neutron stars with moderate magnetic field strengths
($B = 2 \times 10^{11}-10^{13}$G) and hydrogen composition, for an
effective temperature $T_{\rm eff} = 86.2$~eV (Shibanov et al.\ 1992)
{\em (Bottom Left)\/} hot bursting neutron stars with a range of
metallicities as shown with different line styles and colors; each
group of curves corresponds to a different effective temperature
(Suleimanov et al.\ 2011) {\em (Bottom Right)\/} strongly magnetic
neutron stars for a hydrogen compositions and different magnetic field
strengths (\"Ozel 2003). In all cases, the spectra show significant
deviations from the blackbodies at the corresponding effective
temperatures. }
\mbox{}
\label{fig:ns_spec} 
\end{figure}

In weakly magnetic neutron stars and at high effective temperatures,
e.g., in the atmospheres of bursting neutron stars, the thermal energy
$E_{\rm th}$ is much larger than the ionization energy $E_{\rm ion}$
for the lighter elements as well as the majority of the (possibly
present) heavier elements, causing them to be ionized. Only a few
ionization states of heavy elements may survive. In non-magnetic
neutron stars with lower surface temperatures, on the other hand, most
heavy elements remain neutral or at low ionization states, leading to
numerous features in the spectra from atomic transitions (see
Figure~\ref{fig:ns_spec}). In isolated sources with high magnetic
field strengths, such as magnetars, a competition between the high
ionization energies and the high temperatures/densities at the
photosphere determine the ionization equilibrium. Due to suppressed
cross sections, the photosphere reaches deeper in the atmosphere where
temperatures can reach $\sim 1$~keV and particle densities $N \sim
10^{29}~{\rm cm}^{-3}$. Thus, despite the larger energies required for
ionization in strong magnetic fields, the ionized fractions usually
remain high for light elements.

The ionization state of the atmosphere and the magnetic field strength
determine the relevant processes that need to be taken into account in
the solution of the radiative transfer equation. These include
electron scattering, free-free emission and absorption
(Bremsstrahlung), as well as bound-bound and bound-free atomic
transitions where applicable. The cross sections of interactions
depend on the temperature profile of the atmosphere, which, in turn,
needs to be determined self consistently with the radiation field for
the radiative equilibrium condition to be satisfied. Iterative
techniques such as lambda iteration, complete linearization, or
Lucy-Uns\"old scheme are utilized to achieve consistency between the
temperature profile and the radiation field (see, e.g., Mihalas 1978
for a description of these interactions and numerical techniques). \\

\noindent{\bf Spectrum of the Surface Emission} \\
Figure~\ref{fig:ns_spec} shows representative spectra of radiation
emerging from neutron star atmospheres under different physical
conditions. Four cases are shown: non-magnetic, cool neutron stars
($T_{\rm eff} = 10^{4.7-6.5}$~K; Zavlin et al.\ 1996); cool neutron
stars with moderate magnetic field strengths ($B = 2 \times
10^{11}-10^{13}$G; Shibanov et al.\ 1992), hot bursting neutron stars
with different metallicities (Suleimanov et al.\ 2011), and strongly
magnetic neutron stars (\"Ozel 2003).

The characteristic feature of all model neutron star spectra is a
deviation from the blackbody function of the same effective
temperature. Because of the strong energy dependence of the free-free
and bound-free absorption coefficients, the depth of the photosphere
increases with increasing photon energy. As a result, radiation that
emerges at higher photon energies originates deeper in the atmosphere
where the temperature is also higher. This causes a hardening and
broadening of the spectrum compared to a black body at the effective
temperature. Such distortions are often characterized by a parameter
called the color correction factor defined as
\begin{equation}
f_c = \frac{T_{\rm BB}}{T_{\rm eff}}, 
\label{eq:fc}
\end{equation}
where $T_{\rm BB}$ is the temperature obtained from fitting the
spectrum with a blackbody. The hardening and broadening in the spectra
give rise to color correction factors that are in the range $1
\lesssim f_c \lesssim 1.8$. Color correction factors near the low end 
of this range are rare in model spectra and are almost always caused
by the presence of broad and/or numerous absorption lines (see, e.g.,
the upper left panel of Figure~\ref{fig:ns_spec}).

\begin{figure}
\centering
   \includegraphics[width=7.6cm,height=7cm,clip=true]{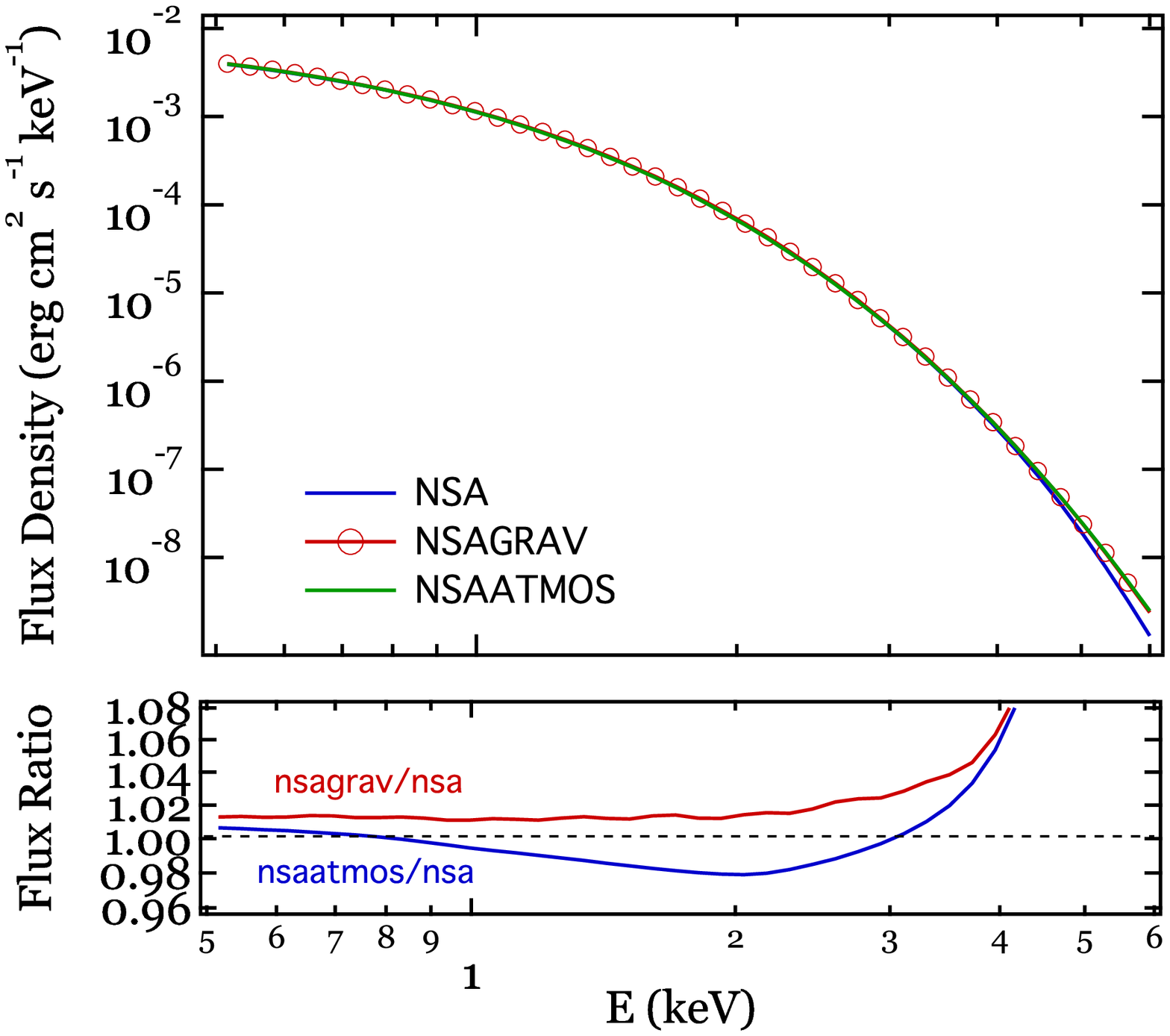}
   \includegraphics[width=7.9cm,height=7cm]{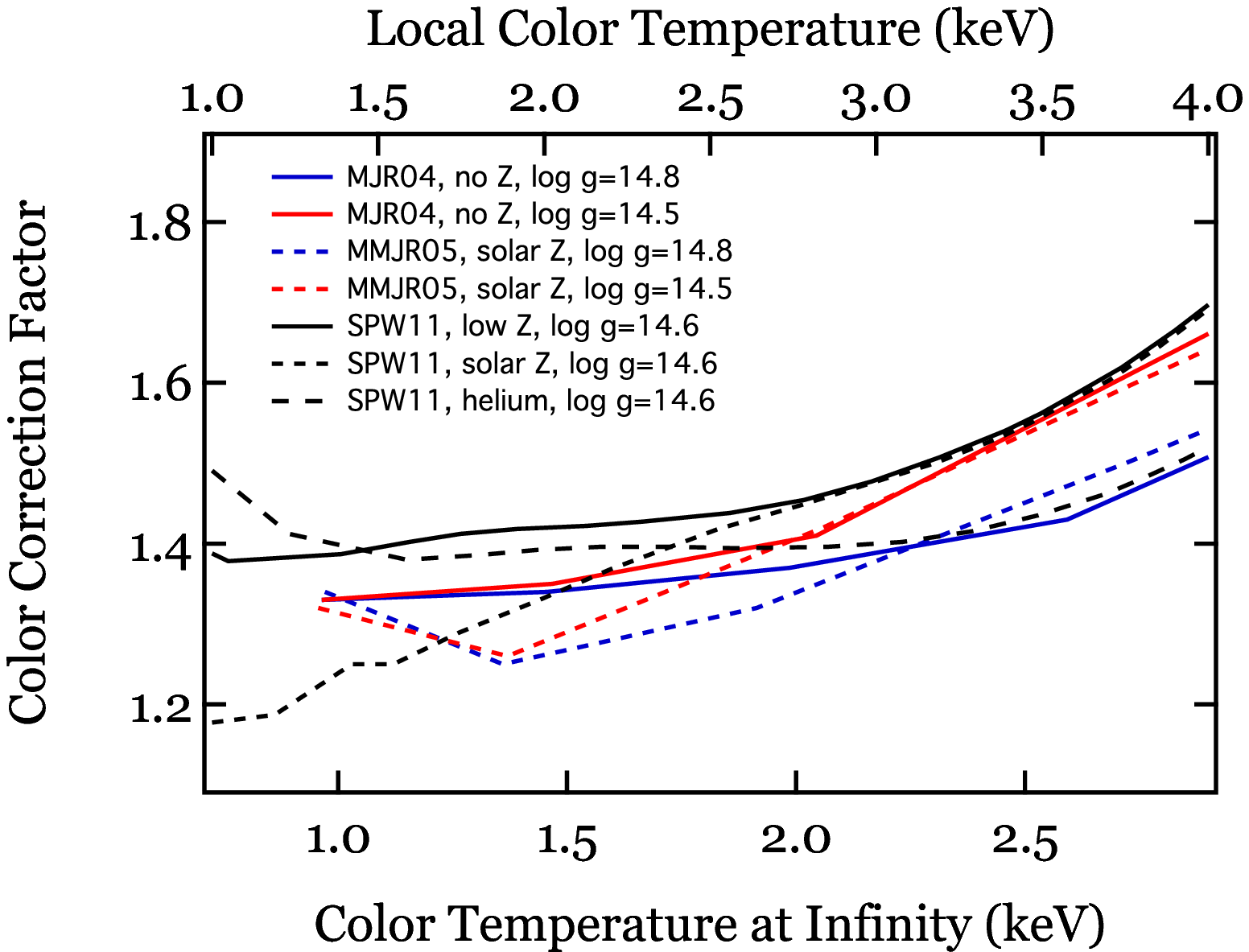}
\caption{\footnotesize {\em (Left)\/} A comparison between three different 
non-magnetic neutron star atmosphere models computed for $T_{\rm eff}
= 120$~eV.  The three models shown are NSA, NSGRAV (Zavlin et al.\
1996), and NSATMOS (Heinke et al.\ 2006). {\em (Right)\/} A comparison
between the color correction factors obtained for atmosphere models
for accreting bursting neutron stars for different effective
temperatures, surface gravities, and atmospheric compositions. The
models shown are from Madej et al.\ 2004 (MJR04), Majczyna et al.\
2005 (MMJR05), and Suleimanov et al.\ 2011 (SPW11). Different model
calculations for the same input model parameters generate predictions
that agree with each other at the 2-7\% level. }
\mbox{}
\label{fig:atmos_models} 
\end{figure}

All atmosphere models satisfy the radiative equilibrium condition by
employing iterative schemes between the radiation field and the
temperature and density in the atmosphere.  Different numerical models
rely on a variety of numerical techniques to achieve this and often
implement convergence criteria with different levels of precision.  In
addition, there are variations between how the effects of neutron star
surface gravity are taken into account, i.e., whether the macroscopic
properties of the neutron star and the surface gravity are allowed to
vary consistently. Finally, there exist some differences in the
physical processes taken into account in separate calculations, which
may involve making simplifying assumptions, e.g., in the treatment of
angular redistribution in scattering, or accounting for additional
physics such as electron conduction. Because of these variations in
technique and the treatment of processes, different model calculations
may produce spectra that deviate from each other when computed for the
same effective temperature, magnetic field strength, and surface
composition.

As an example, Figure~\ref{fig:atmos_models} shows a comparison of
three different non-magnetic hydrogen atmosphere models computed for
$T_{\rm eff} = 120$~eV, for a neutron star with a radius of 10~km and
a $1.4~M_\odot$ mass, assuming no interstellar extinction. All three
models are available on the spectral analysis software XSPEC under
the names NSA, NSGRAV (Zavlin et al.\ 1996), and NSATMOS (Heinke et al.\
2006) and are widely used to fit the observed spectra of neutron
stars. The lower panel shows the fractional differences between the
models, which are most evident at the high energy tail. In general,
however, these model calculations generate predictions that are
consistent at the 2-6\% level, depending on the photon energy.

As a second example, Figure~\ref{fig:atmos_models} shows the color
correction factors obtained for atmosphere models of bursting neutron
stars for different effective temperatures and atmospheric
compositions based on calculations of different groups (MJR04: Madej
et al.\ 2004; MMJR05: Majczyna et al.\ 2005; SPW11: Suleimanov et al.\
2011). These models are, on average, hotter than those shown in the
left panel, and the spectral shapes are dominated by the effects of
Compton scattering. The color correction factors shown in the figure
provide only a gross characterization of the spectral shapes and the
actual values depend on the technique used to calculate them (see
Suleimanov et al.\ 2011 for a discussion). Nevertheless, the
dependence of the color correction factors on the physical conditions
and on the particular numerical techniques is weak and the differences
are limited to $\lesssim 7\%$.

Absorption features in the spectra result from bound-free atomic
transitions, cyclotron absorption, or vacuum polarization. As
expected, at higher metallicities and at lower temperatures, where
higher fraction of neutrals may be present, there are large numbers of
bound-free absorption lines. Light elements such as hydrogen and
helium can also give rise to spectral features at low effective
temperatures, especially at high magnetic field strengths ($B \gtrsim
10^{12}$~G), where their ionization energies are significantly higher
than in the non-magnetic case.

In the presence of strong magnetic fields, electron or proton
cyclotron features also appear in neutron star surface spectra. When
the field strength is in the $10^{11-12}$~G range, the electron
cyclotron energy falls in the soft X-ray band, while in the
$10^{14-15}$~G range, it is the protons that produce the absorption
features in the spectra. Proton cyclotron features at different
magnetic field strengths are shown in the lower right panel of
Figure~\ref{fig:ns_spec}. At these strong fields, vacuum polarization
resonance also plays an important role in shaping the neutron star
surface spectra and affects the features in two ways. First, the
conversion of photon polarization modes as they propagate outward to
lower plasma densities and the enhanced interaction cross sections
associated with this resonant conversion give rise to very broad
absorption features in the tail of the spectra and reduce the overall
hardening. Second, vacuum polarization dramatically suppresses the
equivalent widths of cyclotron lines (\"Ozel 2003; Ho \& Lai 2003; see
Figure~\ref{fig:ns_spec}).

There are a number of mechanisms that determine the widths of atomic
features in the thermal spectra of neutron stars. Pressure broadening
due to the Stark effect is caused by the shifting and splitting of
atomic lines from an atom in the presence of the static electric
fields from the neighboring ions. Considering the first-order effect
that is linear in the electric field, Paerels (1997) and Chang,
Bildsten, \& Wasserman (2005) estimated the amount of broadening due
to this effect as
\begin{equation}
\Delta E_{\rm Stark} \simeq \frac{\hbar^2 n^2}{m_e Z r_0^2} \simeq 1.5 \; 
\left(\frac{Z}{26}\right)^{-1} \left(\frac{n_i}{10^{23}~{\rm cm}^{-3}}\right)^{2/3}
\left(\frac{n}{3}\right)^2~{\rm eV},
\end{equation}
where $Z$ is the atomic number of the atom, $r_0$ is the mean ion
spacing, $n_i$ is the number density of ions, and $n$ is the principal
quantum number of the upper level of the transition. This can be
compared to the thermal Doppler broadening
\begin{equation}
\Delta E_D = \left(\frac{2k_B T}{A m_p c^2}\right)^{1/2} E_0 
\simeq 0.20 
\left(\frac{k_B T}{1~{\rm keV}}\right)^{1/2}
\left(\frac{A}{56}\right)^{-1/2}
\left(\frac{E_0}{1~{\rm keV}}\right)~{\rm eV},  
\end{equation}
where $A$ is the mass number of the atom and $E_0$ is the energy of
the transition. For rapidly spinning neutron stars, both of these
effects can be negligible compared to rotational broadening (\"Ozel \&
Psaltis 2003; Chang et al.\ 2006), which has a characteristic scale of
\begin{equation}
\Delta E_{\rm rot} = \frac{4 \pi \nu_s R}{c} E_0 \simeq
240 \left(\frac{\nu_s}{600~{\rm Hz}}\right)
\left(\frac{R}{10~{\rm km}}\right)
\left(\frac{E_0}{1~{\rm keV}}\right)~{\rm eV},  
\end{equation}
where $\nu_s$ is the spin frequency of the neutron star. Finally, in
the presence of a magnetic field, Zeeman splitting of the atomic levels 
is responsible for additional broadening with a scale given by (see Loeb 
2003)
\begin{equation}
\Delta E_Z \simeq \left(\frac{e \hbar}{m_e c}\right) (M_L + 2 M_S) B
\simeq 5.8 (M_L + 2 M_S)\left(\frac{B}{10^9~{\rm G}}\right)~{\rm eV},  
\end{equation}
where $M_L$ and $M_S$ are the quantum numbers of the orbital angular
momentum and the spin of the electron. \\

\noindent{\bf Angular Distribution (Beaming) of the Surface Emission}

\begin{figure}
\centering
   \includegraphics[width=7.6cm,height=7cm,clip=true]{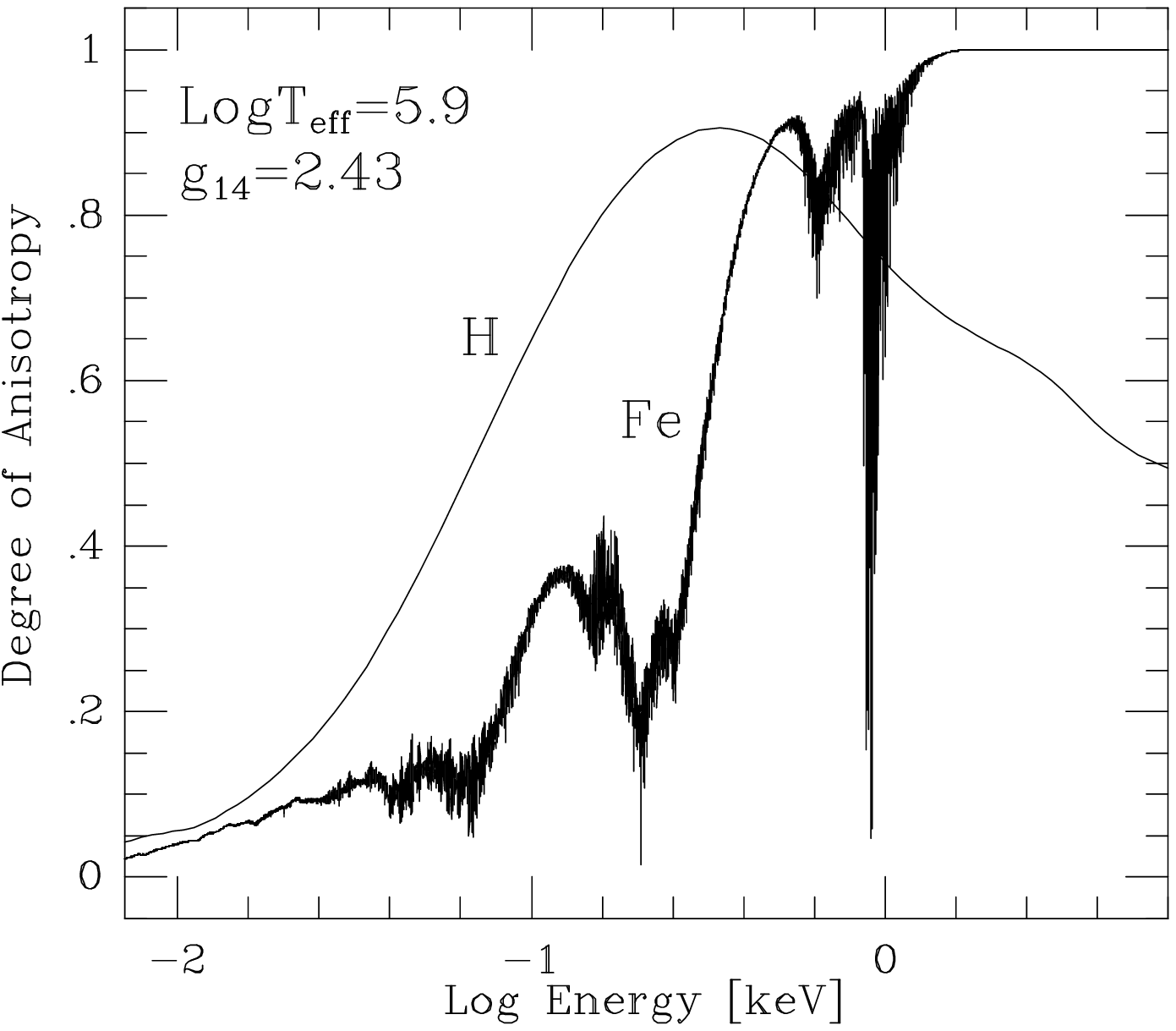}
   \includegraphics[width=7.6cm,height=7.3cm,clip=true]{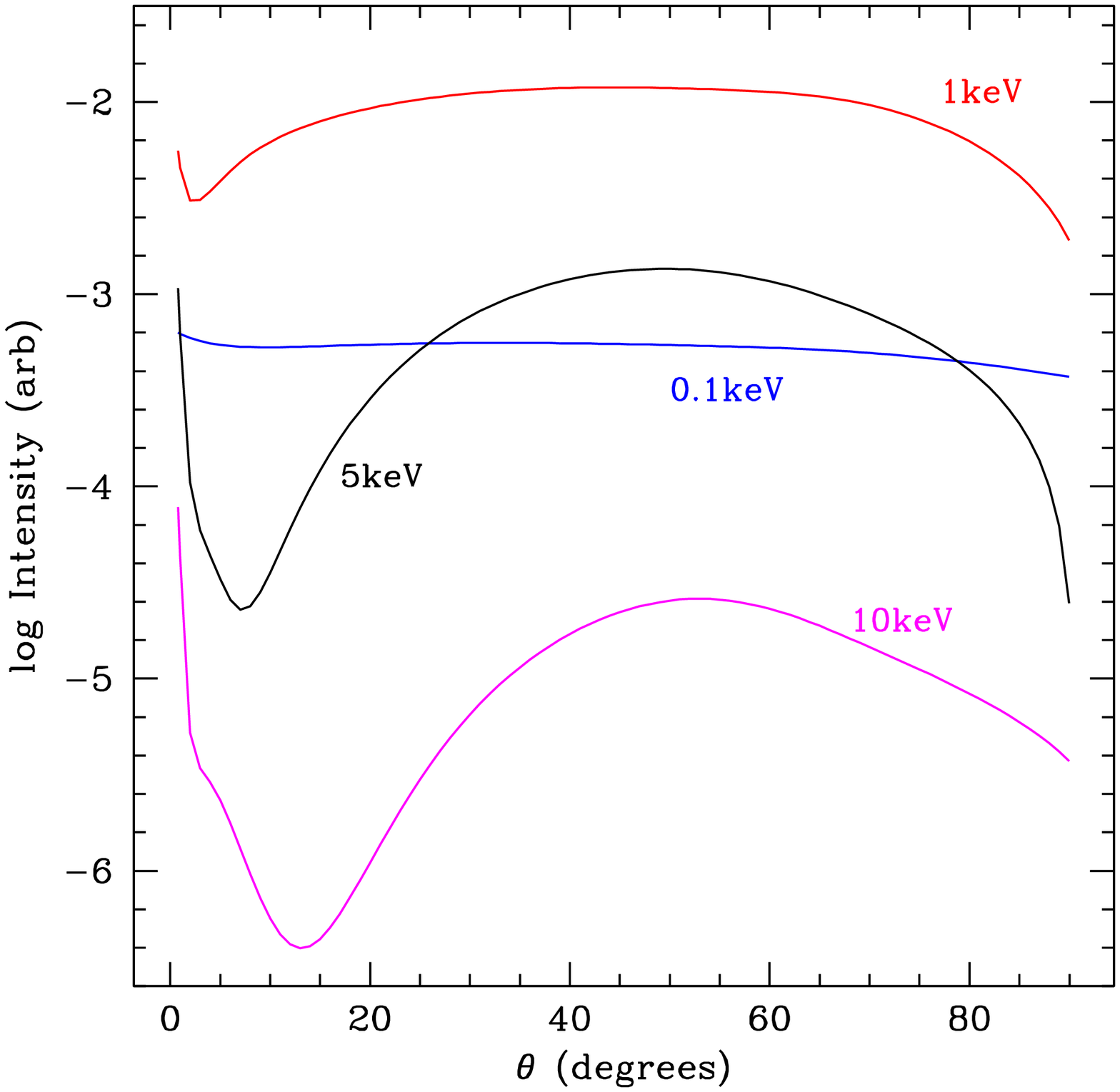}
\caption{\footnotesize {\em (Left)\/} The beaming of radiation emerging
from the surface of a non-magnetic neutron star for different
atmospheric compositions (Zavlin et al.\ 1996). The degree of
anisotropy is defined in equation~(\ref{eq:doa}) in the text. The
complex structure of the degree of anisotropy in the iron atmosphere
reflects the strong and non-monotonic energy dependence of the
interaction cross sections, which causes a variable degree of limb
darkening as a function of photon energy. {\em (Right)\/} The beaming
of radiation emerging from the surface of a $10^{14}$~G magnetar at
four different photon energies (\"Ozel 2001). The characteristic
``pencil'' and ``fan'' components are visible, especially at the
higher photon energies.}
\mbox{}
\label{fig:ns_beam} 
\end{figure}

Various physical phenomena render the intensity of radiation emerging
from the neutron star surface anisotropic. The large temperature
gradient in the atmosphere, in combination with energy dependent cross
sections for the interactions between the photons and matter,
scattering processes, as well as the presence of a strong magnetic
field can cause this anisotropy, also referred to as
beaming. Figure~\ref{fig:ns_beam} shows the beaming of emission from a
non-magnetic and a strongly magnetic isolated neutron star. For the
non-magnetic case, the degree of anisotropy, defined as
\begin{equation}
a_\nu = \frac{I_\nu(\mu=1)-I_\nu(\mu=0)}{I_\nu(\mu=1)+I_\nu(\mu=0)}
\label{eq:doa}
\end{equation}
is plotted against the photon energy for two compositions of the
atmosphere (Zavlin et al.\ 1996). Here, $\mu$ is the cosine of the
angle between the photon direction of propagation and the surface
normal and $I_\nu(\mu=1)$ and $I_\nu(\mu=0)$ are the
frequency-dependent intensities of the emerging radiation along and
perpendicular to the surface normal, respectively. The complex
structure of the degree of anisotropy in the iron atmosphere reflects
the strong and non-monotonic energy dependence of the interaction
cross sections, which causes a variable degree of limb darkening as a
function of photon energy.

In the strongly magnetic atmosphere case, the figure shows the angular
dependence of the emerging radiation for four different photon
energies (\"Ozel 2001).  The beaming is characterized by the presence
of a narrow ``pencil'' component, along the magnetic field direction,
as well as a broad ``fan'' component, at large angles from the field
direction (see Meszaros 1992). The orientations with the highest
intensity in the two components correspond to the directions of
propagation with the reduced opacities for the two normal modes of
polarization.

The anisotropy of radiation emitted from neutron star surfaces has a
direct effect on the observed properties of spinning neutron stars: it
alters the broadening of the spectrum and the amplitude of the
pulsations due to general relativistic effects, which are discussed
below.

\subsection{Observed Physical Quantities}

The observational appearance of a neutron star is affected by its
spacetime, since photons that originate from the neutron star surface
get redshifted and lensed as they propagate to a distant observer.
Even though the combined effects of redshift and lensing, in general,
can only be calculated numerically (see Pechenick, Ftaclas, \& Cohen
1983), several observable properties of a slowly spinning neutron star
that is emitting uniformly depend only on a single element of its
metric, $g_{tt}$, evaluated on its surface (Psaltis 2008). This is
true for any general metric theory of gravity. In particular, in the
Scwarzschild geometry, the time-time component of the metric is given
by $g_{tt}=-(1-2GM/Rc^2)$, where $M$ and $R$ are the mass and the
coordinate radius of the neutron star, respectively. As a result, the
gravitational effects on the observational appearance of a slowly
spinning neutron star can be characterized by the single parameter
$p\equiv 2GM/Rc^2$, which is often called the compactness. For
different equations of state of neutron-star matter, the predicted
compactness of a neutron star ranges from $\sim 0.25$ to $\sim 0.65$,
with the higher values corresponding to the stars that are more
massive and smaller in size.

Photons that propagate from the surface of a slowly spinning neutron
star to a distant observer are gravitationally redshifted by an amount
\begin{equation}
1+z=\left\vert g_{tt}\right\vert^{-1/2}=\left(1-\frac{2GM}{Rc^2}\right)^{-1/2}\;.
\end{equation}
At the same time, they are lensed by the strong gravitational field of
the neutron star, causing its apparent area at infinity to appear
larger by an amount equal to (see Psaltis, \"Ozel, \& DeDeo 2000)
\begin{equation}
\frac{S_\infty}{4\pi R^2}=(1+z)^2=\left(1-\frac{2GM}{Rc^2}\right)^{-1}\;,
\end{equation}
and its apparent radius to appear larger by 
\begin{equation}
R_{\rm app} =R \left(1-\frac{2GM}{Rc^2}\right)^{-1/2}\;.
\label{eq:rapp}
\end{equation}
If we denote by $T_{\rm eff, s}$ the effective temperature of radiation
on the neutron star surface, then the effective temperature measured
by a distant observer is also redshifted, i.e.,
\begin{equation}
T_{{\rm eff}, \infty}=\frac{T_{\rm eff, s}}{1+z}
\end{equation}
so that the total flux observed from the neutron star at a large
distance $D$ becomes
\begin{equation}
F_\infty= \sigma_{\rm B} T_{{\rm eff}, \infty}^4
\left(\frac{R}{D}\right)^2\left(1+z\right)^2 =\sigma_{\rm B} T_{{\rm
eff},\infty}^4\left(\frac{R}{D}\right)^2
\left(1-\frac{2GM}{Rc^2}\right)^{-1}\;.
\end{equation}
Finally, the Eddington luminosity at infinity is also affected by both
the redshift and the gravitational lensing of the photons and is equal
to
\begin{equation}
L_{{\rm E},\infty}=\frac{8\pi m_{\rm p}c}{(1+X)\sigma_{\rm T}}R^2 g_{\rm eff}
\left(1+z\right)^{-2}\;,
\end{equation}
were, $g_{\rm eff}$ is the effective gravitational acceleration on the
neutron star surface. In general relativity, $g_{\rm eff}=(1+z)GM/R^2$
and the Eddington luminosity becomes
\begin{equation}
L_{{\rm E},\infty}=\frac{8\pi G M m_{\rm p}c}{(1+X)\sigma_{\rm T}}
\left(1-\frac{2GM}{Rc^2}\right)^{1/2}\;.
\label{eq:ledd}
\end{equation}

Increasing the spin of a neutron star affects its observable
properties in at least three ways. First, Doppler boosting of the
emission makes the apparent image of the neutron star asymmetric, with
the approaching side appearing brighter than the receeding side, and
broadens any spectral features that originate on its surface. Second,
frame dragging alters the propagation of photons in the neutron star
spacetime and contributes to enhancing the asymmetry of
emission. Finally, as the spin frequency of a neutron star approaches
the breakup frequency, the star becomes significantly oblate.  Several
authors have explored, at different levels of approximation, the
effects of increasing the spin of a neutron star on several
observables such as the lightcurves that arise when the surface
emission on a spinning neutron star is not uniform (Miller \& Lamb
1998; Braje, Romani \& Rauch 2000; Muno, \"Ozel \& Chakrabarty 2003;
Poutanen \& Gierlinski 2003; Cadeau, Leahy \& Morsink 2005; Cadeau et
al. 2007; Morsink et al. 2007), the rotational broadening of atomic
lines that originate on the stellar surfaces (\"Ozel \& Psaltis 2003;
Bhattacharyya, Miller \& Lamb 2006; Chang et al. 2006), as well as
their apparent surface areas (Baub\"ock et al.\ 2012).

\section{A Census of Neutron Star Sources with Surface Emission}

Surface emission has been detected from a large variety of neutron
star sources to date. The spectra of this emission are thermal or have
a distinguishable thermal component that peaks at wavelengths from the
optical regime to the X-rays. The sources include isolated neutron
stars and accreting neutron stars in binaries with ages ranging from
hundreds to billions of years, as summarized in Table~\ref{tab:main}.
In this section, I will discuss the properties of the sources that
show emission from their surfaces.

%%%%%%%%%%%%%%%%%%%%%%
%%%%   Table 0  %%%%%%
%%%%%%%%%%%%%%%%%%%%%%

\begin{table} 
\caption{\label{tab:main}Categories of Neutron Stars with Surface Emission} 
\begin{indented} 
\item[]
\begin{tabular}{@{}lcccccc} 
\br 
Source Type                  & Other Names$^a$ & Age&Temp.&Mag.\ Field& Companion? & Pulsations? \\
	                     &                 &(yr)&($10^6$~K) &    (G)       &            &       \\    
\mr 
% Table: Basic Properties of Neutron Star Source Types
%
%  Source Type	      Age              
%  	            	               
%---------------------------------------------------------------------
Quiescent Accreting NS & qLMXB &$\gtrsim 10^9$ &$0.6-2$  & $\leq 10^{9}$ &  Y & N \\
Bursters               & LMXB  &$\gtrsim 10^9$ &$10-35$  & $\leq 10^{9}$ &  Y & S$^b$ \\
Accreting msec Pulsars & AMSP  &$\gtrsim 10^9$ &$5-10 $  & $10^{8-9}$ &  Y & Y \\
Rotation Powered MSP   & MSP   &$\sim 10^{9-10}$    &$2-8  $  & $10^{8-9}$ &  S & Y \\            
Isolated Cooling NS    & PSR   &$10^{3-5.5} $    &$0.6-2$  & $10^{11-12}$ &  N & Y \\
     -''-              & DINS, CCO &$10^{5.5-6.6}$ &$0.6-1.3$ & $?$ &  N & S \\
Magnetars              & AXP, SGR &$10^{5.5-6.6}$ &$3-8$ & $10^{14-15}$ &  N & Y \\                     
\br 
\end{tabular} 
Notes: (a) Typical related acronyms often used in the literature for these categories;  
qLMXB: quiescent low-mass X-ray binaries, LMXB: low-mass X-ray binaries, AMSP: accretion 
powered milisecond pulsars, MSP: millisecond pulsars, PSR: pulsating sources in the radio, 
DINS: dim isolated neutron stars, CCO: central compact objects, AXP: anomalous X-ray 
pulsars, SGR: soft gamma-ray repeaters. (b) S: sometimes.
\end{indented} 
\end{table}

\subsection{Accreting Neutron Stars in Quiescence}

About half of known neutron stars accreting from a low-mass companion
star manifest themselves as X-ray transients (see Liu et al.\ 2007).
X-ray transients experience several different recurrent accretion
phases, from quiescence to outburst, which are distinguished by
varying flux levels and spectral characteristics. The outburst
luminosities of these accreting neutron stars are typically
$10^{36-38}$~erg~s$^{-1}$, compared to their quiescent luminosities,
which are of the order $10^{36-38}$~erg~s$^{-1}$.  At the high mass
accretion rates during the outbursts that last $\sim$ months, the
emission is dominated by the accretion disk. On the other hand, during
quiescence when mass accretion onto the neutron star either ceases or
continues at very low levels, the observed X-ray emission originates
primarily from the neutron star surface. Other general properties of
this class of sources is shown in Table~\ref{tab:main}.

%%%%%%%%%%%%%%%%%%%%%%
%%%%   Table 1  %%%%%%
%%%%%%%%%%%%%%%%%%%%%%

\begin{table} 
\caption{\label{tab:quies}Accreting Neutron Stars in Quiescence}
\begin{indented} 
\item[]
\begin{tabular}{@{}lcccc} 
\br 
Name & Temperature   & Distance &  Luminosity$^a$  & References$^b$ \\
     &   log($T$/K)    &  (kpc) & ($10^{32}$~erg~s$^{-1}$) &                \\
\mr 
% Table: Accreting Neutron Stars in Quiescence
%
%  Common Name	      kT         Distance & Luminosity & References
%   or Prefix	            	               
%---------------------------------------------------------------------
NS in $\omega$~Cen&  6.0           &  5.3  & 4.9   & 1  \\	
NS in M13&	     6.0           &  7.7  & 5.1   & 1  \\	
NS in NGC~2808&	     6.0           &  9.6  & 10.0  & 1  \\	
U24 in NGC~6397&     6.0           &  2.5  & 1.1   & 2  \\
X7 in 47 Tucanae&    6.1-6.2       &  4.9  & 15.0  & 3  \\
\hline
KS 1731$-$260&	     6.08$-$5.86   &  7.0$^c$  & 24-4 & 4, 5  \\	
MXB 1659$-$29&       6.15$-$5.80   &  10.0$^c$ & 49-2 & 4, 6  \\	
XTE J1701$-$462&     6.28$-$6.15   &  8.8$^c$  & 166-50 & 7  \\
EXO 0748$-$676&      6.15$-$6.1    &  7.4$^c$  & 100-60 & 8  \\	
\hline
\br 
\end{tabular}
\newline
Notes: (a) The unabsorbed bolometric luminosity of the thermal
component of the observed emission. (b) References. 1. Webb \& Barret
2007; 2. Guillot et al.\ 2011; 3. Heinke et al.\ 2006; 4. Cackett et
al. 2006; 5. Cackett et al. 2010; 6. Cackett et al. 2008;
7. Fridriksson et al. 2011; 8. Degenaar et al. 2011. (c) Approximate
distances.\\ 
\end{indented} 
\end{table}

\begin{figure}
\centering
   \includegraphics[scale=0.60,trim=20pt 430pt 200pt 30pt,clip=true]{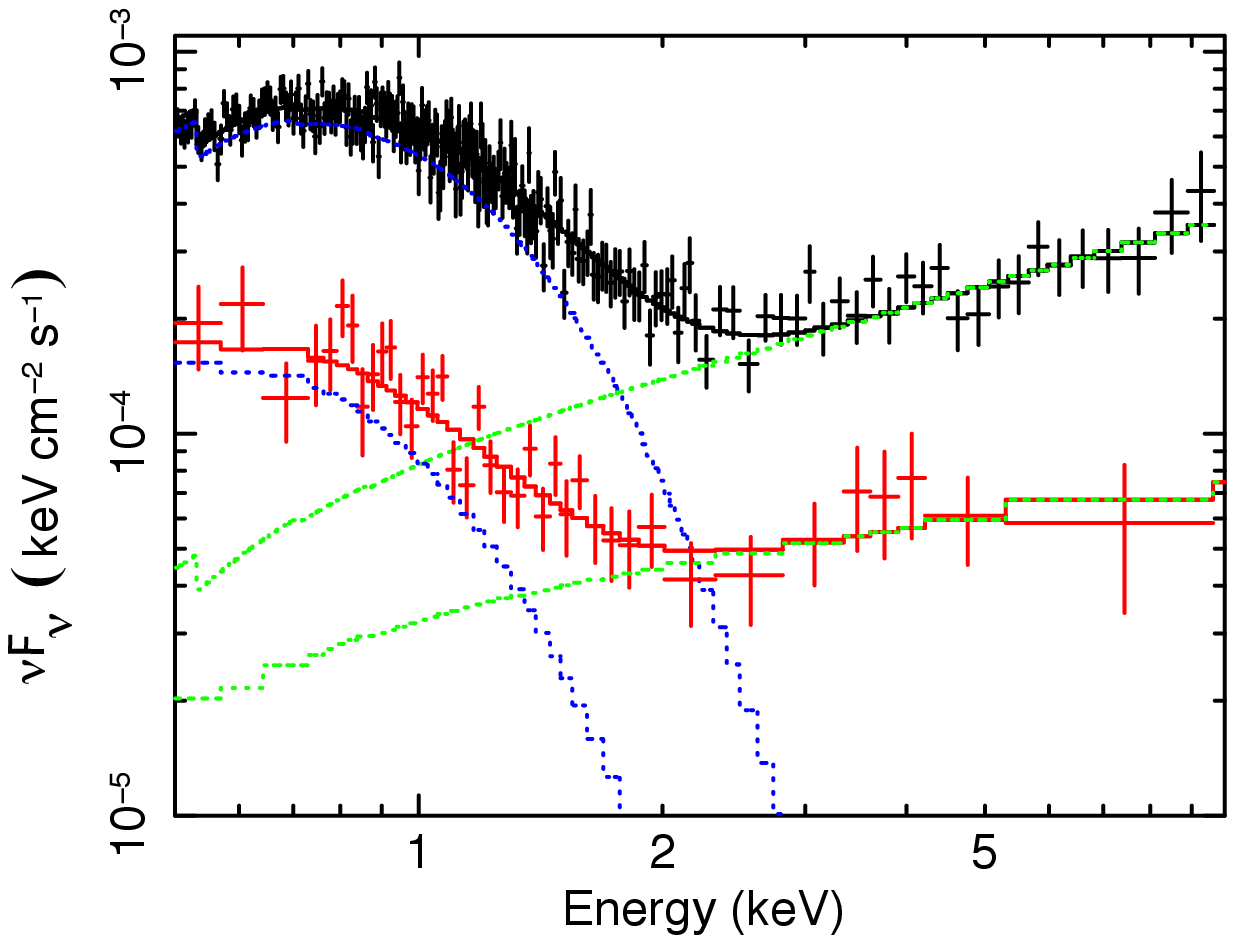}
   \includegraphics[scale=0.35]{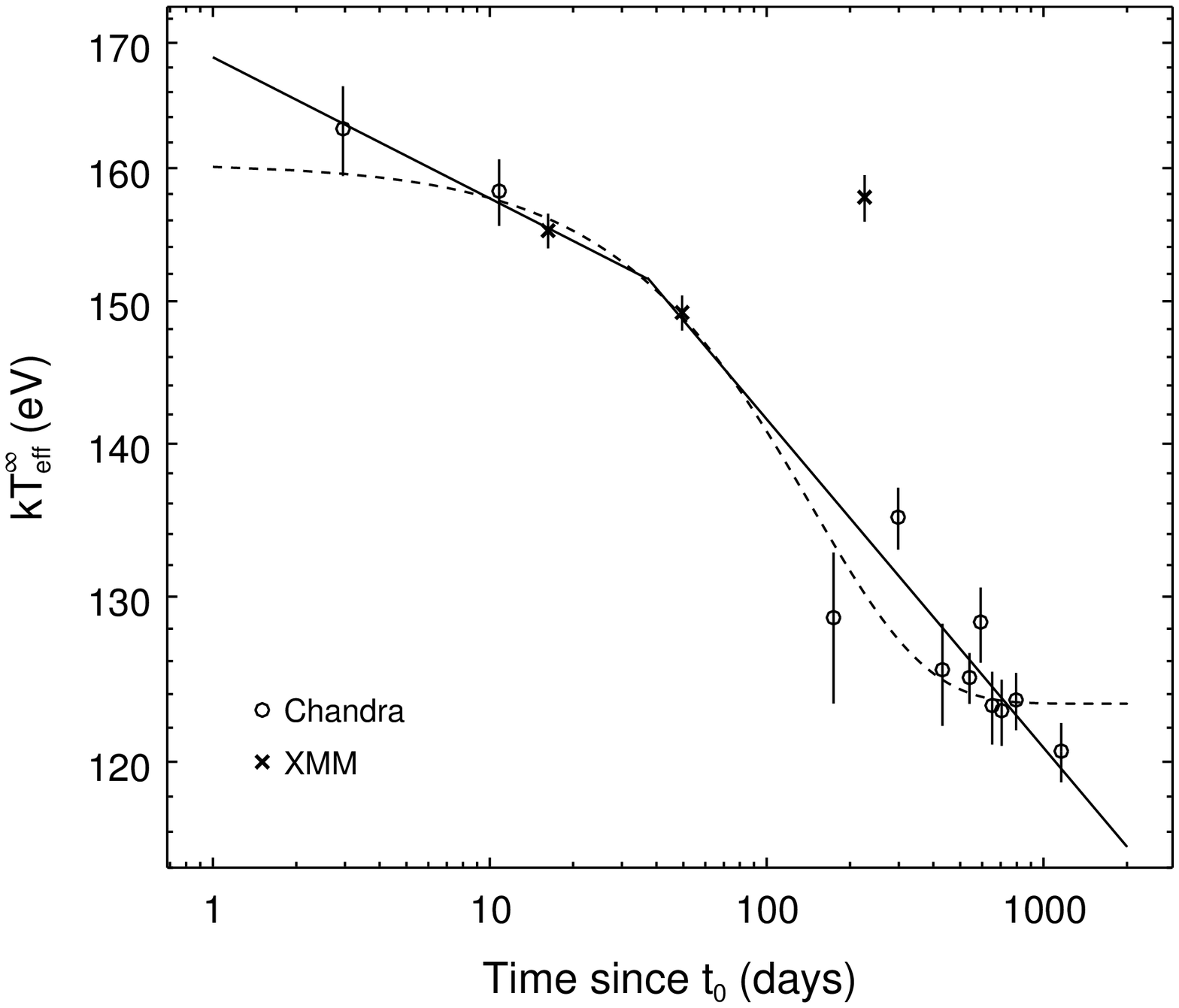}
\caption{\footnotesize {\em (Left)\/} The spectrum of Cen~X-4 at two 
different quiescent epochs (Cackett et al.\ 2010b). The spectra
demonstrate two frequently observed properties of neutron star
transients in quiescence: {\it (i)} the presence of a thermal
component (shown as blue dashed curves) as well as a power-law
component (green dotted line, here dominant above $\sim 2$~keV.  {\it
(ii)} the variability in the flux between different quiescent epochs.
The black spectrum shows the brightest and red the faintest quiescent
observations of Cen X-4. The spectra have been rebinned for visual
clarity. {\em (Right)\/} The effective neutron star surface
temperature of XTE~J1701$-$462 during quiescence, with best-fit
cooling curves shown (Fridriksson et al.\ 2011). The curves correspond
to different fits of the observed cooling trend.}
\mbox{}
\label{fig:quies_spec} 
\end{figure}

The launch of X-ray telescopes with good spatial and energy resolution
and low background, such as the Chandra X-ray Observatory and
XMM-Newton, have enabled detailed studies of these sources. About a
dozen neutron star X-ray transient systems have been detected in
quiescence as well as in outburst. There is, by now, strong evidence
that the quiescent emission of neutron star sources is brighter by
more than an order of magnitude than the quiescent emission of black
hole transients in similar binary systems (Garcia et al.\ 2001),
pointing to the role of the neutron star surface in producing the
quiescent emission. Numerous observations during quiescent periods
also revealed that there is a soft thermal component in the X-ray
spectra, characterized by blackbody temperatures of $0.1-0.3$~keV, in
a fraction of sources (see Figure~\ref{fig:quies_spec} and
Table~\ref{tab:quies}). This indicates that some of the accretion
energy may be deposited as heat in the stellar interior during the
accretion phases and reradiated from the surface during quiescence
(Brown et al.\ 1998; see Section 2.1). However, there have also been
numerous detections of non-thermal X-ray emission, typically described
by a power-law component above $\sim$ few keV, as well as of
variability in the flux levels during quiescent episodes. These
strongly suggest that at least a fraction of the emission arises from
continued accretion (Garcia et al.\ 2001) or coronal emission from the
companion star (Bildsten \& Rutledge 2000; Campana \& Stella
2000). Changing crust properties of the neutron stars leading to time
variability has also been discussed as an alternative (see, e.g.,
Cackett et al.\ 2010b).

Observations of neutron stars in quiescence have been geared towards:
{\it (i)} obtaining the apparent radii of neutron stars by fitting
their spectra with detailed atmosphere models and {\it (ii)} inferring
the properties of the neutron star crust and core by tracking their
flux and temperature evolution during quiescence (see the right panel
of Figure~\ref{fig:quies_spec}). Table~\ref{tab:quies} summarizes the
observations of sources which have been used in spectral and cooling
studies in order to measure the neutron star radius or to infer the
properties of their crusts and cores.

Other transient sources that have been monitored over long periods of
time and studied spectroscopically include Cen X$-$4 (Cackett et al.\
2010b), Aql X$-$1 (Cackett et al.\ 2011), 4U 1608$-$522 (Rutledge et
al.\ 1999; Wachter et al.\ 2002), EXO 1745$-$248 (Degenaar \& Wijnands
2012), and RX J170930.2$-$263927 (Jonker et al.\ 2003). The majority
of these sources show variability in their quiescent flux, pointing to
a low level of continued accretion. The ratio between their quiescent
luminosity and accretion luminosity has been inferred through
long-term averages and used to place constraints on the storage
efficiencies in the cores of these neutron stars (see Section 5).

A number of accretion powered millisecond pulsars, which are also
X-ray transients, have been observed in quiescence. In the case of
SAX~J1808.4$-$3658 and XTE~J1751$-$305, the quiescent spectra are
non-thermal, with a possible contribution from a thermal component
limited to $< 30$~eV and $< 71$~eV, respectively (Campana et al.\
2002; Heinke et al.\ 2009). The quiescent flux of IGR J00291+5934,
observed at multiple epochs with Chandra and XMM-Newton, shows some
variability. Its spectrum is hard and, as in the case of other
quiescent millisecond pulsars, can be modeled by a power-law (Jonker et
al.\ 2005), with a possible contribution from a thermal component at
64~eV (Campana et al.\ 2008). Note that while their quiescent
properties are summarized here, accretion powered millisecond pulsars
are discussed in more detail in section 3.3.

\begin{figure}
\centering
   \includegraphics[scale=0.66,trim=0pt 0pt 0pt 0pt,clip=true]{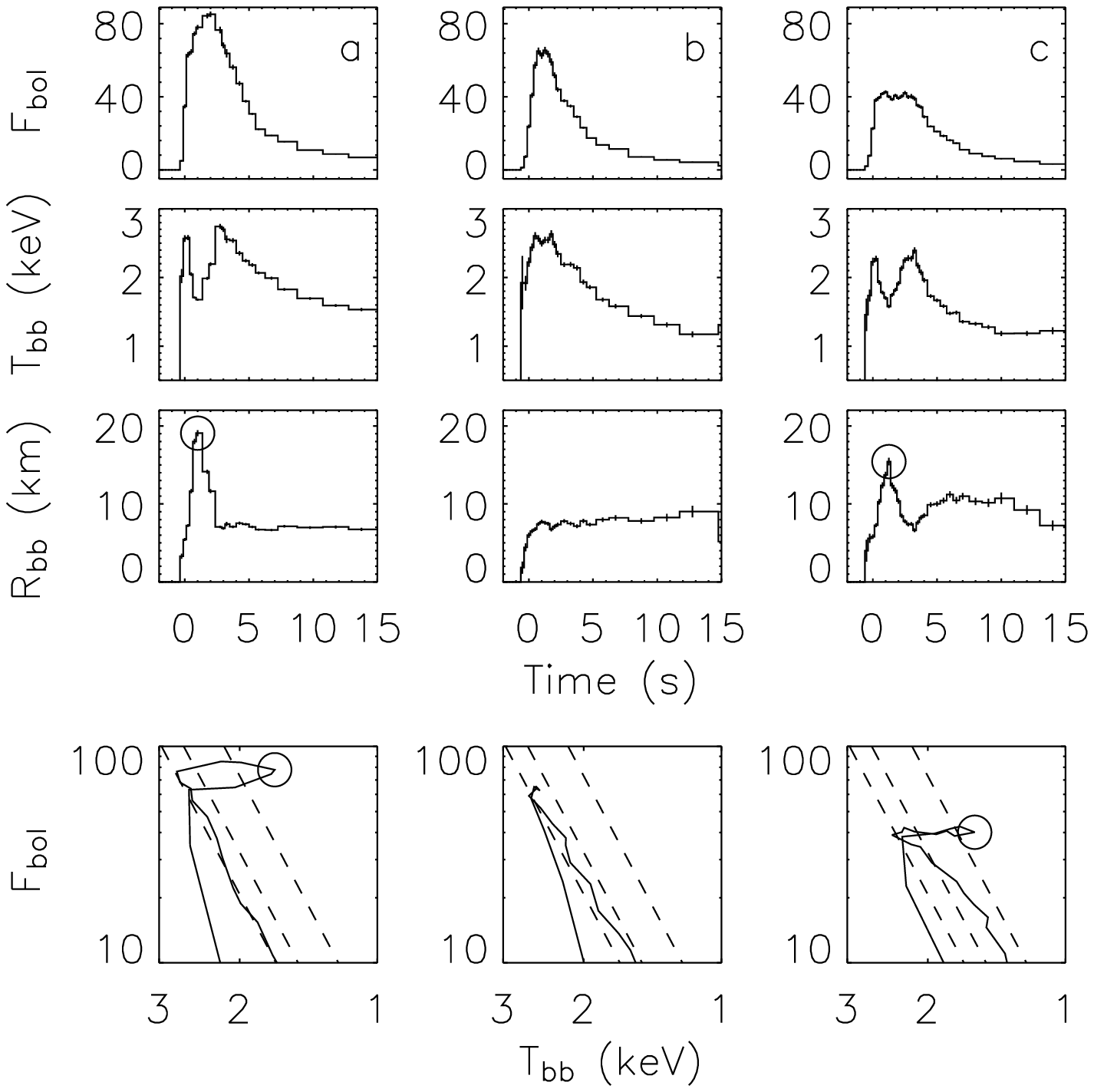}
   \includegraphics[scale=0.35]{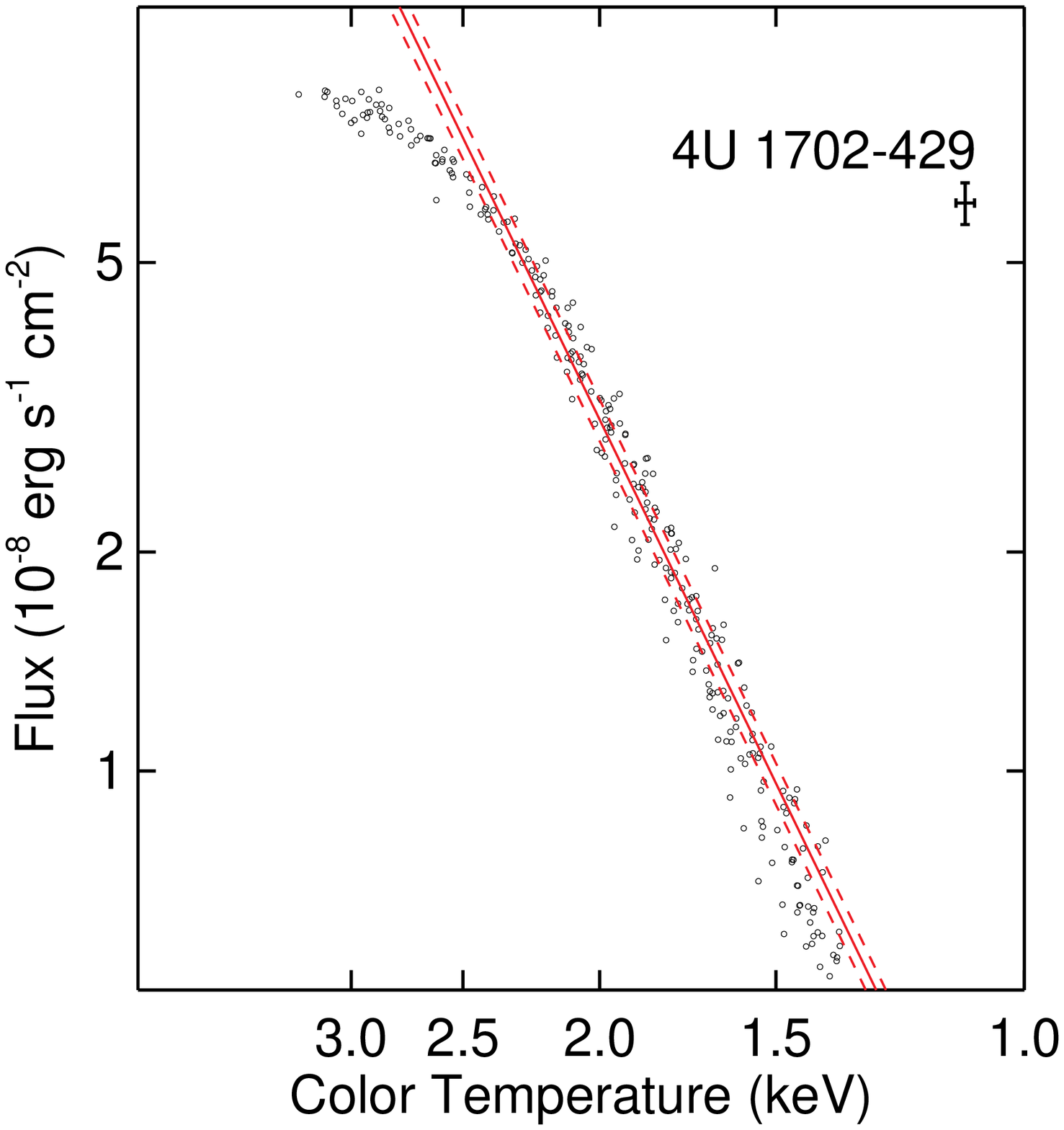}
\caption{\footnotesize {\em (Left)\/} A thermonuclear burst from 
4U 1636+536 showing a photospheric radius expansion burst in panel (a)
compared to an ordinary burst in panel (b). The three panels shown for
each burst are the bolometric flux in units of
$10^{-9}$~erg~s$^{-1}$~cm$^{-2}$, the blackbody temperature, and the
radius of the blackbody assuming that the source lies at 6~kpc
(Galloway et al.\ 2006).  {\em (Right)\/} The flux-temperature diagram
for the bursts from 4U~1702$-$429 observed with the RXTE (G\"uver et
al.\ 2012b). The diagonal lines show the best fit blackbody
normalization and its uncertainty. This figure shows that the apparent
surface areas during the cooling tails of many X-ray bursts are highly
reproducible. }
\mbox{}
\label{fig:burst} 
\end{figure}

\subsection{Accreting Bursting Neutron Stars}

Approximately hundred out of $\sim 150$ neutron stars accreting from
low-mass companions have been observed to exhibit thermonuclear
bursts, which manifest themselves as a sudden rise in the X-ray flux
that lasts $\sim 10-100$~s, accompanied by a characteristic evolution
of the inferred temperature, as shown in Figure~\ref{fig:burst} (see
Galloway et al.\ 2008 for a catalog of burst observations). The bursts
are caused by the unstable burning of the helium layer (sometimes
mixed with hydrogen) accreted onto the neutron star surface (see
Section 2.1).

Even though their discovery dates back to 1975 (Grindlay et al.\ 1976;
Belian, Conner, \& Evans 1976), the study and the statistics of
thermonuclear burst sources have improved greatly with the Rossi X-ray
Timing Explorer (RXTE) in the past decade, which observed well in
excess of 1000 burst events. Spectral studies at high time resolution
revealed thermal spectra throughout each burst (Swank et al.\ 1977;
Galloway et al.\ 2008; G\"uver et al.\ 2012b), with blackbody
temperatures that reach $2-3$~keV at the burst peak. The angular size
corresponding to the emitting area on the neutron star surface is
obtained by
\begin{equation}
\frac{R}{D} = \left(\frac{F_{\rm bol}}{\sigma_{\rm B} T_{\rm BB}^4}\right)^{1/2},
\end{equation}
where $D$ is the distance to the star, the color temperature $T_{\rm
BB}$ is inferred from fitting blackbody functions to the spectra, and
$F_{\rm bol}$ is the observed bolometric flux. (Note that this angular
size needs to be corrected by the color correction factor for
atmospheric effects, see Equation~[\ref{eq:fc}]). The angular size
shows a rapid increase in the burst rise (see Figure~\ref{fig:burst};
Strohmayer, Zhang, \& Swank 1997), as the burst spreads throughout the
neutron star surface, and remains nearly constant during the burst
decay, also referred to as the cooling tail. A statistical study of
the cooling tails of bursters that showed multiple bursts in the RXTE
database revealed remarkable consistency in the angular size obtained
from different bursts of the same source (G\"uver et al.\ 2012b). An
example of a flux-temperature diagram with a highly reproducible
angular size is shown in the right panel of
Figure~\ref{fig:burst}. The angular sizes of the bursting neutron
stars included in that study are given in Table~\ref{tab:bursters}.

A subset of bright bursts shows a particular evolution of the color
temperature and the angular size throughout the burst, where a dip in
the temperature accompanies a rise of the photospheric radius to
values that are significantly larger than the angular size observed in
the cooling tail (Lewin, van Paradijs, \& Taam 1993). The two examples
in the left panel of Figure~\ref{fig:burst} compare such a burst with
an ordinary one. In these photospheric radius expansion episodes, the
burst flux reaches the local Eddington limit and provides a measure of
this quantity for each source (Paczynski 1983; Kato 1983). For sources
that show repeated photospheric radius expansion bursts, the
uncertainties in the Eddington limit can be quantified (Damen et al.\
1990; Kuulkers et al.\ 2003; G\"uver et al.\
2012a). Table~\ref{tab:bursters} shows the bolometric Eddington flux
corrected for interstellar extinction for bursters that have been
reported to have two or more photospheric radius expansion episodes in
the catalog of Galloway et al.\ (2008). The Eddington limit measured
in bursts have been used as distance indicators (Kuulkers et al.\
2003) as well as for measuring the properties of the neutron star (van
Paradijs 1978, 1979; Damen et al.\ 1990; \"Ozel et al.\ 2009).

Periodic flux oscillations have been observed in both the rise and the
cooling phases of thermonuclear bursts (Strohmayer et al.\ 1996;
Galloway et al.\ 2008). In the burst rise, spreading of the burning
front in the accreted layer is modulated at the neutron star spin
period, causing a periodic signal (Strohmayer et al.\ 1997). In the
cooling phases, modes excited on the neutron star surface are thought
to give rise to burst oscillations (Muno et al.\ 2002, 2003; Heyl
2004; Piro \& Bildsten 2005; Narayan \& Cooper 2007). Burst
oscillations have been used to measure the spin period of neutron
stars (Chakrabarty et al.\ 2004). Moreover, the shapes and amplitudes
of the pulses occurring on the neutron star surface have been used to
measure the stellar gravity, as will be discussed in Section~4.2.

%%%%%%%%%%%%%%%%%%%%%%
%%%%   Table 2  %%%%%%
%%%%%%%%%%%%%%%%%%%%%%

\begin{table} 
\caption{\label{tab:bursters}Neutron Stars with Thermonuclear Bursts}
\begin{indented} 
\item[]
\begin{tabular}{@{}lcccc} 
\br 
Name  &  Angular Size     &    Eddington Flux                  &   Distance &  References$^a$ \\
      &  (km/10~kpc) & ($10^{-8}$~erg~s$^{-1}$~cm$^{-2}$) &    (kpc)   &             \\
\mr 
% Table: Bursters
%
%  Common Name	  Angular Size	   Eddington Flux        Distance        References 
%   or Prefix	            	               
%---------------------------------------------------------------------
4U~1608$-$52    &    18.0$\pm$0.1  &  15.4$\pm$0.65   & 5.8$^{+2.0}_{-1.9}$ &  1  \\
4U~1636$-$53    &    11.4$\pm$1.0  &   6.93$\pm$0.64  &      ---            & 2,3 \\
4U~1702$-$429   &    13.3$\pm$0.4  &       ---        &      ---            & 2,3 \\
4U~1705$-$44    &    9.2$\pm$0.5   &       ---        &      ---            & 2,3  \\
4U~1724$-$307   &    10.7$\pm$0.7  &   5.29$\pm$0.70  &      ---            & 2,3  \\
4U~1728$-$34    &    11.6$\pm$0.7  &   8.63$\pm$0.46  &      ---            & 2,3  \\
KS~1731$-$260   &    9.8$\pm$0.4   &   4.71$\pm$0.13  &      5-11           & 2,3,4  \\
4U~1735$-$44   & 8.5$^{+0.08}_{-0.06}$ & 3.15$\pm$0.11 &     ---            & 2,3  \\
EXO~1745$-$248  &   10.8$\pm$1.2   &   6.25$\pm$0.2   &   6.3$\pm$0.6       & 5    \\
4U~1746$-$37    &    4.0$\pm$0.3   &       ----       &      ---            & 2,3  \\
SAX~J1748.9$-$2021 & 9.5$\pm$0.5   &   4.03$\pm$0.54  &      ---            & 2,3  \\
SAX~J1750.8$-$2900 & 9.6$\pm$0.5   &   5.61$\pm$0.17  &      ---            & 2,3  \\
4U~1820$-$30    &    9.6$\pm$0.1   &   5.39$\pm$0.12  &    6.8$-$9.6        & 6    \\
GS~1826$-$24    &   10.2$\pm$0.03  &       ---        &      ---	    & 7    \\	
Aql~X$-$1       &       ---        &  10.44$\pm$0.22  &      ---            & 3	   \\
\br 
\end{tabular} 
Notes: References. (a) 1. G\"uver et al.\ 2010a; 2. G\"uver et al.\ 2012a; 
3. G\"uver et al.\ 2012b; 4. \"Ozel et al.\ 2012; 5. \"Ozel et al.\ 2009; 
6. G\"uver et al.\ 2010b; 7. Galloway \& Lampe 2012.
\end{indented} 
\end{table}

\subsection{Accreting Millisecond X-ray Pulsars}

Neutron stars accreting steadily over a long period of time from a
low-mass companion are believed to have their surface magnetic fields
reduced from $\sim 10^{12}$~G to $\sim 10^8$~G and their spin periods
reduced to milliseconds by accretion torques (Alpar et al.\ 1982;
Radhakrishnan \& Srinivasan 1982). Observations with RXTE revealed
persistent millisecond pulsations in the X-ray flux from a number of
sources (Wijnands \& van der Klis 1998), confirming this
expectation. The known sources belonging to this category are shown in
Table~\ref{tab:msp}. Note that there is some overlap between this
category and the previous two categories (see Table~\ref{tab:main}),
as all accreting millisecond pulsars are transients (Section 3.1),
some of which have been followed into quiescence, and some show
thermonuclear X-ray bursts (Section 3.2).

Among the sources that have shown thermonuclear X-ray bursts, there
have been detections of burst oscillations in six of them. The
asymptotic burst oscillation frequency is very similar to the
frequency of persistent pulsations, providing strong evidence that
both frequencies are equal to the spin frequency of the neutron star
(Chakrabarty et al.\ 2003). 

%%%%%%%%%%%%%%%%%%%%%%
%%%%   Table 3  %%%%%%
%%%%%%%%%%%%%%%%%%%%%%

\begin{table} 
\caption{\label{tab:msp}Millisecond X-ray Pulsars}
\begin{indented} 
\item[]
\begin{tabular}{@{}lcccc} 
\br 
Source Name & Spin Frequency & Orbital Period & Comments & References$^a$ \\
	    &	(Hz)         &    (min)       &          &       \\    
\mr 
% Table: Accretion Powered Millisecond X-ray Pulsars
%
%  Common Name	      Frequency    Orbital Period           References
%   or Prefix	            	               
%---------------------------------------------------------------------
IGR J00291+5934	         & 599    & 147    &    &  1  \\
XTE J0929$-$314	         & 185    & 43.6   &    &  2  \\
NGC6440 X$-$2            & 206    &  57    &    &  3  \\
SAX J1748.9$-$2021$^b$   & 442    & 522    &    &  4  \\
Swift J1749.4$-$2807     & 518    & 529    &    &  5  \\
IGR J1749.8$-$2921       & 11     & 1275.1 & BO$^c$  &  6  \\
XTE J1751$-$305	         & 435    & 42.4   &    &  7  \\
IGR J1751$-$30.57        & 245    & 208    & BO &  8  \\
SWIFT J1756.9$-$2508     & 182    & 54.7   &    &  9  \\
XTE J1807$-$294	         & 191    & 40.1   &    &  10  \\
SAX J1808.4$-$3658       & 401    & 121    & BO &  11 \\
XTE J1814$-$338	         & 314    & 257    & BO &  12 \\
HETE J1900.1$-$2455$^b$  & 377    & 83.3   & BO &  13 \\
Aql X-1$^b$              & 550    & 1194   & BO &  14 \\
\br 
\end{tabular} 
Notes: (a) References. (1) Galloway et al. 2005; 
(2) Galloway et al.\ 2002; 
(3) Altamirano et al.\ 2010a;
(4) Patruno et al.\ 2009;
(5) Altamirano et al.\ 2011;
(6) Altamirano et al.\ 2010b; 
(7) Markwardt et al.\ 2002; 
(8) Papitto et al.\ 2011;
(9) Krimm et al. 2007; 
(10) Kirsch et al. 2004; 
(11) Wijnands \& van der Klis 1998; 
(12) Watts et al. 2005; 
(13) Galloway et al. 2007;
(14) Casella et al.\ 2008; 
(b) Intermittent pulsations; 
(c) Burst Oscillations observed.
\end{indented} 
\end{table}

The pulsed X-ray emission from millisecond pulsars is thought to
originate from the footpoints of the accretion column onto the polar
cap of the neutron star. The spectrum and the angular distribution of
emission in this case is not well understood. It requires modeling of
both the thermal emission from the boundary layer where the accretion
column is stopped at the neutron star surface as well as of the
reprocessing of this radiation through the accretion
column. Simplifying assumptions about the geometry and the
temperature/density structure of the polar cap region and the
accretion column have often been employed (see, e.g., Poutanen \&
Gierlinski 2003; Leahy, Morsink, \& Cadeau 2008). Even though the
observed spectra do not appear to have a dominant blackbody component,
the Comptonization of the surface blackbody photons in the accretion
column is believed to generate the power-law spectra that are observed
up to $\sim 100$~keV (see Poutanen \& Gierlinski 2003). The angular
distribution of the emerging radiation, in this model, is peaked along
the radial direction, with a broad fan beam component due to
Comptonization.

The observed persistent pulsations have amplitudes in the 1$-$12\%
range (see Table~1 of \"Ozel 2009), where the lower value reflects a
typical sensitivity for short duration searches. (Deeper searches have
been performed in a small number of persistent sources and yielded
either upper limits below 1\% or the discovery of small amplitude
intermittent pulsations; see Dib et al.\ 2005 and Casella et al.\
2008). As in the case of burst oscillations, the persistent pulsations
have been used to constrain the masses and radii of neutron stars (see
Section 4.2). The shape of the footpoint of the accretion column,
where the thermal emission originates from, as well as its location
with respect to the rotation axis, present challenges in modeling the
pulse shapes. Indeed, there are theoretical calculations that point to
non-circular footpoints (Bachetti et al.\ 2010) as well as
observational evidence that the footpoints show long-term motions on
the stellar surface (Papitto et al.\ 2007; Hartman et al.\ 2008).

\subsection{Rotation-Powered Millisecond Pulsars with Thermal Emission}

In the standard recycling paradigm (Alpar et al.\ 1982; Radhakrishnan
\& Srinivasan 1982), rotation-powered millisecond pulsars 
emerge as the descendants of accreting millisecond pulsars when
accretion ceases. As in the case of other rotation-powered pulsars,
these sources show predominantly non-thermal emission from their
magnetospheres. However, recent X-ray studies have shown that a number
of rotation-powered millisecond pulsars also exhibit thermal soft
X-ray emission (Grindlay et al. 2002; Zavlin 2006, 2007; Bogdanov et
al. 2006a). Deep observations of globular clusters with Chandra and
XMM-Newton have been particularly successful in identifying the
presence and characteristics of the soft thermal emission from these
sources (see, e.g., Becker et al.\ 2003; Webb, Olive, \& Barret 2004;
Bassa et al.\ 2004; Elsner et al.\ 2008; Bogdanov \& Grindlay 2009;
Bogdanov et al.\ 2010, 2011).

The thermal emission in these rotation-powered pulsars originates from
the polar caps on the neutron star surface and is modulated at the
spin frequency of the star. The polar caps are believed to be heated
as a result of bombardment by the relativistic pairs traveling in the
magnetosphere (Ruderman \& Sutherland 1975; Arons 1981; Harding \&
Muslimov 2001). The energy from the relativistic charges is deposited
deep in the atmosphere and, therefore, the spectrum of emerging
radiation has all the typical characteristics of surface emission
discussed in Section 2.3. A power-law component is also detected in
the X-ray spectra of these sources and has been attributed {\it (i)}
to Comptonization of the thermal spectrum by the particles in the
magnetosphere, or {\it (ii)} to non-thermal emission by pairs in the
magnetosphere or by particles in the pulsar wind nebula (see Bogdanov
et al.\ 2006 for a discussion).

As in the case of pulsed surface emission from accreting millisecond
pulsars, the pulse profiles and the energy spectrum of
rotation-powered millisecond pulsars have been used to constrain the
compactness of neutron stars (see Section 4.2 below).

\subsection{Isolated Cooling Neutron Stars}

%%%%%%%%%%%%%%%%%%%%%%
%%%%   Table 4  %%%%%%
%%%%%%%%%%%%%%%%%%%%%%

\begin{table} 
\caption{\label{tab:isolated}Isolated Neutron Stars with Thermal Emission}
\begin{indented} 
\item[]
\begin{tabular}{@{}lcccccc} 
\br 
Name & Distance &  SD Age$^a$ &  Kin. Age$^b$ &    Temp.\ & Luminosity$^c$ & Refs.$^d$ \\
     &  (kpc)   &  log($t$/yr) &  log($t$/yr) &   log($T$/K) & log($L$/erg~s$^{-1}$) &  \\
\mr 
% Table: Pulsars and Nearby Isolated Neutron Stars with Thermal Spectra
%
%  Common Name	  Distance      SD Age   Kin. Age   Temp    Luminosity     References
%   or Prefix	   (kpc)       log(t/yr) log(t/yr)  log(T/K) log(L/erg/s)     
%---------------------------------------------------------------------------------
PSR B0538+2817     &    1.2      & 4.47  &   ---    &   6.1$^d$  &  32.6$-$33.6  &  1  \\     
PSR B0633+1748$^e$ & 0.12$-$0.22 & 5.53  &   ---    &   5.8$^d$  &  30.9$-$31.5  &  1  \\
PSR B0656+14       & 0.26$-$0.32 & 5.04  &   ---    &   5.7$^d$  &  32.2$-$33.0  &  1  \\
PSR J0822$-$4247$^e$ & 1.9$-$2.5 & 3.90  &   3.6    &   6.2      &  33.9$-$34.0  &  1  \\ 
PSR B0833$-$45$^e$ & 0.22$-$0.28 & 4.05  &   4.3    &   5.8      &  33.4$-$33.7  &  1  \\
PSR B1055$-$52     & 0.5$-$1.5   & 5.43  &   ---    &   5.9$^d$  &  32.1$-$33.2  &  1  \\   
PSR 1119$-$6127    & $\sim$8.4   & 3.2   &   ---    &   6.2      &  33.0$-$33.4  & 2,3 \\
1E~1207$-$52       & 1.3$-$1.9   & ---   &   3.9    &   6.2$^f$  &  33.3$-$33.8  &  1  \\   
PSR B1706$-$44     & 1.8$-$3.2   & 4.24  &   ---    &   5.8      &  31.8$-$32.9  &  1  \\  
PSR B2334+61       & $< 3.1$     & 4.6   & $\sim$4  &   5.8      &  31.7$-$32.7  &  3  \\
\hline
RX J0420.0-5022  &     ---   &  6.3  &   ---    &   5.7$^{d,f}$ &    ---        &  4,5 \\
RX J0720.4-3125& 0.27$-$0.53 &  6.2  &   5.8    & 5.6$-$6.0$^f$ &  31.3$-$32.5  &  6   \\
RX J0806.4-4123  &     ---   &  ---  &   ---    &   6.0$^{d,f}$ &    ---        &  4   \\
RX J1308.6+2127  &     ---   &  6.2  &   5.7-6.1&   6.0$^{d,f}$ &    ---        &  7,8, 9 \\
RX J1605.3+3249  &     ---   &  ---  &   5.7    &   6.0$^{d,f}$ &    ---        &  10, 11   \\
RX J1856.5-3754  &   0.12    & 	6.6  &   5.6    &   5.8       &    31.6       & 12,13,14 \\
RX J2143.0+0654  &    ---    &  6.6  &   ---    &   6.1       &    ---        & 15, 16    \\ 
\hline
\br
\end{tabular} 
Notes: (a) Spindown age; (b) Kinematic age; (c) The bolometric
luminosity of the thermal component. (d) References. 1. Page et al.\
2004; 2. Safi-Harb \& Kumar 1998; 3. Page et al.\ 2009; 4. Haberl et
al.\ 2004; 5. Kaplan \& van Kerkwijk 2011; 6. Kaplan et al.\ 2002,
2003, 2007; 7. Haberl et al.\ 2003; 8. Airhart et al.\ 2008; 9. Motch
et al.\ 2009; 10. van Kerkwijk et al.\ 2004; 11. Tetzlaff et al.\
2012; 12. Walter et al.\ 2010; 13. van Kerkwijk \& Kaplan 2008;
14. Drake et al.\ 2002; 15. Kaplan \& van Kerkwijk 2009; 16.\ Zane et
al.\ 2005; (d) Temperature obtained through blackbody fit; (e)
Alternate names: PSR~B0633+1748 = Geminga, PSR~J0822$-$4247 =
Puppis~A, PSR B0833$-$45 = Vela; (f) broad absorption lines in the
X-ray spectrum.
\\
\end{indented} 
\end{table}

Radiation from isolated neutron stars originates predominantly from
non-thermal processes in the pulsar magnetosphere, which often dwarfs
any emission from the neutron star surface. Nevertheless, in a small
number of young-to-middle age pulsars, as well as in some nearby
isolated neutron stars, a soft thermal component in the optical/X-ray
spectra has been unequivocally detected. Earlier convincing results of
this detection of surface emission came from ROSAT observations of
isolated pulsars (see \"Ogelman 1995; Becker \& Tr\"umper 1997 for a
summary of ROSAT results). Subsequently, the ensemble of such sources
has been significantly increased through observations with modern
instruments (see Page et al.\ 2009 for a discussion of the most recent
observations; see also the reviews by van Kerkwijk \& Kaplan 2007 and
Mereghetti 2011a).

The sources in this category have been subclassified under several
different names, based often on the method of detection, as also shown
in Table~\ref{tab:main}. For example, young neutron stars in supernova
remnants are often called central compact objects, isolated neutron
stars with low fluxes are called dim isolated neutron stars,
etc. Table~\ref{tab:isolated} summarizes the properties of surface
emission in these sources, divided into two major groups: sources that
were known to be rotation-powered pulsars and sources that were
discovered through their thermal emission. Sources with only upper
limits on thermal emission are not included in this table.

\begin{figure}
\centerline{\includegraphics[scale=0.3]{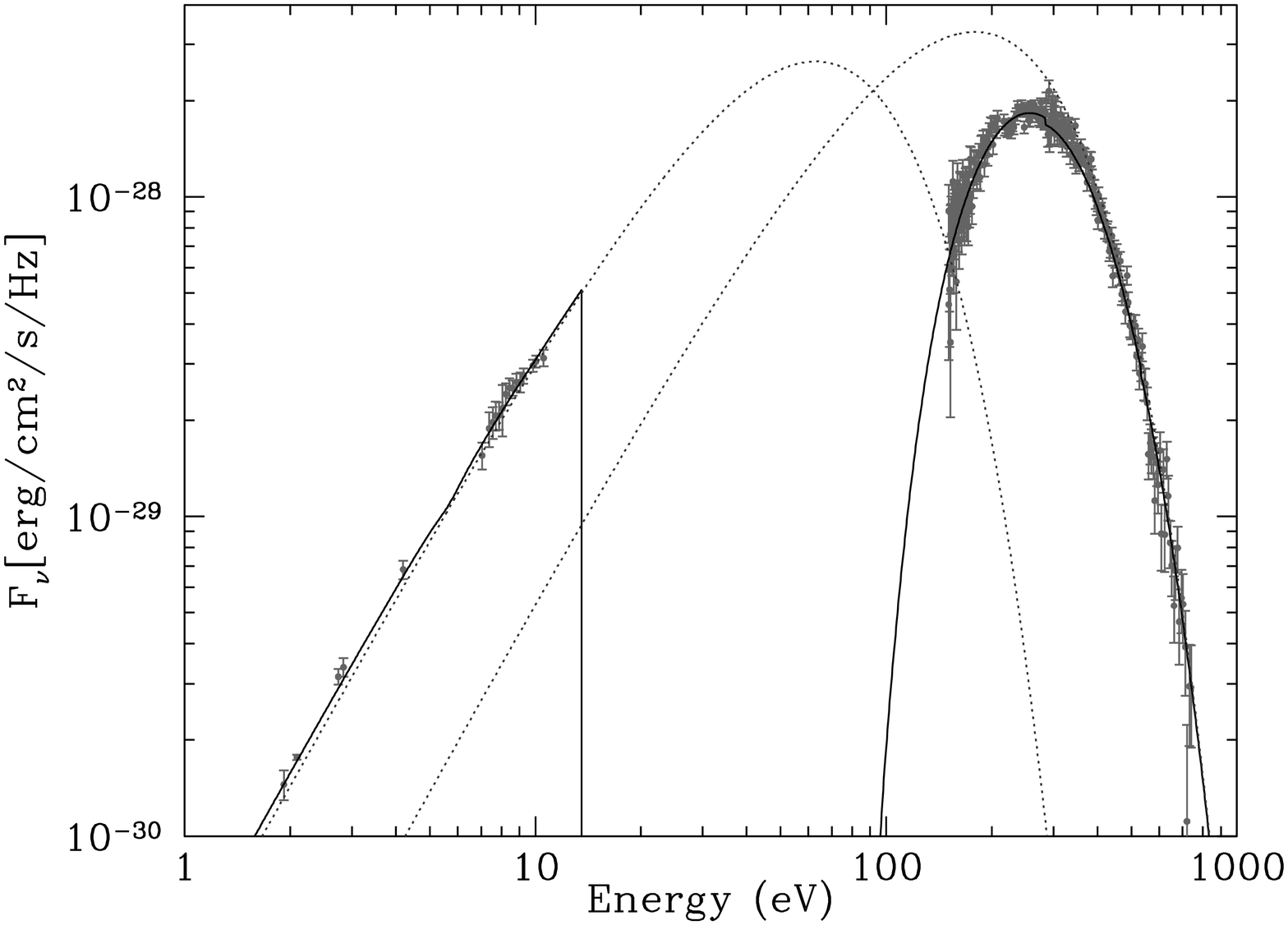}
   \includegraphics[scale=0.33]{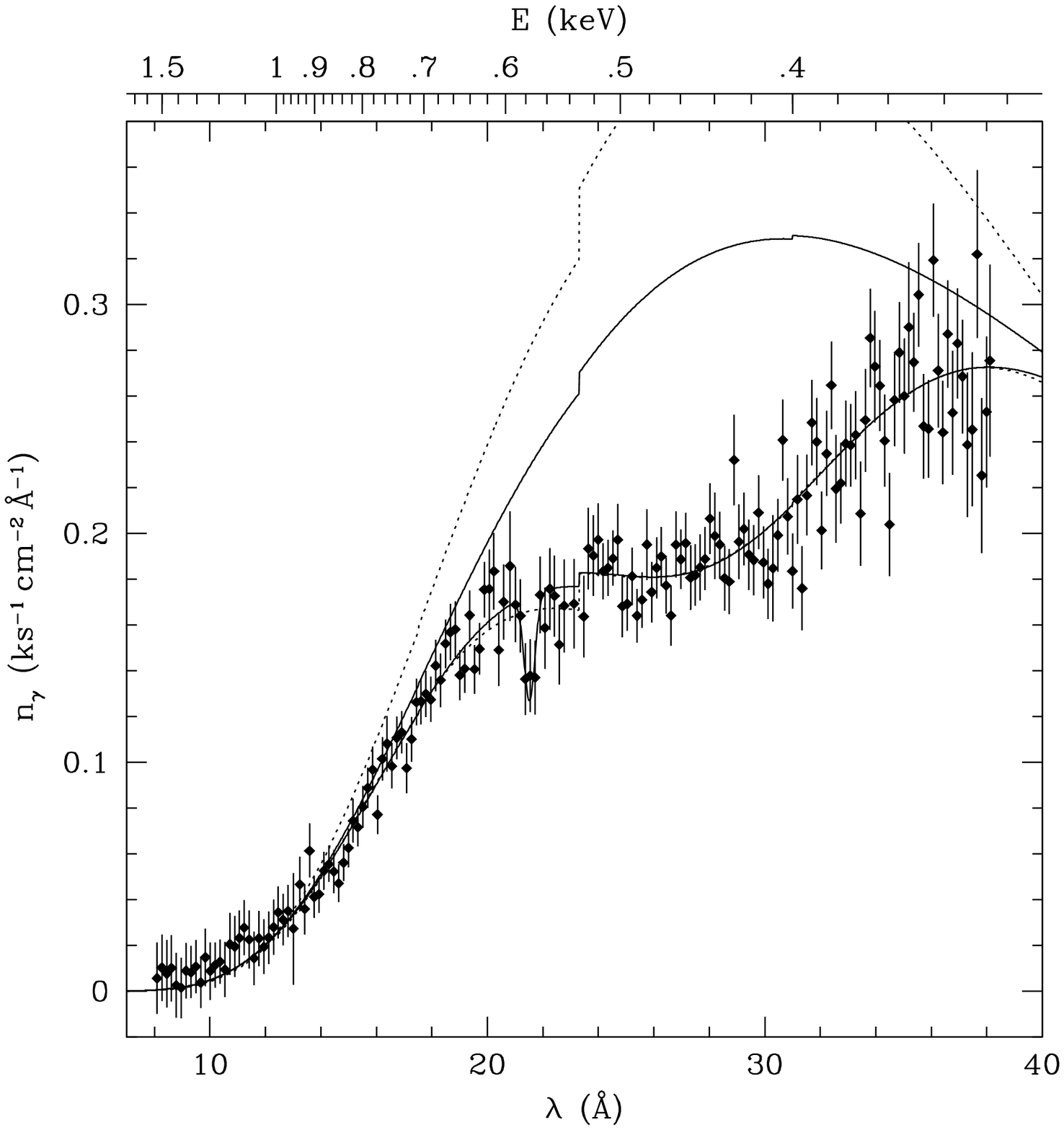}}
\caption{\footnotesize {\em (Left)\/} Broadband spectrum of RX J1856.5$-$3754 
from Braje \& Romani (2002) modeled with two blackbodies at $kT =
61$~eV and $kT= 20$~eV.  {\em (Right)\/} A broad absorption feature
observed in the spectrum RX J1605.3+3249 with RGS onboard XMM-Newton
(van Kerkwijk et al.\ 2004). The two overplotted curves correspond to
blackbodies with one (lower dotted line) and two (lower continuous
line) Gaussian absorption features. Such features have been observed
in the thermal spectra of several other nearby isolated neutron stars
but have not been conclusively identified.}
\mbox{}
\label{fig:dins_spec} 
\end{figure}

In radio pulsars, thermal radiation from the surface is superimposed
on the non-thermal magnetospheric emission and contributes a fraction
of the flux that decreases with age between $\sim
10^3-10^6$~yr. Typical temperatures obtained from fitting blackbodies
and/or hydrogen atmospheres range from $5 \times 10^5 - 1.5 \times
10^6$~K (see Table~\ref{tab:isolated}), where the lower end reflects
the sensitivity of the X-ray instruments as well as the effect of
interstellar extinction.

In nearby isolated neutron stars, the emission appears to be
predominantly from their surfaces. The spectra of sources that are
studied through the deepest observations reveal the presence of
multiple components. For example, the spectrum of RX~J1856$-$3754
shown in Figure~\ref{fig:dins_spec} has been modeled by a completely
featureless blackbody in the X-rays (Drake et al.\ 2002) but the
emission observed in the optical is higher than the extrapolation of
the X-ray blackbody. Therefore, the broadband spectrum has been
interpreted as the sum of two blackbodies with different temperatures
originating from hot and cold regions on the neutron star (Pons et
al.\ 2002; Braje \& Romani 2002; see, however, Ho 2007 for a
discussion of limitations of this interpretation and alternative
models). On the other hand, the optical emission from several other
neutron stars, which has been studied with the Hubble Space Telescope,
is also higher than the extrapolation of the X-ray blackbody but has a
spectral slope that is inconsistent with a Rayleigh-Jeans tail (Kaplan
et al.\ 2011). Moreover, in several sources such as PSR~1E1207$-$52
and RX~J1605+3249, broad spectral features are clearly visible in
their X-ray spectra, as shown in Figure~\ref{fig:dins_spec}. The
origin of these features is not well established. They have been
interpreted as cyclotron features (Sanwal et al.\ 2002; see also
Halpern \& Gotthelf 2011) as well as atomic features (see, e.g.,
Hailey \& Mori 2002; Mori \& Hailey 2006).  The feasibility and
potential problems of each interpretation is discussed in detail in
van Kerkwijk et al.\ (2007).

Distances to several of the nearby sources have been determined
accurately either via parallax measurements or via their associations
with supernova remnants (see Table~\ref{tab:isolated}). Combined with
the spectroscopic measurement of their angular sizes, this information
has led to measurements of the neutron star radii (see, e.g., Pons et
al.\ 2002). The difficulty in these measurements arises from the
presence of multi-temperature components of emission from the neutron
star surface that are inferred either from the detection of pulsed
emission or from spectral modeling. The systematic uncertainties in
these studies can be significantly reduced by combining the spectral
information with the amplitudes of pulsed emission (Psaltis, \"Ozel,
\& DeDeo 2000; Drake et al.\ 2002; Braje \& Romani 2002).

The ages of some of these isolated neutron stars can be inferred via
their associations with supernova remnants, their kinematic
properties, or their spindown rates (Table~\ref{tab:isolated}). In
combination with the distance and spectroscopic measurements discussed
above, the time evolution of the X-ray luminosity from these sources
can be studied. As it will be discussed in Section 5, this provides
one of the best tools for probing the cooling and, hence, the interior
composition, of neutron stars.

\subsection{Anomalous X-ray Pulsars and Soft Gamma Ray Repeaters}

\begin{figure}
\centerline{\includegraphics[scale=0.45]{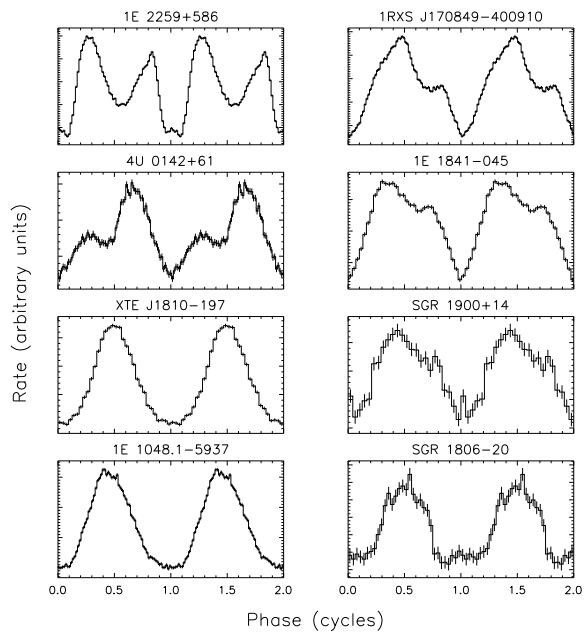}}
\caption{\footnotesize Pulse profiles of magnetars 
observed in the X-rays showing significant harmonic structure (Woods
\& Thompson 2004 using data from F. Gavriil and V. Kaspi.)}
\mbox{}
\label{fig:pulse_prof} 
\end{figure}

One last class of neutron stars from which surface emission has been
observed includes the isolated sources referred to as Anomalous X-ray
Pulsars (AXPs) and Soft Gamma-ray Repeaters (SGRs). These X-ray
sources have persistent luminosities of $10^{34-36}$~erg~s$^{-1}$ and
temperatures in the range $kT \sim 0.3-0.6$~keV. They all show pulsed
X-ray emission, from which their spin periods and period derivatives
can be measured. The high observed spindown rates, $ \dot{P} \sim
10^{-11}$~s~s$^{-1}$, yield dipole magnetic field strengths in excess
of $10^{13}$~G (Kouveliotou et al.\ 1998; see
Equation~\ref{eq:dip_field} as well as a discussion of SGR~0418+5729
below) and have led to the designation of these sources as
magnetars. The presence of such strong magnetic fields and their role
in powering the emission from AXPs and SGRs can also be inferred from
a number of other arguments, as discussed in Section 2.1. The spectral
and timing properties of all sources belonging to this category can be
found in the online catalog maintained by the pulsar group at McGill
University\footnote{\texttt{http://www.physics.mcgill.ca/~pulsar/magnetar/main.html}}.

Even though SGRs were discovered through the recurrent bursts they
produce in hard X-rays and soft gamma-rays (Mazets \& Golenetskii
1981), and the AXPs were discovered through their persistent emission
in the X-rays, by now, bursting behavior has been observed from nearly
all of these sources (Gavriil, Kaspi, \& Woods 2002). Within the
magnetar model, the bursts are believed to be powered by the
reconfiguration and dissipation of the magnetic field. The highly
super-Eddington energy release of $10^{44-45}$~erg in less than a
second timescale points to a magnetospheric origin for these events
(see Woods \& Thompson 2004). On the other hand, the persistent
emission originates deep in the crust and seems to be enhanced on year
timescales following bursting activity (Kouveliotou et al.\ 2003) and
glitches (Dib, Kaspi, \& Gavriil 2008) and to be often correlated with
changes in pulse profiles (Woods et al.\ 2004). In fact, the
persistent X-ray emission in some sources shows such large excursions
that they become detectable only during outbursts (Ibrahim et al.\
2004). Long-term monitoring observations of these transient sources as
well as during post-burst cooling of persistent sources has led to
better understanding of crust cooling and energy injection mechanisms
(e.g., Kouveliotou et al.\ 2003; G\"uver et al.\ 2007; Rea et al.\
2009).

The surface emission from magnetars is primarily observable in the
X-rays. The spectra are soft and broader than a blackbody, which has
led to their empirical description by a blackbody plus power-law or
two blackbody components. Hydrogen atmosphere models with high
magnetic field strengths naturally generate such broad spectra (see
Section 2.3). Reprocessing of the surface emission by mildly
relativistic charges in the magnetosphere further broadens the
spectrum and suppresses the equivalent widths of cyclotron features
(Gavriil \& Lyutikov 2006; G\"uver et al.\ 2007, 2008; Rea et al.\
2008). Detailed comparisons of these models to observations have been
successful and provide a spectroscopic measurement of the neutron star
magnetic field strength (see Section 6).

Persistent emission has also been detected in the hard X-rays
as well as in the optical and infrared wavebands (Hulleman, van Kerkwijk, 
\& Kulkarni 2000; Kuiper, Hermsen, \& Mendez 2004; Wang, Chakrabarty, 
\& Kaplan 2008; see Mereghetti 2011b for a review of multiwavelength 
observations). Furthermore, there has also been a detection of
transient pulsed radio emission from two AXPs (Camilo et al.\ 2004,
2006). All but the soft X-ray emission is thermodynamically
inconsistent with originating from the neutron star surface (\"Ozel
2004; Wachter et al.\ 2004) and has been attributed to non-thermal
emission from the neutron star magnetosphere (Heyl \& Hernquist 2005;
Thompson \& Beloborodov 2005; Baring \& Harding 2007).

The X-ray pulse profiles of AXPs and SGRs have been studied in detail
both during persistent emission and in connection to bursting activity
(Woods et al.\ 2001; Gavriil \& Kaspi 2002; Woods et al.\ 2004; Dib,
Kaspi, \& Gavriil 2007).  The peak-to-peak amplitudes of pulsations
range from 10\%$-$80\% and have been used to constrain the magnetar
emission geometry (\"Ozel, Psaltis, \& Kaspi 2001). The pulse profiles
have significant harmonic structure (see Figure~\ref{fig:pulse_prof}),
depend on photon energy, and evolve following bursting and glitching
activity (Woods et al.\ 2004). These characteristics point to a
non-dipolar magnetic field topology as well as magnetic field
reconfiguration during bursts and glitches, as will be discussed in
Section 6.

\section{Neutron Star Radii and Compactness}

One of the main goals of observing and modeling the surface emission
from neutron stars has been to measure their radii and masses. These
two macroscopic properties serve as direct probes of the microphysics
of the neutron star interiors, which are characterized by densities
significantly larger than the nuclear saturation density $\rho_s
\simeq 2.7 \times 10^{14}$~g~cm$^{-3}$ and low temperatures. The physical 
conditions in the centers of neutron stars occupy a distinct region in
the QCD phase diagram that cannot be probed by other cosmological
observations or terrestrial experiments (e.g., Fukushima \& Hatsuda
2011). Owing to the difficulties in determining the equation of state
of neutron star matter from first principles, neutron star
observations serve as an important constraint for determining the
ultradense matter equation of state and for guiding calculations of
the microphysics.

\begin{figure}
\centerline{\includegraphics[scale=0.40]{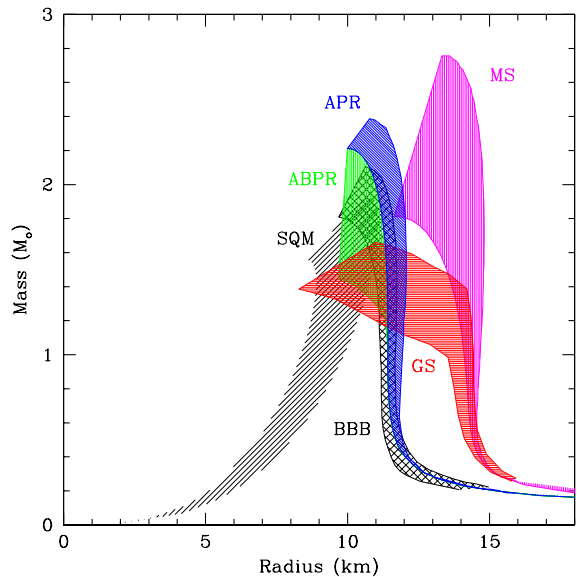}}
\caption{\footnotesize Mass-radius relations for a selection of 
neutron star equations of state. Each color-shaded region corresponds
to a different calculation and represents a range of model parameters
investigated in the corresponding study. APR is the nucleonic
equations of state of Akmal et al.\ (1998) with the expansion in terms
of 2- and 3- body interactions. MS is a field theoretical calculation
with meson exchange interactions (M\"uller \& Serot 1996). GS
represents field theoretical calculations that incorporate a condensate
of kaons (Glendenning \& Schaffner-Bielich 1999). ABPR is a hybrid
model based on the APR equation of state but incorporates a transition
to quark matter at densities larger than $\sim 2-3 \rho_s$ (Alford et
al.\ 2005). BBB represents a Brueckner-Hartree-Fock model. The
equations of state that include strange quark matter are shown as the
shaded region labeled SQM (Prakash et al.\ 1995). }
\mbox{}
\label{fig:eos} 
\end{figure}

\subsection{Neutron Star Structure and Equation of State}

A number of different approaches have been followed in the
calculations of the equation of state of neutron star matter. A
comprehensive overview of the methods and the details of the nuclear
physics can be found in Glendenning 2000, Baldo \& Burgio (2012) and
Lattimer \& Prakash (2001). Here, we provide a summary of the basic
methods as well as a compilation of a few representative model
equations of state.

One approach relies on determining the two-body potentials in the
vicinity of $\rho_s$ using nucleon-nucleon scattering data below
350~MeV and the properties of light nuclei, in addition to
incorporating the contributions from the three-body potentials (Akmal,
Pandharipande, \& Ravenhall 1998; Morales, Pandharipande, Ravenhall
2002; Gandolfi, Carlson, \& Reddy 2012). The expansion in terms of
many body interactions, however, breaks down at densities larger than
$\rho_s$. A second approach is based on field-theoretical calculations
of constituents interacting via meson exchange (M\"uller \& Serot
1996). A third approach involves microscopic ab initio calculations
based on the Brueckner-Hartree-Fock model (Baldo, Bombaci, Burgio
1997) or its relativistic counterpart, the
Dirac-Brueckner-Hartree-Fock model (M\"uther, Prakash, Ainsworth
1987).

In all of these calculations, the presence of additional components
such as hyperons, mesons, or quark matter can be incorporated. For
example, field-theoretical models have been developed that include
hyperons (Glendenning \& Moszkowski 1991) or kaon condensates
(Glendenning \& Schaffner-Bielich 1999). Non-relativistic potential
models that allow quark degrees of freedom to appear at high densities
in the cores of hybrid stars have also been investigated in great
detail (e.g., Alford et al.\ 2005). Finally, there are models based on
the assumption that the strange quark matter is the ultimate ground
state of matter, which predict entire self-bound stars of
up-down-strange quark matter, with stellar masses that increase with
radius (Alcock, Farhi, \& Olinto 1986; Prakash, Cook, \& Lattimer
1995).

The equation of state of neutron star matter determines the
macroscopic properties of the stars and, in particular, their masses
and radii. In fact, there is a unique map between the microscopic
pressure-density ($P-\rho$) relation and the macroscopic mass-radius
($M-R$) relation of stars (Lindblom 1992). In principle, the $P-\rho$
relation can be obtained from astrophysical measurements of neutron
star masses and radii by inverting this mapping. In practice, however,
this requires a measurement of radii for neutron stars that span the
entire range of masses between, e.g., $0.2-2 M_\odot$. Neutron stars
with masses much smaller than the Chandrasekhar mass of the progenitor
cores cannot be formed astrophysically (see the discussion in \"Ozel
et al.\ 2012), severely limiting the applicability of this direct
inversion.

Even though the full functional form of the $P-\rho$ relation cannot
be mapped out from astrophysical observations, it has been shown that,
for most model equations of state, neutron star masses and radii allow
us to infer the pressure of ultradense matter at a few appropriately
chosen densities above $\rho_s$. In particular, the radii at
$1.4~M_\odot$ lead primarily to the determination of the pressure at
$2 \rho_s$ (Lattimer \& Prakash 2001), the slope of the mass-radius
relation is most strongly affected by the pressure at $4 \rho_s$, and
the maximum mass of neutron stars is dictated by the pressure at $\sim
8 \rho_s$ (Read et al.\ 2009; \"Ozel \& Psaltis 2009). Therefore,
measuring the masses and radii of even a small number of neutron stars
can provide significant input to the microphysics calculations
(\"Ozel, Baym, \& G\"uver 2010; Steiner, Lattimer, \& Brown 2010).

Figure~\ref{fig:eos} shows the mass-radius relations for a number of
equations of state representing the different approaches discussed
above. For each equation of state, the shaded region represents the
range of uncertainty in the mass-radius relations that are obtained
for different input parameters within those calculations. The curves
labeled APR correspond to a nucleonic equations of state with the
expansion in terms of 2- and 3- body interactions and are
characterized by radii that are nearly independent of the stellar mass
(Akmal et al.\ 1998). The MS region is an example of a field
theoretical calculation with meson exchange interactions; as in the
case of APR relations, the radii are very weakly dependent on mass
(M\"uller \& Serot 1996). BBB is a representative
Brueckner-Hartree-Fock model based on similar potentials as those
incorporated in the APR equation of state and predict comparable
dependence of mass on radius. The equations of state by Glendenning \&
Schaffner-Bielich (1999), which are the field theoretical calculations
that incorporate a condensate of kaons, possess an inflection point at
a characteristic density in the $P-\rho$ relation. This manifests
itself as a characteristic kink in the mass-radius relations of GS as
shown in Figure~\ref{fig:eos} and reduces the predicted maximum mass
neutron stars can support. Moreover, for the mass range of
astrophysical interest, the radii become smaller with increasing
stellar mass. Hybrid neutron stars with quark matter cores are
represented by the region labeled ABPR, which is based on the APR
equation of state but incorporates a transition to quark matter at
densities larger than $\sim 2-3 \rho_s$ (Alford et al.\
2005). Finally, the equations of state that include strange quark
matter are shown as the shaded region SQM, characterized by a positive
slope in the mass-radius relation (Prakash et al.\ 1995).

\begin{figure}
\centerline{\includegraphics[scale=0.50]{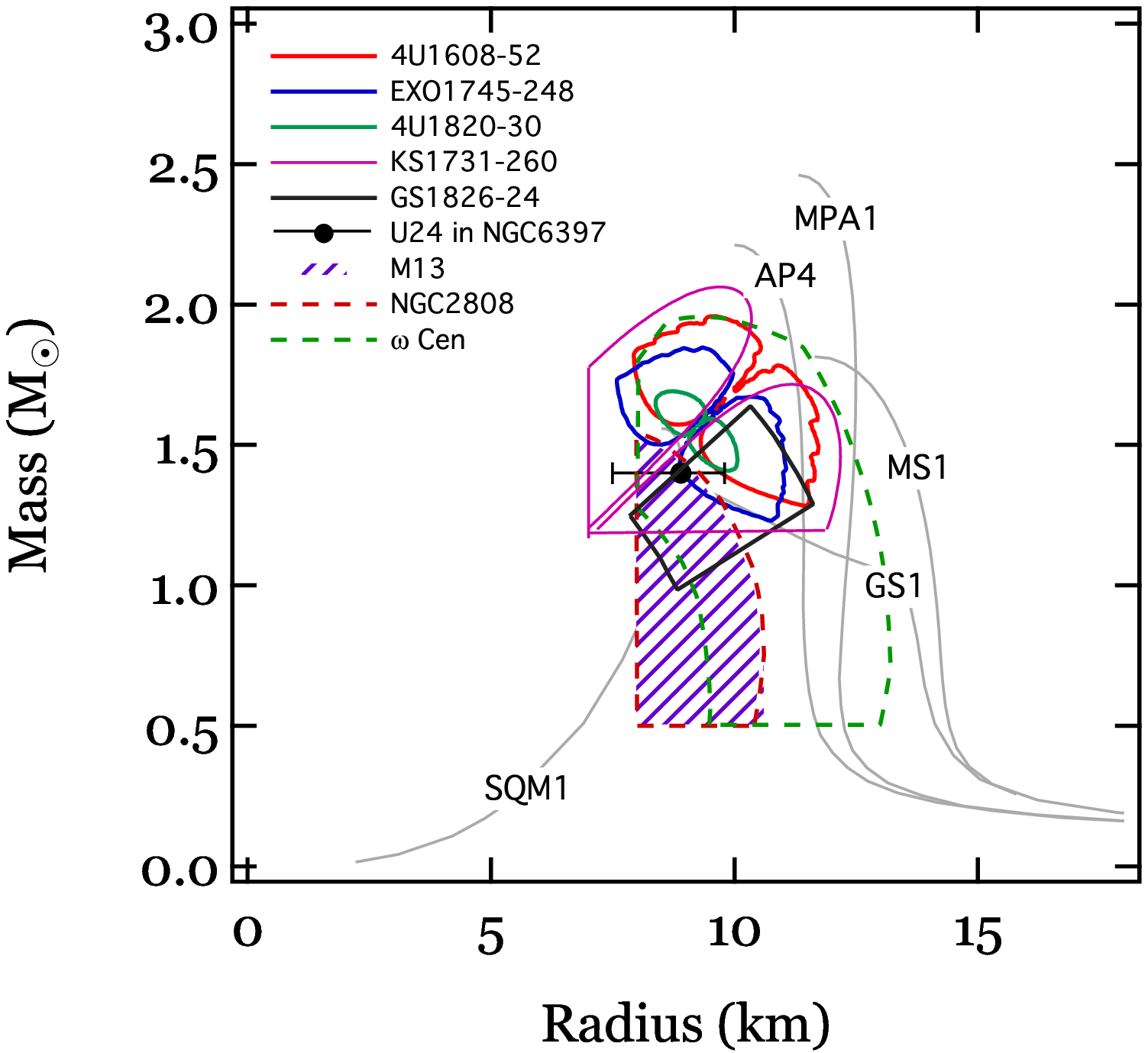}
   \includegraphics[scale=0.50]{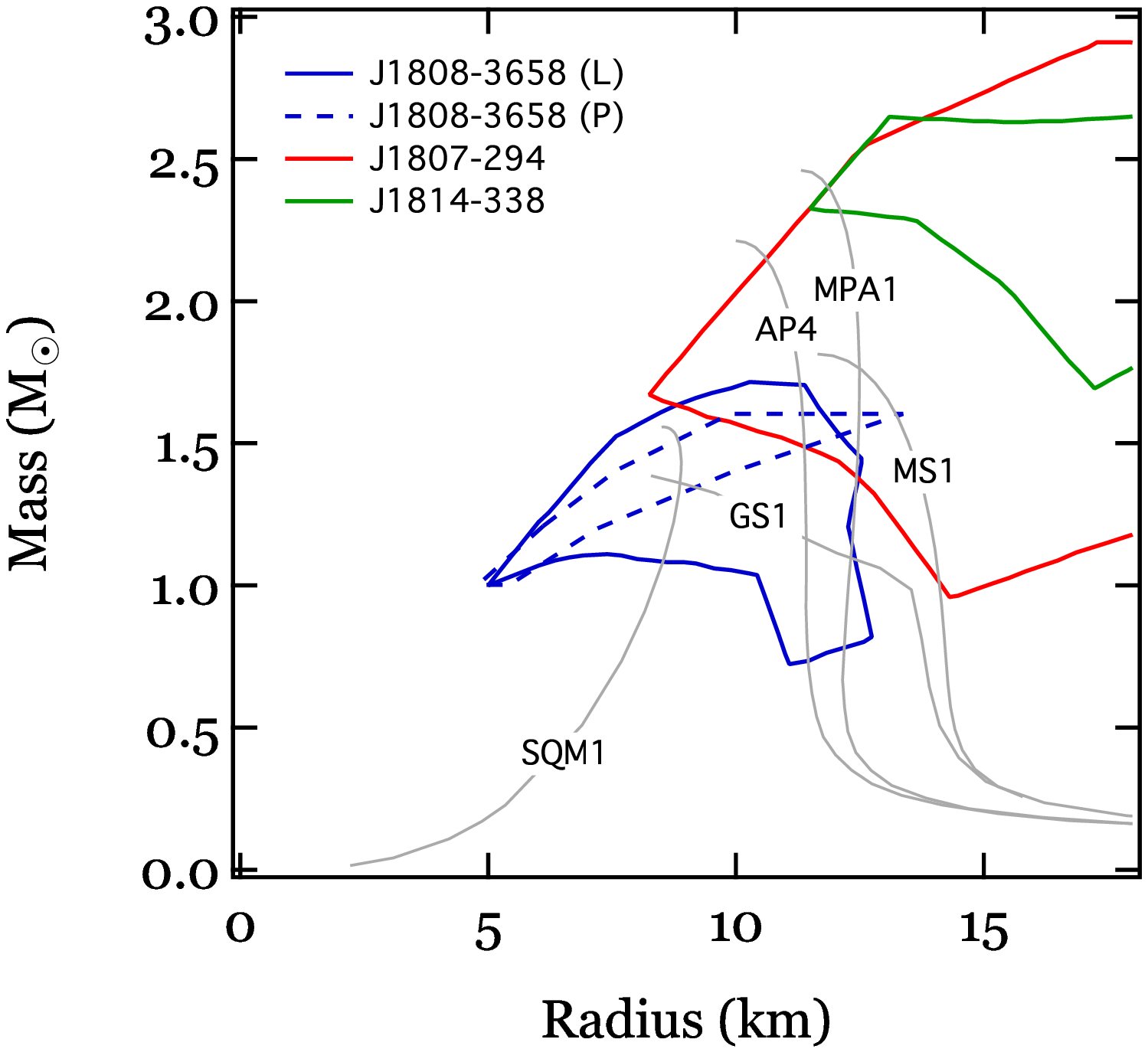}}
\caption{\footnotesize {\em (Left)\/} Radii measurements of 
neutron stars in quiescence and from thermonuclear bursts. All current
measurements are consistent with radii in the range $8-12$~km and
disfavor neutron stars with $\sim 15$~km radii. {\em (Right)\/}
Constraints on neutron star masses and radii obtained from fitting
pulse profiles of millisecond X-ray pulsars. The constraints for
SAX~J1808$-$3658 are from the independent analyses of Leahy et al.\
(2008, labeled L) and Poutanen \& Gierlinski (2003, labeled P). In
both panels, the labels follow the designation of Lattimer \& Prakash
(2001): AP4 represents the nucleonic equation of state of Akmal \&
Pandharipande (1997), MPA1 is a relativistic
Dirac-Brueckner-Hartree-Fock model (M\"uther, Prakash, Ainsworth
1987), GS1 represents a field-theoretical calculation with a kaon
condensate (Glendenning \& Schaffner-Bielich 1999); MS1 is a typical
meson exchange model (M\"uller
\& Serot 1996); and SQM1 represents a strange quark matter model 
(Prakash et al.\ 1995).}
\mbox{}
\label{fig:radii} 
\end{figure}

Measurements of neutron star radii and compactness have been achieved
through observations and modeling of several different classes of
sources. The number of independent observables, the uncertainties
associated with each observable, and those in the theoretical models
all play a role in determining the overall accuracy of the radii
determinations. Below, I discuss the neutron star radius and
compactness determinations by different techniques in various groups
of sources. The first technique utilizes spectral data and includes
radii measurements in accreting neutron stars during quiescence and
during thermonuclear bursts. The second technique is based on modeling
pulse profiles obtained from timing data and leads primarily to
constraints on the neutron star compactness (mass-to-radius ratio) in
accreting millisecond pulsars, millisecond radio pulsars, and stars
that show flux oscillations during thermonuclear bursts.

\subsection{Spectral Measurements}

Numerous observations of accreting neutron stars during quiescence
led, so far, to constraining determinations of apparent radii in a
handful of sources. In particular, sources in globular clusters, to
which distances can be measured through independent means, have
primarily been selected for these studies. In addition, sources that
show modest or negligible non-thermal components in their quiescent
spectra and exhibit little variability between different quiescent
episodes serve as the ideal targets for the radii and compactness
measurements. In all the studies, atmosphere modeling yields apparent
neutron star radii (see Equation~\ref{eq:rapp}), which are represented
as correlated contours on the mass-radius plane. If a fixed (often
arbitrary) neutron star mass, is assumed, the measurement of the
apparent radius can be converted into a value for the neutron star
radius. A compilation of the measurements performed to date is shown
in Figure~\ref{fig:radii}.

The quiescent emission from the source X7 in the globular cluster 47
Tucanae was modeled by Heinke et al.\ (2006) using non-magnetic
hydrogen atmosphere models and the multiple observations obtained with
the ACIS detector on board the Chandra X-ray Observatory. They found a
neutron star radius of $R_{\rm NS} = 14.5^{+1.6}_{-1.4}$~km, where the
errors denote 90\% confidence, when they kept the neutron star mass
fixed at 1.4~$M_\odot$. A caveat with this particular observation is
that it suffered from severe pile-up ($\sim 15\%$ level), which occurs
when two or more photons arrive on the same pixel of the CCD within a
read-out time and get recorded as a single photon, causing distortions
in the observed spectrum. While pile-up corrections were applied in
the analysis, the temperature inferred from the spectrum and the radii
results are very sensitive to pile-up modeling. For this reason, this
source is not included in Figure~\ref{fig:radii}.

Webb \& Barret (2007) applied a variety of hydrogen atmosphere models
to three quiescent neutron stars in globular clusters. Their results
on the neutron stars in M13, NGC~2808, and $\omega$Cen are shown in
Figure~\ref{fig:radii}. The absence of solutions with masses below
0.5~$M_\odot$ and radii less than 8~km reflects the range of model
parameters that were not explored in the analysis. Webb \& Barret
found the most constrained radii in two of the sources: $\leq 11$~km
for M13 and $\leq 10.5$~km for the neutron star in NGC~2808, where the
uncertainty range reflects the formal errors at 90\% confidence level
in each fit; Webb \& Barret also explore the small differences in the
radii arising from fitting different atmosphere models to the spectra.

Figure~\ref{fig:radii} also shows the radius measurement for U24 in
the globular cluster NGC 6397 (Guillot, Rutledge, \& Brown 2011).
Fitting hydrogen atmosphere models to the source spectra obtained
during five different epochs, Guillot et al.\ (2011) found evidence
for little flux variability between the epochs and $\lesssim 5\%$
contribution from a power-law component. They reported an apparent
radius of $R_{\rm app} = 11.9^{+1.0}_{-0.8}$~km, which corresponds to
a neutron star radius of $R_{\rm NS} = 8.9^{+0.9}_{-0.6}$~km if its
mass is assumed to be $1.4 M_\odot$.

A second category of sources in which radii measurements have been
performed consists of neutron stars that show thermonuclear bursts
(see Section 3.2). Because the bursts are observed in non-magnetic
neutron stars as a result of unstable ignition of helium at the bottom
of the accreted layer, the burning propagates rapidly ($< 1$~s) across
the surface of the neutron star and bright thermal emission is
observed from the stellar surface for the duration of the
burst. Several observables can be derived, in principle, from the
spectra during bursts and can be used to determine the neutron star
radius (see van Paradijs 1979; Damen et al.\ 1990; \"Ozel 2006). They
are {\it (i)} the redshifts of atomic lines in burst spectra {\it
(ii)} the apparent radii of thermal emission during burst cooling,
which can be fit with atmosphere models for accreting, bursting
neutron stars, and {\it (iii)} the Eddington limit in bright,
so-called photospheric radius expansion bursts.

The only measurement of gravitationally redshifted atomic lines with a
high resolution instrument was reported by Cottam, Paerels, \& Mendez
(2002), who analyzed combined spectra from 28 bursts from the neutron
star in EXO~0748$-$676. \"Ozel (2006) made use of the redshift
measurement in conjunction with the apparent radii determined during
the cooling phases of the thermonuclear bursts from this source and
obtained a neutron star radius of $R = 13.8 \pm 1.8$~km. However, a
subsequent discovery of the spin period of the neutron star called
into question the consistency between the width of the observed lines
and the width expected from stellar rotation (Galloway et al.\ 2010;
Lin et al.\ 2010), rendering the redshift measurement in
EXO~0748$-$676 untenable.

The uniformity and the reproducibility of emission from the stellar
surface are key ingredients for performing radius measurements from
thermonuclear bursts. The large number of bursts observed with RXTE,
BeppoSAX, Chandra, and XMM-Newton has allowed these hypotheses to be
observationally tested. Indeed, an analysis of the large database of
burst observations found that the spectra are thermal and that, for
numerous sources, the apparent radius is consistent to within $3-8\%$
during the cooling tail in each burst and between bursts observed from
each source (G\"uver et al.\ 2012b; a small number of outlier bursts
with larger variations have been observed in some sources and are
discussed in detail in G\"uver et al.\ 2012b; Bhattacharyya, Miller,
\& Galloway 2010; Zhang, Mendez,
\& Altamirano 2011). In addition, the Eddington flux was also shown to
be reproducible to $10\%$ (G\"uver et al.\ 2012a; see also Kuulkers et
al.\ 2003).

The apparent radii of neutron stars are measured during the cooling
tails of thermonuclear bursts by making use of bursting neutron star
atmosphere models for a range of compositions and depend on the
stellar mass and radius through Equation~(\ref{eq:rapp}).  The
Eddington fluxes, measured in bright bursts that show photospheric
radius expansion, have a different dependence on mass and radius, as
shown in Equation~(\ref{eq:ledd}). The combination of these
measurements, together with an estimate of the distance to each
source, has led to the determination of the neutron star radii in a
number of bursters to date, with weakly correlated errors (\"Ozel,
G\"uver, \& Psaltis 2009; G\"uver et al.\ 2010a; G\"uver et al.\
2010b; \"Ozel, Gould, \& G\"uver 2012). The results for 4U~1608$-$52,
KS~1731-260, EXO~1745$-$248, and 4U~1820-30 are shown in
Figure~\ref{fig:radii}.  The radii are tightly clustered, with a 90\%
confidence range spanning $R = 8-11$~km.

A different approach to measuring radii using burst spectra was
employed by Majczyna \& Madej (2005) and Zamfir, Cumming, \& Galloway
(2011). Majczyna \& Madej (2005) modeled the distortions in the burst
spectra of 4U 1728$-$34 obtained with RXTE as a function of surface
gravity and redshift of the neutron star and converted the constraints
on these two parameters to constraints on the neutron star radius,
which yielded, at 90\% confidence level, $R \lesssim 11$~km for this
source. Because the distortions to the spectrum due to changing
surface gravity are marginal and the RXTE energy resolution is not
adequate to detect such small deviations, this measurement is highly
uncertain. Zamfir et al.\ (2011), on the other hand, studied the
evolution of the spectra as a function of the declining flux during
the cooling tail of the bursts from GS~1826$-$24. The radius
constraints they obtained, which span $R \simeq 8-12$~km, are shown in
Figure~\ref{fig:radii}. Suleimanov et al.\ (2011) attempted a similar
study on one burst from 4U~1724$-$307; however, the spectra from this
burst are inconsistent with thermal spectra, showing evidence for
atomic edges and a reflection component (in't Zand \& Weinberg 2010),
making the radius derived in this study also highly uncertain (see the
discussion in the appendix of G\"uver et al.\ 2012b).

Spectral modeling of surface emission from a number of isolated
sources such as RX~J1856.5$-$3754 (e.g., Drake et al.\ 2002; Braje \&
Romani 2002; Ho 2007) and Cas~A (Ho \& Heinke 2009) has also been
performed. Several of these studies incorporate different elemental 
abundances (e.g., C atmospheres, see Ho \& Heinke 2009) and 
physical conditions (e.g., magnetized condensed surfaces, see Turolla, Zane, 
\& Drake 2004). The inconclusive phenomenology of their X-ray emission,
uncertainties in the importance of their magnetic fields, as well as
the low count rates have led only to weak constraints on the radii of
these neutron stars. However, these sources still provide important
information about the interiors of the neutron stars through their
inferred cooling histories, as will be discussed in Section~5.1.

It is remarkable that radii measurements obtained on nine different
sources, using at least three distinct spectroscopic methods, during
bursts and in quiescence, result in a narrow range of values, with a
clear upper limit of 12~km.

\subsection{Pulse Profile Fitting} 

In this category, there are constraints on neutron star radii and/or
compactness ratio that result from modeling lightcurves of
millisecond X-ray or millisecond radio pulsars. In the case of
accreting millisecond X-ray pulsars, the quasi-thermal emission
originates from the base of the accretion column, which is then
Comptonized in the column (Poutanen \& Gierlinski 2003; see section
3.3). Similarly, in addition to non-thermal emission from their
magnetospheres, rotation-powered millisecond pulsars show a
predominant thermal component in the soft X-rays (Grindlay et al.\
2002; Zavlin 2006; 2007; Bogdanov et al. 2006a). By modeling the
shapes and amplitudes of the pulses from these surface hotspots, and
taking into account the gravitational bending of light and Doppler
boosting due to the stellar rotation, both of which depend on the
stellar compactness, constraints on this parameter can be
obtained. The lightcurves also depend on the location of the hotspots
and the line of sight of the observer with respect to the rotation
axis, as well as on the size of the hotspot. Therefore, lightcurve
fits constrain a combination of all of these parameters and result in
measurements of these parameters with correlated uncertainties.

This method has been applied to three accreting millisecond X-ray
pulsars: SAX~J1808.4$-$3658 (Poutanen \& Gierlinski 2003; Leahy,
Morsink \& Cadeau 2008; Morsink \& Leahy 2011), XTE J1814$-$338 (Leahy
et al.\ 2009), and XTE J1807$-$294 (Leahy, Morsink, \& Chou 2011).
The constraints on the mass and radius of the neutron stars inferred
in these studies are shown together in Figure~\ref{fig:radii}. The
contours depict 99\% confidence levels. The large uncertainties in the
mass-radius measurements shown in this figure reflect the influence
of the various geometric factors discussed above, which are difficult
to constrain.

X-ray data from a number of millisecond radio pulsars, obtained with
ROSAT, Chandra, and XMM-Newton, have been analyzed using hydrogen
atmosphere models for the thermal emission from a polar cap (see,
e.g., Zavlin \& Pavlov 1998). In these models, pulse profiles are
calculated based on the beaming of radiation predicted by the
theoretical models. For PSR J0437$-$4715, Bogdanov, Rybicki, \&
Grindlay (2007) obtain R=6.8-13.8 km (90\% confidence) for a neutron
star mass of 1.4~$M_{\odot}$ (see also Pavlov \& Zavlin
1997). However, a subsequent measurement of a $1.76~M_\odot$ mass
(Verbiest et al.\ 2008) set a lower limit on its radius to $R>8.3$~km
(99.9\% confidence). For pulsars PSR J2124$-$3358 and PSR J0030+0451,
similar analyses lead to a lower limit on their radii of 7.8~km (68\%
confidence) and 10.7~km (95\% confidence), respectively, assuming a
stellar mass of 1.4~$M_\odot$.

Strong X-ray pulsations originating from the surfaces of neutron stars
have also been detected during thermonuclear X-ray bursts (Strohmayer
1996). Modeling the pulse profiles of these burst oscillations can
also lead to constraints on the neutron star compactness (see
Weinberg, Miller \& Lamb 2001 and references therein). In particular,
the amplitude of the oscillations, the deviations from a sinusoidal
waveform, and their dependence on photon energy can be probes of the
neutron star radius and mass (see, e.g., Weinberg et al.\ 2001; Muno
et al.\ 2002, 2003). This technique was applied to oscillations
observed from 4U~1636$-$536 (Nath, Strohmayer, \& Swank 2002) and from
XTE J1814$-$338 (Bhattacharyya et al.\ 2005). For the case of
4U~1636$-$536, assigning the oscillation frequency to the neutron star
spin frequency, as has been later demonstrated, led to no significant
constraints on the compactness of the neutron star. On the other hand,
Bhattacharyya et al.\ (2005) report a limit of $R>4.2 GM/c^2$ (90\%
confidence level) for the neutron star in XTE J1814$-$338, based on
fitting its pulse profile.

As I will discuss in Section 8, pulse profile modeling is a
potentially powerful probe of the neutron star compactness and radius,
which will be exploited by future X-ray timing satellites with high
signal-to-noise capabilities.

\section{The Composition of the Neutron Star Interior and Its Crust}

Observations of thermal emission from young, cooling neutron stars as
well as from transiently accreting neutron stars in quiescence have
been used to track neutron star cooling and probe their internal
composition and equation of state.

As discussed in Section 2.1, neutron star cooling depends sensitively
on the properties and interactions of the dense neutron star interior
as well as on the composition of the stellar crust (see
Figure~\ref{fig:cooling}). Cooling models are divided into three broad
categories: the standard cooling models, which incorporate modified
Urca and nucleon-nucleon bremsstrahlung processes for neutrino
emission and predict the slowest cooling rates; the minimal cooling
models, which take into account additional neutrino emission from the
breaking and formation of Cooper pairs in the superfluid core and lead
to moderate cooling rates in a limited temperature range; and enhanced
cooling models, which include direct Urca processes and lead to the
lowest surface temperatures and the shortest thermal emission
timescales. The cooling rate also depends on the composition of the
envelope and, in particular, of a thin layer within the envelope in
which the ions are in liquid phase (see Figure~\ref{fig:cooling}). As
a general trend, neutron stars with higher central densities (which
occurs in the cores of higher mass stars) and with light element
envelopes cool faster. Furthermore, in neutrino cooling calculations
with superfluid baryons, models with higher critical temperatures
allow larger neutrino emissivity and cool more rapidly. Therefore,
observations of the thermal emission of cooling neutron stars with
different ages can potentially provide information about the neutron
star interior at supernuclear densities as well as the composition of
the neutron star crust.  Finally, the critical temperature at which
the transition to superfluidity occurs can be probed by comparison of
cooling curves with data (Yakovlev \& Pethick 2004).

Making a meaningful comparison between observations and cooling curves
requires the measurement of three quantities for each neutron star:
the effective temperature, the distance, and the time since the
cooling commences. For isolated neutron stars, the latter is the age
of the neutron star, while in X-ray transients, it is the time since
the last outburst.

The temperature of the surface emission is obtained from fitting
atmospheric models to the observed spectra.  The angular sizes
inferred from these fits are converted into an emitting area using
additional information regarding the distance to the source. For young
isolated sources, the non-uniformity of surface emission as well as
the significant contribution from non-thermal magnetospheric emission,
and in several cases, the emission from a supernova remnant,
complicate the temperature determination. In addition, for some
sources, realistic (light element) atmosphere models do not yield
emitting areas that can be obtained with reasonable neutron star radii
(Chang \& Bildsten 2003; Page et al.\ 2004). This could be due to
systematic uncertainties in the source distances, low signal-to-noise
in sources with faint thermal components, which introduces
uncertainties in the temperature measurement, or uncertain subtraction
of the non-thermal emission components. Finally, owing to the higher
magnetic field strengths in pulsars, which affects cooling rates as
well as heat transport in the crust, it may be challenging to map the
surface temperature distribution from the spectra. In accreting
sources, similar concerns with non-thermal components arise due to
accretion or emission from the binary companion. In that case,
selecting sources with the smallest level of non-thermal components in
the spectra and the smallest amount of variation in the quiescent flux
levels between different outburst episodes help reduce the
uncertainties in the measured effective temperatures.

Distance measurements come from a variety of techniques. For a handful
of nearby sources, parallax measurements yield reliable distances (see
the References in Table~\ref{tab:isolated}).  Pulsar dispersion
measures are also a common way to obtain distance estimates to
isolated sources. For sources associated with supernova remnants, the
distance estimates to supernova remnants are taken as the source
distances. Finally, in the absence of direct distance determinations,
interstellar absorption measurements to other nearby stars are
utilized. For neutron star X-ray transients that are located within
globular clusters, optical observations of the stars in the cluster
provide some of the better constrained source distances. Distance
measurements are used in conjunction with the thermal fluxes to
compare them with cooling curves that track the evolution of
luminosity with time.

For isolated sources, the ages of neutron stars are inferred via their
associations with supernova remnants, their kinematic properties, or
their spindown rates (Table~\ref{tab:isolated}). It is well known that
the measurements via these different methods seldom agree (e.g., Gaensler
\& Frail 2000). For pulsars, ages are estimated from the observed
spindown rate $t_{\rm sd} \equiv P/2 \dot{P}$, which assumes that
pulsars are born with spin periods much smaller than their currently
measured period $P$ and that the spindown torque is due only to a
dipole magnetic field. The kinematic ages in a handful of sources are
obtained from the combination of pulsar transverse velocities and
their distances to a likely site of origin, such as the center of an
associated supernova remnant or a nearby massive star cluster. There
are unquantified uncertainties associated with each method, which
remain the primary challenge in comparing young cooling neutron stars
to theoretical cooling curves. In X-ray transients, cooling is tracked
from the moment when the outburst ceases and the source enters a
quiescent period, which is determined primarily by X-ray monitoring
observations and introduces less uncertainties in cooling times.

\begin{figure}
\centerline{\includegraphics[scale=0.55]{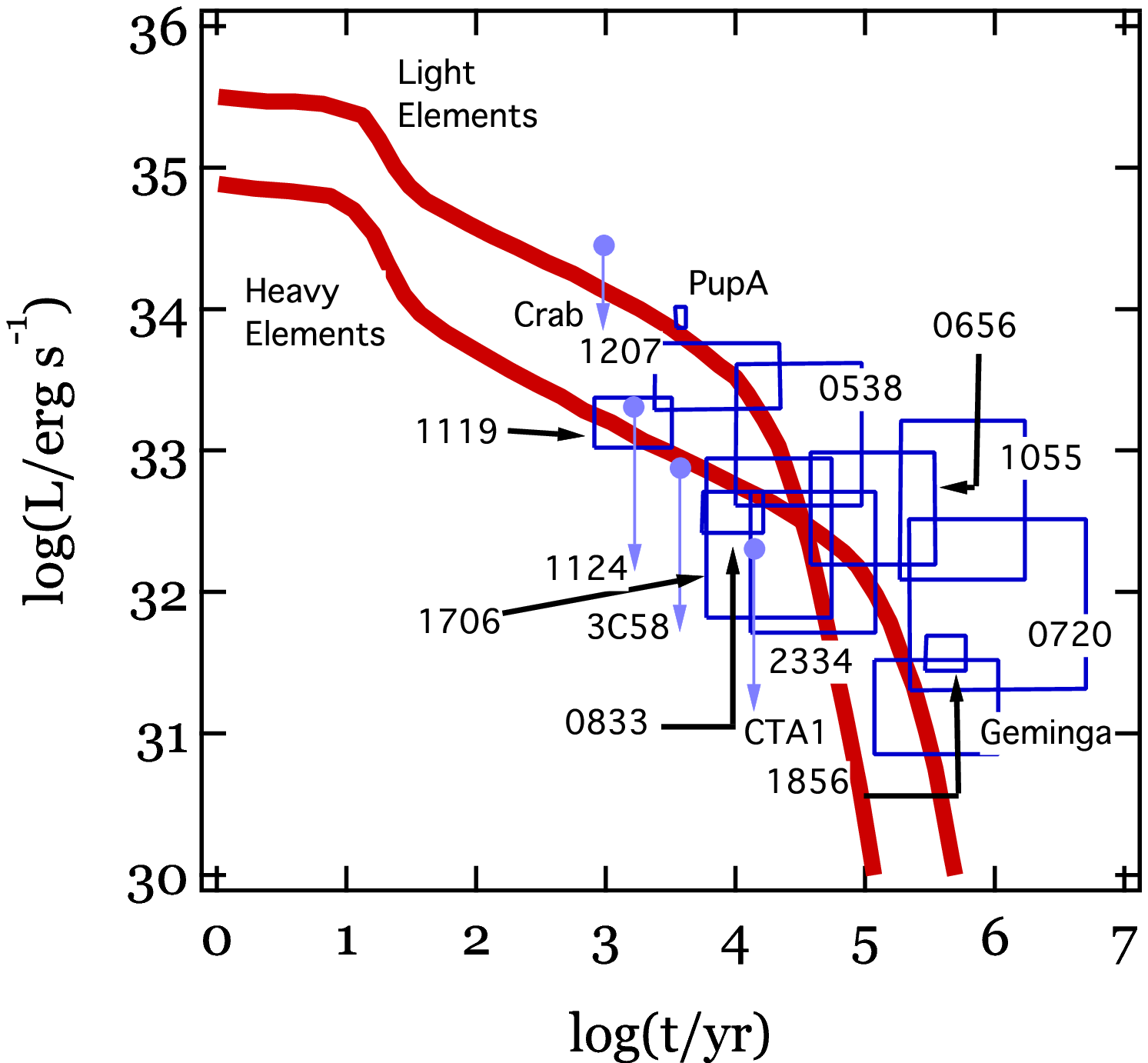}
   \includegraphics[scale=0.55]{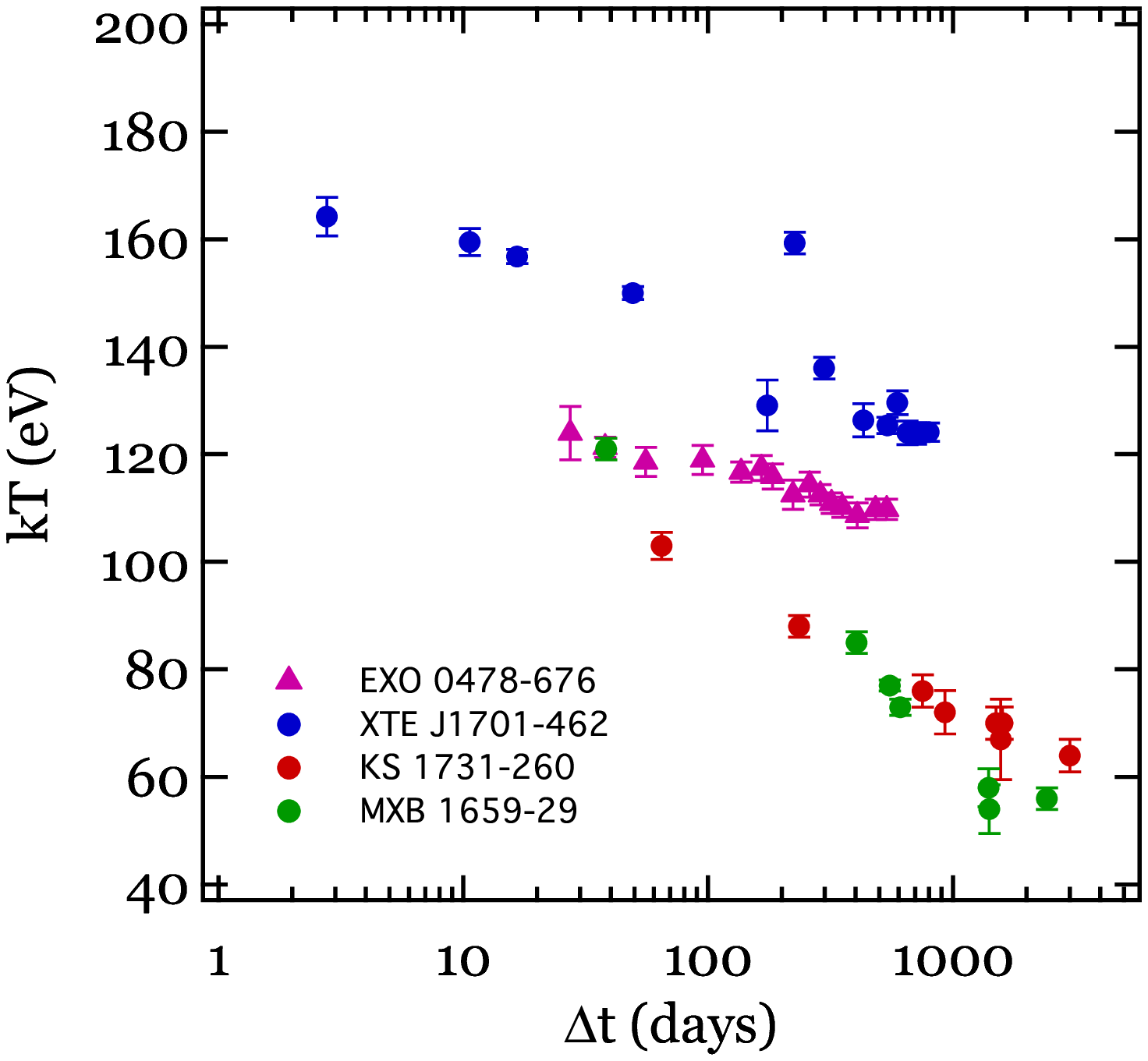}}
\caption{\footnotesize {\em (Left)\/} Cooling curves for different compositions 
of the neutron star envelope and data on young cooling neutron stars,
adapted from Page et al.\ (2009). For visual clarity, source names
have been abbreviated in the figure. {\em (Right)\/} The evolution of
the inferred neutron star temperature in four neutron-star X-ray
transients during quiescence, adapted from Degenaar et al.\ (2011b). }
\label{fig:cooling_data}
\end{figure}

\subsection{Comparison of Cooling Models to Thermal Evolution of Isolated Neutron Stars}

Isolated neutron star sources for which measurements of a surface
temperature, a characteristic age, and distance exist are shown in
Figure~\ref{fig:cooling_data} and discussed in
Table~\ref{tab:isolated}. In the left panel, the isolated sources are
compared with two minimal cooling curves obtained for different
envelope compositions. In these particular models, the neutron star is
assumed to have a mass of 1.4~$M_\odot$ with the APR equation of
state, and the neutron $^3P_2$ gap in superfluidity is chosen such
that the neutrino cooling rate due to Cooper pair formation and
breaking is high (case ``a'' in Page et al.\ 2004). The sources
include four known pulsars with only upper limits on their thermal
emission. Even though more stringent upper limits exist for compact
central sources in supernova remnants (see Page et al.\ 2009), their
neutron star nature is not established because no pulsations have been
detected. For this reason, they have been excluded from this figure.

It is evident from the comparison of the data with the cooling curves
in this figure as well as in Figure~\ref{fig:cooling} that fast
cooling, as would be predicted by direct Urca processes, is
inconsistent with the high temperatures observed in neutron stars in
the $10^4-10^6$~yr age range. In addition, the high surface
temperatures observed in the young neutron stars PSR~1E1207$-$52 and
the PSR~J0822$-$4247 (in remnant Puppis A) require light element
envelopes with no fast neutrino emission processes in the
cores. However, the upper limits in the thermal emission from three
compact objects known to be neutron stars (PSR~J1124$-$5916, the
pulsar in 3C58, and the pulsar in CTA1) as well as the emission
observed from the Vela pulsar are barely consistent with the
predictions of minimal cooling with heavy element envelopes.

With the data currently available, it becomes evident that there
exists no single universal cooling curve followed by all neutron
stars. This is not surprising given that different neutron stars may
have a range of masses, different envelope compositions, or a range of
magnetic field strengths. It is not clear from the current data which
of all these parameters produces the dominant difference in cooling
histories. A range of neutron star masses can account for some of the
observed differences (e.g., Yakovlev \& Pethick 2004) and has, in
fact, been suggested as a way of distinguishing between neutron stars
of different masses (Kaminker et al.\ 2001). Heterogeneity in the
composition of neutron star envelopes also allows cooling models to
account for the bright young pulsars as well as for pulsars with
significant thermal emission at ages greater than a million years
(Page et al.\ 2009).

Because the cooling timescales are very long, the comparison of the
cooling curves to data discussed above has been made possible by
combining the thermal properties of neutron stars at different ages
instead of tracking the temperature evolution of individual neutron
stars with time. The only exception in isolated cooling sources is the
observations by Chandra of the very young source in supernova remnant
Cas A, which showed a temperature decrease of $\sim 80000$~K, from
$2.12 \times 10^6$~K to $2.04 \times 10^6$~K, over a time of $\sim
10$~years (Heinke \& Ho 2010). The main evidence that this source
might be a neutron star is the carbon atmosphere modeling of its
surface spectrum (Ho \& Heinke 2009). Within this interpretation, the
temperature at 330 years of age is too high to be driven by direct
Urca processes but the time evolution of the temperature is too steep
for the standard cooling scenario. This steep temperature gradient has
been attributed to a transition to superfluidity in the core of the
neutron star within the minimum cooling paradigm (Page et al.\ 2011;
Shternin et al.\ 2011).

\subsection{Results for Neutron Star Transients in Quiescence}

In contrast to isolated cooling neutron stars, in the case of
accreting neutron stars, theoretical cooling curves have been compared
to the temperature evolution of individual sources as they enter
quiescence.  This is because both the thermal relaxation timescale of
the crust as well as the recurrence timescale of quiescent episodes
are in general short enough to allow detectable changes in the source
temperature over the timescale of the observations.

Clear evidence for the thermal evolution of the crust is present in
the observations of four neutron stars out of approximately a dozen
sources that have been monitored in quiescence (see
Table~\ref{tab:quies}). These sources are KS~1731$-$260 (Cackett et
al.\ 2010a), MXB~1659$-$29 (Cackett et al.\ 2008), XTE J1701$-$462
(Fridriksson et al.\ 2011), and EXO~0748$-$676 (Degenaar et al.\
2011b; Diaz-Trigo et al.\ 2011) and are distinguished from the other
neutron stars in quiescence for two reasons. First, they have
prolonged periods of outbursts, which deposit sufficient energy to
drive their crusts out of thermal equilibrium with their
cores. Second, their quiescent emission is characterized by relatively
low variability and, in general, a weak contribution from non-thermal
processes, allowing the surface temperature to be measured more
accurately. The evolution of the inferred surface temperature with
time is shown in Figure~\ref{fig:cooling_data} for these four
sources. Detailed modeling of the rapid decline in the surface
temperature in the first two sources in this list revealed that there
is efficient conduction of heat in the crust, which, in turn,
indicates low levels of impurities in the ion lattice (Shternin et
al.\ 2007; Brown \& Cumming 2009). The same is true for the rapid
cooling observed in XTE~J1701$-$462 (Fridriksson et al.\ 2010; see,
however, Page \& Reddy 2012 for an alternate interpretation). In
contrast, EXO~0748$-$676 shows a very mild decline in its temperature,
which may be related to its different outburst characteristics
compared to those of the other three sources or the presence of a
hotter core (Degenaar et al.\ 2011b). Monitoring of the thermal
evolution of the more recently discovered sources during their
quiescence, such as IGR~J17480$-$2446 (Degenaar et al.\ 2011a) may
help increase this sample.

While tracking the temperature decline during quiescence reveals the
thermal relaxation of the crust, studies of the long-term time
averaged outburst luminosities and the asymptotic quiescent
luminosities may probe the thermal properties of the stellar core
(see, e.g., Yakovlev et al.\ 2003). The core temperature is sensitive
only to the long-term time-averaged mass accretion rate because the
thermal relaxation timescale of the core is $\sim 10^4$~years. A
comparison of the asymptotic quiescent luminosity, which accounts for
the energy stored in the stellar interior and is subsequently
reradiated in photons, to the average outburst luminosity, which
represents the total accretion energy available, yields estimates of
the storage efficiency $f$ (see Equation~\ref{eq:lq}). Even though
there are large variations in the observed storage efficiencies, they
are in general low (see, e.g., Page \& Reddy 2006) and indicate that a
large fraction of the accretion energy is radiated efficiently from
the stellar interior via neutrinos.

\section{Neutron Star Magnetic Field Geometry and Evolution}

Studies of the thermal emission and the atmospheres of neutron stars
that aim to constrain the strength, geometry, and the evolution of
their magnetic fields have so far been performed for isolated radio
and X-ray pulsars, as well as AXPs and SGRs. Other lines of evidence
that do not rely on the properties of surface emission, such as the
occurrence of thermonuclear bursts or the lack of pulsations, have
been used to infer the magnetic field strengths of accreting neutron
stars but will not be reviewed here.

A general result that is common to the studies of the different types
of moderate to high magnetic field strength sources is that modeling
the pulse profiles in nearly all cases points to a complex,
non-dipolar magnetic field geometry. 

The pulsed thermal emission that has been observed from a wide variety
of neutron stars, such as isolated and millisecond radio pulsars and
isolated X-ray pulsars, has been modeled by one or two circular
hotspots that correspond to the magnetic poles on a neutron star, the
rest of which is taken to be too cold to contribute to the surface
emission. The pulse profiles are then computed for varying sizes and
temperatures of the magnetic poles, the observer's line of sight,
taking into account the general relativistic bending of light and
compared to the thermal emission from pulsars (e.g., Page 1995; Zane
et al.\ 2006). Further complexities in the models include offset
dipolar geometry, where the magnetic poles are not taken to be
antipodal, temperature gradients in the polar caps, which may be due
to a variation in the field strength, different temperatures on the
two magnetic poles, or emission geometries that are motivated by
different multipole fields. The application of these models to pulsar
pulse profiles yielded evidence for complex field topology and
emission geometries for a number of sources, such as non-uniform
temperature on polar caps in PSR J0437$-$4715 (Zavlin \& Pavlov 1998),
significantly different temperatures on two antipodal caps in
PSR~J0822$-$4247 (Gotthelf, Perna, \& Halpern 2010), offset dipole
field in the millisecond radio pulsar PSR J0030+0451 (Bogdanov \&
Grindlay 2009), as well as non-dipolar field topology in the isolated
neutron star RX~J0720.4$-$3125 (P{\'e}rez-Azor{\'{\i}}n et al.\ 2006),
in the ``three musketeers'', i.e., the three radio pulsars that showed
clear evidence for surface emission in ROSAT observations (De Luca et
al.\ 2005), and in the Geminga pulsar (Page, Shibanov, \& Zavlin
1995).

Modeling pulse profiles of magnetars, which have been monitored both
during quiescence and in connection to bursting activity, also point
to a complex and evolving magnetic field topology. Pulse profiles of
AXPs and SGRs, which can be stable over a timescale of years, show
multiple peaks, a high pulsed fraction, as well as significant
substructure (e.g., Gavriil \& Kaspi 2002). However, both long-term
evolution (Dib et al.\ 2007) and sudden changes in the pulse
morphology have been observed. The latter point to magnetic field
reconfiguration that accompany or are triggered by bursts and
outbursts in these sources: for example, a sudden change in the pulse
profile in SGR~1900+14 accompanied a flux increase of nearly three
orders of magnitude and persisted as the flux decayed (Woods et al.\
2001).  In AXP 1E~2259.1+586, more than 80 individual bursts were
detected at the onset of an outburst that included major changes in
the pulse profile and persistent flux (Woods et al.\ 2004). Pulse
profile modeling with one or two magnetic poles help constrain the
magnetic field geometry as well as the size and the temperature of the
polar caps in magnetars (\"Ozel 2002).

\begin{figure}
\centerline{\includegraphics[scale=0.5]{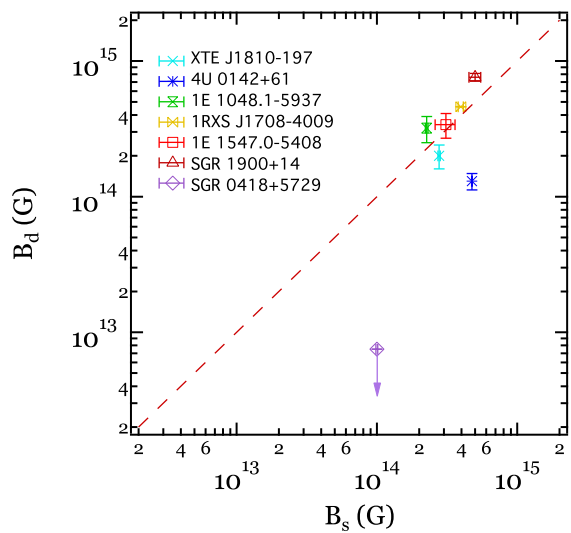}}
\caption{\footnotesize The dipole magnetic
field strengths of AXPs and SGRs inferred from their rate of spindown
using Equation~\ref{eq:dip_field} plotted against their surface field
strengths measured by fitting their thermal continuum spectra with
highly magnetic neutron star atmosphere models. }
\label{fig:mag_field}
\end{figure}

Magnetars are unique in that, owing to their very strong magnetic
fields, physical processes that take place in this regime imprint
particular characteristics on their thermal continuum spectra, as
discussed in Section 2.3. Therefore, in the case of AXPs and SGRs, the
spectra provide a handle on the magnetic field strength on the neutron
star surface. The magnetic field strengths of numerous sources have
been measured by modeling their spectra with strongly magnetic neutron
star atmosphere models (G\"uver et al.\ 2007, 2008; \"Ozel et al.\
2008; G{\"o}{\u g}{\"u}{\c s} et al.\ 2011; Ng et al.\ 2011; G\"uver
et al.\ 2011). The dipole magnetic field strength $B_{\rm d}$ inferred
from their measured rate of spindown using Equation~\ref{eq:dip_field}
is plotted against the spectroscopic magnetic field strengths $B_{\rm
s}$ obtained for these sources in Figure~\ref{fig:mag_field}. In the
majority of the sources, the two magnetic field strengths are
comparable to each other, lying close to the $B_{\rm s} = B_{\rm d}$
line shown in the figure. Furthermore, the spectroscopic field
strengths determined from spectra obtained during different epochs
from XTE~J1810$-$197, when the source progressed through significantly
different flux levels, agree remarkably well (G\"uver et
al. 2007). Interestingly, detailed modeling of the pulse profiles of
this AXP, which has the pulse profile with the least structure (see
Figure~\ref{fig:pulse_prof}), reveals that its magnetic field topology
is consistent with a dipolar geometry (Perna \& Gotthelf 2008).

The notable exception to the similarity between the spindown and
spectroscopic field strengths is SGR~0418+5729. Despite a
spectroscopically inferred surface field of $10^{14}$~G and other
properties that closely resemble all other SGRs, its period derivative
is so small that the current upper limit on $\dot{P}$ yields an upper
limit on its dipole magnetic field strength of $7.5
\times 10^{12}$~G (Rea et al.\ 2010). This has been interpreted as 
evidence for a highly non-dipolar field that is dominant on the
neutron star surface and that shapes the surface emission and the
characteristics of the energetic bursts (Rea et al.\ 2010; G\"uver et
al.\ 2011).

\section{Complementary Approaches to Neutron Star Physics}

This review focused on the physics of the surface emission from
neutron stars and the implications of the current observations for the
neutron star interior and crust. There are, however, numerous other
astrophysical probes of neutron star properties that have been pursued
during the last decade and provide complementary information on
neutron star physics. I will provide a here a brief discussion of
these additional tools.

Neutron star masses have been measured with the highest precision
using pulsar timing techniques (see the review by Kramer 2008 for
details). The most constraining aspect of the mass measurements from
the point of view of neutron star physics is finding the heaviest
pulsar, which sets a lower bound on the maximum mass of neutron stars.
The maximum mass is a key constraint on the dense matter equation of
state. It is a consequence of general relativity and is determined
primarily by the stiffness of dense matter at densities larger than
$\sim 4 \rho_{\rm sat}$. The current record holder is PSR~J1614$-$2230
with a mass of $M=1.97 \pm 0.04~M_\odot$ (Demorest et al.\ 2010).
Such a large mass already argues against significant softening of the
equation of state by the emergence of new degrees of freedom at high
densities (\"Ozel et al.\ 2010). Mass measurements of a large sample
of neutron stars also reveal their intrinsic mass distribution and
probe the physics of the supernova explosions and neutron star
formation (Thorsett \& Chakrabarty 1999; Kiziltan et al.\ 2011; Zhang
et al.\ 2011; \"Ozel et al.\ 2012).

Long-term monitoring of radio pulsars has also shown that they undergo
secular spindown interrupted by sudden glitches (see Espinoza et al.\
2011 and references therein). These glitches have been interpreted as
sudden transfer of angular momentum from superfluid neutrons in the
interior of the star to the outer crust (Anderson \& Itoh 1975). The
observed magnitudes of the glitches have been used to place constraints
on the relative moment of inertia between the crust and the core. In
the case of the Vela pulsar, which has shown prolific glitches,
measurements indicate that $\ge 1.4\%$ of the neutron star's moment of
inertia drives these glitches. This translates into a lower bound on
the radius of 8.9~km for a neutron star with mass $ > 1.35~M_\odot$
(Link, Epstein, \& Lattimer 1999).

An independent constraint on the masses and radii of accreting neutron
stars arises from the observations of fast quasi-periodic oscillations
(QPOs) in their X-ray brightness (van der Klis 2000). The frequencies
of these oscillations are variable but can be as high as $\simeq
1330$~Hz, strongly suggesting that they occur in the accretion flow
very close to the neutron star surfaces. Requiring that an observed
QPO frequency is lower than the Keplerian frequency at the
neutron star surface, which is the largest possible dynamical
frequency in the exterior spacetime of the neutron star, places an
upper bound on the stellar radius (Miller, Lamb, \& Psaltis 1998).
For most equations of state, however, the neutron star radii are
smaller than the radius of the innermost stable circular orbit in
their spacetimes. For this reason, the maximum possible stable
dynamical frequency in the exterior of a neutron star is the Keplerian
frequency at the location of the innermost stable circular orbit,
which depends primarily on the neutron star mass. Therefore, requiring
that an observed QPO frequency is lower than this maximum stable
frequency also places an upper bound on the neutron star mass
(Kluzniak, Michelson, \& Wagoner 1990; Miller et al.\ 1998). The
maximum observed QPO frequency currently stands at $\simeq 1330$~Hz
and provides only weak upper limits on the neutron star radius of $\le
15$~km (van Straaten et al.\ 2000; see also Boutelier et al.\ 2009).

Precise measurements of neutron star masses are, in principle,
possible via observations of QPO frequencies but only within a
particular model for their interpretation. For example, the rapid
decline of the amplitude and coherences of QPOs observed at high
frequencies have been interpreted as signatures of the innermost
stable circular orbit. Identifying the QPO frequency with the
Keplerian frequency at that radius leads to a direct measurement of
the neutron star mass (e.g., Barret, Olive, \& Miller 2006). On the
other hand, modeling the observed correlations (Psaltis, Belloni, \&
van der Klis 1999) between the frequencies of different QPOs in the
same source within models that assign their frequencies to epicyclic
motions in the neutron star spacetimes lead to different inferences
for the neutron star masses (e.g., Stella, Vietri,
\& Morsink 1999). All such mass measurements are, by construction, model
dependent and can only be considered as preliminary at this time, pending
validation of the theoretical model within which they were obtained.

A different bound on the inner radius of the accretion disk, and
therefore, on the radius and mass of the neutron star, comes from
observations of relativistically broadened iron fluoresence lines in
the X-ray spectra of accreting neutron stars (Cackett et al.\
2010c). Interpreting these observations requires detailed models of the
fluorescence yield as a function of radius in the accretion flow as
well as assumptions about the ionization state and the origin of the
incident radiation on the disk. Current analyses impose a lower limit
on the inner edge of the accretion disk of $6 \; GM/c^2$ and find that in
at least two neutron stars (4U~1636$-$53 and HETE~J1900.1$-$2455), the
neutron star radius has to be smaller than this limit (Cackett et al.\
2010c).

\section{Future Outlook}

During the last decade, tremendous progress has taken place in the
understanding of neutron stars through observations and modeling of
their surface emission. Not only the number but also the classes of
neutron stars that show thermal emission have dramatically increased
with the discovery of sources such as nearby isolated neutron stars,
magnetars, and millisecond accreting X-ray pulsars. These discoveries
were made possible primarily through X-ray satellites such as the
Rossi X-ray Timing Explorer, the Chandra X-ray Observatory, and
XMM-Newton that improved the timing, spectral, and imaging
capabilities of earlier instruments by orders of magnitude.

An equal amount of progress has been achieved in theoretical
calculations of neutron star atmospheres and their emission
characteristics. The new studies significantly widened the parameter
space in neutron star properties that have been explored and
incorporated different regimes of magnetic field strengths from
unmagnetized stars to magnetars, different compositions and states of
matter from cold lattices to fully ionized plasmas, and a variety of
mechanisms that power the surface emission from particle bombardment
to magnetic field decay. Moreover, the connections between the
macroscopic properties of neutron stars and the physics of their
interiors have been strengthened to the point that neutron star
observations can provide tight constraints on basic parameters that
describe the physics of ultradense matter such as the pressure at
several times the nuclear saturation density and the critical
temperature for the transition to superfluidity in the neutron star
core.

Current observations provide tantalizing hints of spectral lines from
the atmospheres of different types of neutron stars. Identification of
these lines either with atomic transitions or cyclotron features has
not been conclusive. Future observations with upcoming detectors that
have larger collecting areas and high spectral energy resolution, such
as Astro-H (Takahashi et al.\ 2010) and proposed missions including
ATHENA have the potential of detecting a multitude of lines that can
be identified and used to measure neutron star properties. More
specialized detectors that will measure the polarization properties of
neutron star surface emission can provide additional handles on the
magnetic field structure and geometry of neutron stars.

A second promising avenue towards obtaining independent constraints on
the masses and radii of neutron stars is via modeling of the pulse
profiles generated from non-uniform emission from the surface of a
rotating neutron star. Current investigations have developed a
theoretical framework that takes into account the effects of general
relativistic light bending in the rotating spacetimes of neutron stars
that have been applied primarily to observations of accreting
millisecond pulsars and to magnetic neutron stars. The optimal setting
for such studies is during the first fraction of a second of a
thermonuclear burst on a weakly magnetic neutron star. In this case,
the shape and the temperature profile of the small burning area
immediately after ignition has the least effect on the pulse profile,
which is determined primarily by general relativistic effects.
Observations of neutron stars with the high timing resolution and the
large collecting area afforded by the planned mission LOFT (Feroci et
al.\ 2011) will make such measurements possible in the near future.

The largest uncertainty in several of the measurements discussed in
this review, such as the cooling curves of neutron stars and the
radius determinations from thermally emitting stars, arises from the
poorly constrained distances to these sources. This is especially
problematic for sources in the Galactic disk, for which very few
handles on their distance are available. The situation will
dramatically change with the launch of GAIA (Turon, O'Flaherty, \&
Perryman 2005) that will chart the three-dimensional map of the Galaxy
and measure the distance to a large number of sources, especially to
those with binary companions.

The improvements in the observations and modeling of diverse phenomena
from the surfaces of neutron stars will tighten the uncertainties in
the neutron star properties inferred from each measurement. Moreover,
the detection of multiple phenomena from each neutron star, e.g.,
thermonuclear bursts, quiescent emission, and burst oscillations, will
eventually lead to independent measurements of neutron star masses and
radii. This can then be used to test the consistency of the various
measurements and explore the systematic uncertainties in each
technique.  In addition, as measurements become overconstrained, they
offer the possibility of testing the predictions of general relativity
for the strong gravitational fields in the vicinity of neutron star
surfaces (Cooney, DeDeo, \& Psaltis 2010; Deliduman, Ek{\c s}i,
\& Kele{\c s} 2011; see also Psaltis 2008 for a review). 

Several complementary probes of neutron star physics that do not rely
on the emission from their surfaces will become available in the near
future. Ongoing searches for the fastest spinning neutron stars may
detect pulsars spinning at sub-millisecond periods and, hence, place
strong upper limits on the radii of neutron stars (Cook, Shapiro, \&
Teukolsky 1994). Long-term observations of the double pulsar will
allow for a measurement of the moment of inertia of one of the pulsars
and, therefore, constrain the density profile of its interior (Kramer
\& Wex 2009). Neutrino detectors may observe the burst of neutrinos
that are emitted during the formation of a proto-neutron star in a
supernova explosion and provide information on not only the mechanism
of the explosion itself but also on the equation of state of
ultradense matter (see Kotake, Sato, \& Takahashi 2006). Finally,
detection of gravitational waves during the coalescence of two neutron
stars or the inspiraling of a neutron star into a black hole will open
a new window into the neutron star structure and equation of state
(e.g., Read et al.\ 2009; Pannarale et al.\ 2011).

\ack

I am grateful to Dimitrios Psaltis for numerous useful discussions and
Gordon Baym for helpful guidance. I thank Joel Fridriksson, Dany Page,
and Tolga G\"uver for carefully reading and providing detailed
comments on the manuscript. I also thank numerous authors for allowing
me to reproduce their figures in this review.

\section*{References} 

\begin{harvard}

\item[] Adler, S.~L.\ 1971, Annals of Physics, 67, 599  

\item[] Aguilera, D.~N., Pons, J.~A., \& Miralles, J.~A.\ 2008, A\&A, 486, 255 

\item[] Airhart, C., Woods, P.~M., Zavlin, V.~E., et al.\ 2008, 40 Years of 
Pulsars: Millisecond Pulsars, Magnetars and More, 983, 345

\item[] Akmal, A., Pandharipande, V.~R., \& Ravenhall, D.~G.\ 1998, PRC, 58, 1804 

\item[] Alcock, C., Farhi, E., \& Olinto, A.\ 1986, ApJ, 310, 261 

\item[] Alford, M., Braby, M., Paris, M., \& Reddy, S.\ 2005, ApJ, 629, 969 

\item[] Alpar, M.~A., Cheng, A.~F., Ruderman, M.~A., \& Shaham, J.\ 1982, Nature, 300, 728 

\item[] Altamirano, D., Patruno, A., Heinke, C.~O., et al.\ 2010a, ApJ, 712, L58 

\item[] Altamirano, D., Watts, A., Kalamkar, M., et al.\ 2010b, The Astronomer's Telegram, 2932, 1 

\item[] Altamirano, D., Cavecchi, Y., Patruno, A., et al.\ 2011, ApJ, 727, L18 

\item[] Anderson, P.~W., \& Itoh, N.\ 1975, Nature, 256, 25 

\item[] Arons, J.\ 1981, ApJ, 248, 1099 

\item[] Arras, P., Cumming, A., \& Thompson, C.\ 2004, ApJ, 608, L49 

\item[] Arzoumanian, Z., Chernoff, D.~F., \& Cordes, J.~M.\ 2002, ApJ, 568, 289 

\item[] Bachetti, M.,  Romanova, M.~M., Kulkarni, A., Burderi, L.,  \& 
di Salvo, T.\ 2010, MNRAS, 403, 1193 

\item[] Baldo, M., Bombaci, I., \& Burgio, G.~F.\ 1997, A\&A, 328, 274

\item[] Baldo, M. \& Burgio, G.~F.\ 2012, Rep. Prog. Phys., 75, 026301

\item[] Barret, D., Olive, J.-F., \& Miller, M.~C.\ 2006, MNRAS, 370, 1140 

\item[] Bassa, C., Pooley, D., Homer, L., et al.\ 2004, ApJ, 609, 755 

\item[] Baub{\"o}ck, M., Psaltis, D., {\"O}zel, F., \& Johannsen, T.\ 2012, ApJ, 753, 175 

\item[] Becker, W., Swartz,  D.~A., Pavlov, G.~G., et al.\ 2003, ApJ, 594, 798 

\item[] Becker, W., \& Truemper, J.\ 1997, A\&A, 326, 682 

\item[] Belian,~R.~D., Conner,~J.~P., \& Evans,~W.~D. 1976, ApJ, 206, L135

\item[] Bhattacharyya, S., Miller, M.~C., \& Galloway, D.~K.\ 2010, MNRAS, 401, 2 

\item[] Bhattacharyya, S., Miller, M.~C., \& Lamb, F.~K.\ 2006, ApJ, 644, 1085 

\item[] Bhattacharyya, S., Strohmayer, T.~E., Miller, M.~C., \& Markwardt, C.~B.\ 2005, 
ApJ, 619, 483 

\item[] Bildsten, L., Salpeter, E.~E., \& Wasserman, I.\ 1992, ApJ, 384, 143

\item[] Bogdanov, S., \& Grindlay, J.~E.\ 2009, ApJ, 703, 1557 

\item[] Bogdanov, S.,  Grindlay, J.~E., \& Rybicki, G.~B.\ 2006, ApJ, 648, L55

\item[] Bogdanov, S., Grindlay, J.~E., \& Rybicki, G.~B.\ 2008, ApJ, 
689, 407

\item[] Bogdanov, S., Rybicki, G.~B., \& Grindlay, J.~E.\ 2007, ApJ, 
670, 668 

\item[] Bogdanov, S., van den Berg, M., Heinke, C.~O., et al.\ 2010, ApJ, 709, 241 

\item[] Bogdanov, S., van den Berg, M., Servillat, M., et al.\ 2011, ApJ, 730, 81 

\item[] Boutelier, M., Barret, D., \& Miller, M.~C.\ 2009, MNRAS, 399, 1901 

\item[] Braje, T.~M., \& Romani, R.~W.\ 2002, ApJ, 580, 1043 

\item[] Braje, T.~M., Romani, R.~W., \& Rauch, K.~P.\ 2000, ApJ, 531, 447 

\item[] Brown, E.~F., Bildsten, L., \& Chang, P.\ 2002, ApJ, 574, 920 

\item[] Brown, E.~F., Bildsten, L., \& Rutledge, R.~E.\ 1998, ApJ, 504, L95 

\item[] Brown, E.~F., \& Cumming, A.\ 2009, ApJ, 698, 1020 

\item[] Bulik, T., \& Miller, M.~C.\ 1997, MNRAS, 288, 596

\item[] Cackett, E.~M., Brown, E.~F., Cumming, A., Degenaar, N., 
Miller, J.~M., \& Wijnands, R.\ 2010a, ApJ, 722, L137

\item[] Cackett, E.~M., Brown, E.~F., Miller, J.~M., \& Wijnands, R.\ 2010b, ApJ, 720, 1325

\item[] Cackett, E.~M., Fridriksson, J.~K., Homan, J., Miller, J.~M., 
\& Wijnands, R.\ 2011, MNRAS, 414, 3006 

\item[] Cackett, E.~M., Miller, J.~M., Ballantyne, D.~R., et al.\ 2010c, ApJ, 720, 205 

\item[] Cackett, E.~M., Miller, J.~M., Bhattacharyya, S., et al.\ 2008, ApJ, 674, 415 

\item[] Cackett et al. 2008, ApJ, 687, 87

\item[] Cackett et al. 2006, MNRAS, 372, 479

\item[] Cadeau, C., Leahy, D.~A., \& Morsink, S.~M.\ 2005, ApJ, 618, 451 

\item[] Cadeau, C., Morsink, S.~M., Leahy, D., \& Campbell, S.~S.\ 2007, ApJ, 654, 458 

\item[] Camilo, F., Ransom,  S.~M., Halpern, J.~P., \& Reynolds, J.\ 2007, ApJ, 666, L93 

\item[] Camilo, F., Ransom, S.~M., Halpern, J.~P., et al.\ 2006, Nature, 442, 892 

\item[] Campana, S., Stella, L., Israel, G., \& D'Avanzo, P.\ 2008, ApJ, 689, L129

\item[] Casella, P., Altamirano, D., Patruno, A., Wijnands, R., 
\& van der Klis, M.\ 2008, ApJ, 674, L41 

\item[] Chakrabarty, D., Morgan, E.~H., Muno, M.~P., et al.\ 2003, Nature, 424, 42 

\item[] Chang, P., \& Bildsten, L.\ 2003, ApJ, 585, 464

\item[] Chang, P., Morsink, S., Bildsten, L., \& Wasserman, I.\ 2006, ApJ, 636, L117 

\item[] Cook, G.~B., Shapiro, S.~L., \& Teukolsky, S.~A.\ 1994, ApJ, 424, 823 

\item[] Cooney, A., Dedeo, S., \& Psaltis, D.\ 2010, PRD, 82, 064033 

\item[] Cottam, J., Paerels, F., \& Mendez, M.\ 2002, Nature, 420, 51 

\item[] Cumming, A., Arras, P., \& Zweibel, E.\ 2004, ApJ, 609, 999 

\item[] Damen, E., Magnier, E., Lewin, W.~H.~G., et al.\ 1990, A\&A, 237, 103 

\item[] DeDeo, S., \& Psaltis, D.\ 2003, Physical Review Letters, 90, 141101 

\item[] Degenaar, N., Brown, E.~F., \& Wijnands, R.\ 2011a, MNRAS, 418, L152 

\item[] Degenaar et al. 2011b, MNRAS, 412, 1409

\item[] Degenaar, N., \& Wijnands, R.\ 2012, MNRAS, 422, 581

\item[] Deliduman, C., Ek{\c s}i, K.~Y., \& Kele{\c s}, V.\ 2012, JCAP, 5, 36 

\item[] De Luca, A., Caraveo,  P.~A., Mereghetti, S., Negroni, M., 
\& Bignami, G.~F.\ 2005, ApJ, 623, 1051 

\item[] Demorest, P.~B., Pennucci, T., Ransom, S.~M., Roberts, M.~S.~E., 
\& Hessels, J.~W.~T.\ 2010, Nature, 467, 1081 

\item[] Dib, R., Kaspi, V.~M., \& Gavriil, F.~P.\ 2007, ApJ, 666, 1152 

\item[] Dib, R., Kaspi, V.~M., \& Gavriil, F.~P.\ 2008, ApJ, 673, 1044 

\item[] Dib, R., Ransom, S.~M.,  Ray, P.~S., Kaspi, V.~M., \& 
Archibald, A.~M.\ 2005, ApJ, 626, 333

\item[] Drake, J.~J., Marshall, H.~L., Dreizler, S., et al.\ 2002, ApJ, 572, 996 

\item[] Elsner, R.~F., Heinke,  C.~O., Cohn, H.~N., et al.\ 2008, ApJ, 687, 1019 

\item[] Ertan, {\"U}., Ek{\c s}i, K.~Y., Erkut, M.~H., \& Alpar, M.~A.\ 2009, ApJ, 702, 1309 

\item[] Ertan, {\"U}., Erkut, M.~H., Ek{\c s}i, K.~Y., \& Alpar, M.~A.\ 2007, ApJ, 657, 441 

\item[] Espinoza, C.~M., Lyne, A.~G., Stappers, B.~W., \& Kramer, M.\ 2011, MNRAS, 414, 1679 

\item[] Feroci, M., Stella, L., van der Klis, M., et al.\ 2011, Experimental Astronomy, 100

\item[] Fridriksson, J.~K., Homan, J., Wijnands, R., et al.\ 2011, ApJ, 736, 162 

\item[] Fridriksson et al. 2010, ApJ, 714, 270

\item[] Fujimoto, M.~Y., Hanawa, T., \& Miyaji, S.\ 1981, ApJ, 247, 267

\item[] Fukushima, K., \& Hatsuda, T.\ 2011, Reports on Progress in Physics, 74, 014001 

\item[] Fushiki, I., \& Lamb, D.~Q.\ 1987, ApJ, 323, L55 

\item[] Gaensler, B.~M., \& Frail, D.~A.\ 2000, Nature, 406, 158

\item[]  Galloway, D.~K., Chakrabarty, D., Morgan, E.~H., \& Remillard, R.~A.\ 2002, 
ApJL, 576, L137

\item[] Galloway, D.~K., Markwardt, C.~B., Morgan, E.~H., Chakrabarty, D., 
\& Strohmayer, T.~E.\ 2005, ApJL, 622, L45 

\item[] Galloway, D.~K., Morgan, E.~H., Krauss, M.~I., Kaaret, P., 
\& Chakrabarty, D.\ 2007, ApJL, 654, L73 

\item[] Galloway, D.K., Muno M. P., Hartman J. M., Psaltis D., 
Chakrabarty D.\ 2008, ApJS, 179, 360

\item[] Galloway, D.~K., Psaltis, D., Muno, M.~P., \& Chakrabarty, D.\ 2006, ApJ, 639, 1033 

\item[] Gandolfi, S., Carlson, J., \& Reddy, S.\ 2012, Phys. Rev. C, 85, 
032801 

\item[] Garcia, M.~R., McClintock, J.~E., Narayan, R., et al.\ 2001, 
ApJL, 553, L47

\item[] Gavriil, F.~P., \& Kaspi, V.~M.\ 2002, ApJ, 567, 1067 

\item[] Gavriil, F.~P., Kaspi, V.~M., \& Woods, P.~M.\ 2002, Nature, 419, 142 

\item[] Geppert, U., K{\"u}ker, M., \& Page, D.\ 2004, A\&A, 426, 267 

\item[] Glendenning, N.~K.\ 2000, Compact stars : nuclear physics, 
particle physics, and general relativity / Norman K.~Glendenning.~ New
York : Springer, 2000

\item[] Glendenning, N.~K., \& Moszkowski, S.~A.\ 1991, PRL, 67, 2414

\item[] Glendenning, N.~K., \& Schaffner-Bielich, J.\ 1999, PRC, 60, 025803 

\item[] Gnedin, I.~N., \& Pavlov, G.~G.\ 1974, Zhurnal Eksperimental noi i 
Teoreticheskoi Fiziki, 65, 1806 

\item[] G{\"o}{\u g}{\"u}{\c s}, E., G{\"u}ver, T., {\"O}zel, F., Eichler, D., 
\& Kouveliotou, C.\ 2011, ApJ, 728, 160 

\item[] Goldreich, P., \& Julian, W.~H.\ 1969, ApJ, 157, 869 

\item[] Goldreich, P., \& Reisenegger, A.\ 1992, ApJ, 395, 250 

\item[] Gotthelf, E.~V., Perna, R., \& Halpern, J.~P.\ 2010, ApJ, 724, 1316 

\item[] Greenstein, G., \& Hartke, G.~J.\ 1983, ApJ, 271, 283 

\item[] Grindlay, J.~E., Camilo, F., Heinke, C.~O., et al.\ 2002, ApJ, 581, 470 

\item[] Gudmundsson, E.~H., Pethick, C.~J., \& Epstein, R.~I.\ 1983, ApJ, 272, 286 

\item[] Gudmundsson, E.~H., Pethick, C.~J., \& Epstein, R.~I.\ 1982, ApJ, 259, L19 

\item[] Guillot, S., Rutledge, R.~E., \& Brown, E.~F.\ 2011, ApJ, 732, 88 

\item[] G{\"u}ver, T., G{\"o}{\v g}{\"u}{\c s}, E., \"Ozel, F.\ 2011, MNRAS, 418, 2773 

\item[] G{\"u}ver, T.,  {\"O}zel, F.,  G{\"o}{\u g}{\"u}{\c s}, E.\ 2008, ApJ, 675, 1499 

\item[] G{\"u}ver, T.,  {\"O}zel, F., G{\"o}{\u g}{\"u}{\c s}, E., 
\& Kouveliotou, C.\ 2007, ApJ, 667, L73 

\item[] G\"uver, T., \"Ozel, F., Cabrera-Lavers, A., \& Wroblewski, 
P.\ 2010a, ApJ, 712, 964

\item[] G{\"u}ver, T., {\"O}zel, F., \& Psaltis, D.\ 2012a, ApJ, 747, 77 

\item[] G{\"u}ver, T., Psaltis, D., {\"O}zel, F.\ 2012b, ApJ, 747, 76 

\item[] G\"uver, T., Wroblewski, P., Camarota, L., \"Ozel, F.\ 2010b, ApJ, 719, 1807

\item[] Haberl, F., Motch, C., Zavlin, V.~E., et al.\ 2004, A\&A, 424, 635

\item[] Haberl, F., Schwope, A.~D., Hambaryan, V., Hasinger, G., \& Motch, C.\ 2003, A\&A, 403, L19

\item[] Haensel, P., \& Zdunik, J.~L.\ 2008, A\&A, 480, 459 

\item[] Haensel, P., \& Zdunik, J.~L.\ 1990, A\&A, 227, 431 

\item[] Hailey, C.~J., \& Mori, K.\ 2002, ApJ, 578, L133 

\item[] Halpern, J.~P., \& Gotthelf, E.~V.\ 2011, ApJ, 733, L28 

\item[] Hansen, C.~J., \& van Horn, H.~M.\ 1975, ApJ, 195, 735 

\item[] Harding, A.~K., \& Lai, D.\ 2006, Reports on Progress in Physics, 69, 2631

\item[] Harding, A.~K., \& Muslimov, A.~G.\ 2001, ApJ, 556, 987 

\item[] Hartman, J.~M., Patruno, A., Chakrabarty, D., et al.\ 2008, ApJ, 675, 1468 

\item[] Heger, A., Cumming, A., Galloway, D.~K., \& Woosley, 
S.~E.\ 2007, ApJL, 671, L141

\item[] Heinke, C.~O., \& Ho, W.~C.~G.\ 2010, ApJL, 719, L167 

\item[] Heinke, C.~O., Jonker, P.~G., Wijnands, R., \& Taam, R.~E.\ 2007, ApJ, 
660, 1424

\item[] Heinke, C.~O., Rybicki, G.~B., Narayan, R., \& 
Grindlay, J.~E.\ 2006, ApJ, 644, 1090

\item[] Heyl, J.~S.\ 2004, ApJ, 600, 939 

\item[] Heyl, J.~S., \& Hernquist, L.\ 1998, MNRAS, 300, 599 

\item[] Ho, W.~C.~G.\ 2007, MNRAS 380, 71 

\item[] Ho, W.~C.~G., \& Heinke, C.~O.\ 2009, Nature, 462, 71 

\item[] Ho, W.~C.~G., \& Lai, D.\ 2001, MNRAS, 327, 1081 

\item[] Ho, W.~C.~G., Lai, D., Potekhin, A.~Y., \& Chabrier, G.\ 2003, 
ApJ, 599, 1293 

\item[] Hulleman, F., van Kerkwijk, M.~H., \& Kulkarni, S.~R.\ 2000, Nature, 408, 689 

\item[] Hulleman, F., Tennant,  A.~F., van Kerkwijk, M.~H., et al.\ 2001, ApJ, 563, L49 

\item[] Ibrahim, A.~I., Markwardt, C.~B., Swank, J.~H., et al.\ 2004, ApJ, 609, L21 

\item[] Illarionov, A.~F., \& Sunyaev, R.~A.\ 1975, A\&A, 39, 185 

\item[] Jonker, P.~G., Campana, S., Steeghs, D., et al.\ 2005, MNRAS, 361, 511 

\item[] Jonker, P.~G., M{\'e}ndez, M., Nelemans, G., Wijnands, R., 
\& van der Klis, M.\ 2003, MNRAS, 341, 823 

\item[] Joss, P.~C.\ 1977, Nature, 270, 310

\item[] Kaminker, A.~D., Pavlov, G.~G., \& Shibanov, I.~A.\ 1982, ApSS, 86, 249

\item[] Kaplan, D.~L., Kamble, A., van Kerkwijk, M.~H., \& Ho, W.~C.~G.\ 2011, ApJ, 736, 117 

\item[] Kaplan, D.~L., Kulkarni, S.~R., van Kerkwijk, M.~H., \& Marshall, H.~L.\ 2002, ApJ, 570, L79

\item[] Kaplan, D.~L., \& van Kerkwijk, M.~H.\ 2011, ApJ, 740, L30

\item[] Kaplan, D.~L., \& van Kerkwijk, M.~H.\ 2009, ApJ, 692, L62 

\item[] Kaplan, D.~L., van Kerkwijk, M.~H., \& Anderson, J.\ 2007, ApJ, 660, 1428

\item[] Kaplan, D.~L., van Kerkwijk, M.~H., Marshall, H.~L., et al.\ 2003, ApJ, 590, 1008

\item[] Kato, M.\ 1983, PASJ, 35, 33 

\item[] Kirsch, M.~G.~F., Mukerjee, K., Breitfellner, M.~G., Djavidnia, S., 
Freyberg, M.~J., Kendziorra, E., \& Smith, M.~J.~S.\ 2004, A\&A, 423,
L9

\item[] Kiziltan, B., Kottas, A., \& Thorsett, S.~E.\ 2010, ApJ, submitted, arXiv:1011.4291 

\item[] Kluzniak, W., Michelson, P., \& Wagoner, R.~V.\ 1990, ApJ, 358, 538

\item[] Kotake, K., Sato, K., \& Takahashi, K.\ 2006, Rep. on Prog. in Phys., 69, 971 

\item[] Kouveliotou, C., Dieters, S., Strohmayer, T., et al.\ 1998, Nature, 393, 235 

\item[] Kouveliotou, C., Eichler, D., Woods, P.~M., et al.\ 2003, ApJ, 596, L79 

\item[] Kramer, M.\ 2008, Rev.\ in Mod.\ Astr., 20, 255 

\item[] Kramer, M, \& Wex, N 2009, Classical and Quantum Gravity, 26, 073001 

\item[] Krimm, H.~A., et al.\ 2007, ApJ, 668, L147 

\item[] Kuiper, L., Hermsen, W., \& Mendez, M.\ 2004, ApJ, 613, 1173

\item[] Kuulkers, E., den Hartog, P.~R., in't Zand, J.~J.~M., 
et al.\ 2003, A\&A, 399, 663

\item[] Lamb, D.~Q., \& Lamb, F.~K.\ 1978, ApJ, 220, 291

\item[] Lattimer, J.~M., \& Prakash, M.\ 2001, ApJ, 550, 426 

\item[] Leahy, D. A., Morsink, S. M., \& Chou, Y. 2011, ApJ, 742, 17 

\item[] Leahy, D.~A., Morsink, S.~M., Chung, Y.-Y., \& Chou, Y.\ 2009, 
ApJ, 691, 1235 

\item[] Lin, J., \"Ozel, F., Chakrabarty, D., \& Psaltis, D.\ 2010, 
ApJ, 723, 1053 

\item[] Lindblom, L.\ 1992, ApJ, 398, 569

\item[] Link, B., Epstein, R.~I., \& Lattimer, J.~M.\ 1999, PRL, 83, 3362 

\item[] Lewin, W.\ H.\ G., van Paradijs, J., \& Taam, R.\ E.\ 1993, 
Space Science Reviews, 62, 223

\item[] Lyutikov, M., \& Gavriil, F.~P.\ 2006, MNRAS, 368, 690 

\item[] Madej, J., Joss, P.~C., \& R{\'o}{\.z}a{\'n}ska, A.\ 2004, ApJ, 602, 904

\item[] Majczyna, A., Madej, J., Joss, P.~C., \& R{\'o}{\.z}a{\'n}ska, A.\ 2005, A\&A, 430, 643

\item[] Markwardt, C.~B., Swank, J.~H., Strohmayer, T.~E., Zand, J.~J.~M.~i., 
\& Marshall, F.~E.\ 2002, ApJL, 575, L21 

\item[] Mazets, E.~P., \& Golenetskii, S.~V.\ 1981, ApSS, 75, 47 

\item[] Mereghetti, S.\ 2011a,  in High-Energy Emission from Pulsars and 
their Systems, 345, arXiv:1008.2891

\item[] Mereghetti, S.\ 2011b, Advances in Space Research, 47, 1317 

\item[] M{\'e}sz{\'a}ros, P.\ 1992, High-energy radiation from magnetized neutron stars, 
University of Chicago Press

\item[] Meszaros, P., \& Ventura, J.\ 1979, PRD, 19, 3565 

\item[] Miller, M.~C.\ 1995, ApJ, 448, L29

\item[] Miller, M.~C., \& Lamb, F.~K.\ 1998, ApJ, 499, L37
 
\item[] Miller, M.~C., Lamb, F.~K., \& Psaltis, D.\ 1998, ApJ, 508, 791 

\item[] Morales, J., Pandharipande, V.~R., \& Ravenhall, D.~G.\ 2002, PRC, 66, 054308 

\item[] Mori, K., \& Hailey, C.~J.\ 2006, ApJ, 648, 1139 

\item[] Morsink, S.~M., \& Leahy, D.~A.\ 2011, ApJ, 726, 56 

\item[] Morsink, S.~M., Leahy, D.~A., Cadeau, C., \& Braga, J.\ 2007, ApJ, 663, 1244 

\item[] Motch, C., Pires, A.~M., Haberl, F., Schwope, A., \& Zavlin, V.~E.\ 2009, A\&A, 497, 423 

\item[] M{\"u}ller, H., \& Serot, B.~D.\ 1996, Nucl.\ Phys.\ A, 606, 508 

\item[] Muno, M.~P., \"Ozel, F., \& Chakrabarty, D.\ 2002, ApJ, 581, 550 

\item[] Muno, M.~P., \"Ozel, F., \& Chakrabarty, D.\ 2003, ApJ, 595, 1066 

\item[] M{\"u}ther, H., Prakash, M., \& Ainsworth, T.~L.\ 1987, Phys. 
Lett. B, 199, 469

\item[] Narayan, R., \& Cooper, R.~L.\ 2007, ApJ, 665, 628 

\item[] Narayan, R., \& Heyl, J.~S.\ 2003, ApJ, 599, 419 

\item[] Nath, N.~R., Strohmayer, T.~E., \& Swank, J.~H.\ 2002, \
ApJ, 564, 353

\item[] Ng, C.-Y., Kaspi, V.~M., Dib, R., et al.\ 2011, ApJ, 729, 131 

\item[] \"Ogelman, H.\ 1995 in Lives of Neutron Stars, ed. M.A. Alpar, Ü. Kiziloglu and 
J. van Paradijs, (Kluwer, Dordrecht)

\item[] {\"O}zel, F.\ 2001, ApJ, 563, 276 

\item[] {\"O}zel, F.\ 2002, ApJ 575, 397 

\item[] {\"O}zel, F.\ 2003, ApJ, 583, 402

\item[] {\"O}zel, F.\ 2004, arXiv:astro-ph/0404144 

\item[] {\"O}zel, F. 2006, Nature, 441, 1115

\item[] {\"O}zel, F.\ 2009, ApJ, 691, 1678 

\item[] {\"O}zel, F., Baym, G., \& G\"uver, T.\ 2010, PRD, 82, 101301 

\item[] \"Ozel, F., Gould, A., \& Guver, T.\ 2012, ApJ, 748, 5

\item[] {\"O}zel, F., G{\"u}ver, T., G{\"o}{\u g}{\"u}{\c s}, E.\ 2008, 
40 Years of Pulsars: Millisecond Pulsars, Magnetars and More, 983, 254 

\item[] \"Ozel F., G\"uver T., \& Psaltis D., 2009, ApJ, 693, 1775

\item[] \"Ozel, F., \& Psaltis, D.\ 2003, ApJL, 582, L31 

\item[] {\"O}zel, F., \& Psaltis, D.\ 2009, PRD, 80, 103003 

\item[] {\"O}zel, F., Psaltis, D., \& Kaspi, V.~M.\ 2001, ApJ, 563, 255

\item[] \"Ozel, F., Psaltis, D., Narayan, R., \& Santos Villarreal, A.\ 2012, ApJ, 
in press, arXiv:1201.1006 

\item[] {\"O}zel, F., Psaltis, D., Ransom, S., Demorest, P., \& Alford, M.\ 2010, 
ApJ, 724, L199 

\item[] Paczynski, B.\ 1983, ApJ, 267, 315 

\item[] Paczynski, B.\ 1992, AcA, 42, 145 

\item[] Page, D.\ 1995, ApJ, 442, 273 

\item[] Page, D., Lattimer, J.~M., Prakash, M., \& Steiner, A.~W.\ 2004, ApJ Suppl., 155, 623 

\item[] Page, D., Lattimer, J.~M., Prakash, M., \& Steiner, A.~W.\ 2009, 
ApJ, 707, 1131 

\item[] Page, D., Prakash, M., Lattimer, J.~M.,  \& Steiner, A.~W.\ 2011, PRL, 106, 081101

\item[] Page, D., \& Reddy, S.\ 2006, Annual Review of Nuclear and Particle Science, 56, 327

\item[] Page, D., \& Reddy, S.\ 2012, in Neutron Star Crust, ed.\ C. A. Bertulani and J. Piekarewicz,
arXiv:1201.5602 

\item[] Page, D., Shibanov, Y.~A., \& Zavlin, V.~E.\ 1995, ApJ, 451, L21 

\item[] Pannarale, F., Rezzolla, L., Ohme, F., \& Read, J.~S.\ 2011, PRD, 84, 104017 

\item[] Papitto, A., di Salvo, T., Burderi, L., et al.\ 2007, MNRAS, 375, 971

\item[] Papitto, A., Riggio, A., di Salvo, T., et al.\ 2010, MNRAS, 407, 2575 

\item[] Patruno, A., Altamirano, D., Hessels, J.~W.~T., et al.\ 2009, ApJ, 690, 1856 

\item[] Pechenick, K.~R., Ftaclas, C., \& Cohen, J.~M.\ 1983, ApJ, 274, 846 

\item[] P{\'e}rez-Azor{\'{\i}}n, J.~F., Pons, J.~A., Miralles, J.~A., 
\& Miniutti, G.\ 2006, A\&A, 459, 175 

\item[] Perna, R., \& Gotthelf, E.~V.\ 2008, ApJ, 681, 522 

\item[] Perna, R., Narayan, R.,  Rybicki, G., Stella, L., \& Treves, A.\ 2003, ApJ, 594, 936 

\item[] Pethick, C.~J.\ 1992, Reviews of Modern Physics, 64, 1133 

\item[] Piro, A.~L., \& Bildsten, L.\ 2005, ApJ, 629, 438 

\item[] Pons, J.~A., Walter, F.~M., Lattimer, J.~M., et al.\ 2002, ApJ, 564, 98

\item[] Potekhin, A.~Y., Chabrier, G., \& Yakovlev, D.~G.\ 1997, A\&A, 323, 415 

\item[] Poutanen, J., \& Gierli{\'n}ski, M.\ 2003, MNRAS, 343, 1301 

\item[] Prakash, M., Cooke, J.~R., \& Lattimer, J.~M.\ 1995, PRD, 52, 661

\item[] Psaltis, D.\ 2008, Liv.\ Rev.\ in Rel., 11, 9 

\item[] Psaltis, D.\ 2008, PRD, 77, 064006 

\item[] Psaltis, D., Belloni, T., \& van der Klis, M.\ 1999, ApJ, 520, 262

\item[] Psaltis, D., {\"O}zel, F., \& DeDeo, S.\ 2000, ApJ, 544, 390 

\item[] Radhakrishnan, V., \& Srinivasan, G.\ 1982, Current Science, 51, 1096

\item[] Rea, N., Esposito, P., Turolla, R., et al.\ 2010, Science, 330, 944

\item[] Rea, N., Israel, G.~L., Turolla, R., et al.\ 2009, MNRAS, 396, 2419 

\item[] Rea, N., Zane, S., Turolla, R., Lyutikov, M., G\"utz, D.\ 2008, ApJ, 686, 1245 

\item[] Read, J.~S., Lackey, B.~D., Owen, B.~J., \& Friedman, J.~L.\ 2009, PRD, 79, 124032 

\item[] Read, J.~S., Markakis, C., Shibata, M., et al.\ 2009, PRD, 79, 124033

\item[] Romanova, M.~M., Ustyugova, G.~V., Koldoba, A.~V., 
\& Lovelace, R.~V.~E.\ 2004, ApJ, 616, L151

\item[] Ruderman, M.~A., \& Sutherland, P.~G.\ 1975, ApJ, 196, 51 

\item[] Ruffert, M.\ 1996, A\&A, 311, 817 

\item[] Rutledge, R.~E., Bildsten, L., Brown, E.~F., Pavlov, G.~G., 
\& Zavlin, V.~E.\ 1999, ApJ, 514, 945
 
\item[] Rutledge, R.~E., Bildsten, L., Brown, E.~F., Pavlov, G.~G., 
Zavlin, V.~E., \& Ushomirsky, G.\ 2002, ApJ, 580, 413

\item[] Sanwal, D., Pavlov, G.~G., Zavlin, V.~E., \& Teter, M.~A.\ 2003, ApJ, 574, L61 

\item[] Schatz, H., et al.\ 2001, Physical Review Letters, 86, 3471 

\item[] Shapiro, S.~L., \& Teukolsky, S.~A.\ 1986, Black Holes, White Dwarfs 
and Neutron Stars: The Physics of Compact Objects, Wiley

\item[] Shternin, P.~S., Yakovlev, D.~G., Haensel, P., \& Potekhin, A.~Y.\ 2007, 
MNRAS, 382, L43 

\item[] Shternin, P.~S., Yakovlev, D.~G., Heinke, C.~O., Ho, W.~C.~G., 
\& Patnaude, D.~J.\ 2011, MNRAS, 412, L108 

\item[] Slane, P.~O., Helfand, D.~J., \& Murray, S.~S.\ 2002, ApJL, 571, L45 

\item[] Steiner, A.~W., Lattimer, J.~M., \& Brown, E.~F.\ 2010, 
ApJ, 722, 33

\item[] Stella, L., Vietri, M., \& Morsink, S.~M.\ 1999, ApJ, 524, L63 

\item[] Strohmayer, T., \& Bildsten, L.\ 2006, Compact stellar X-ray sources, 113

\item[] Strohmayer, T.\ E., Zhang, W., \& Swank, J.\ H.\ 1997, ApJ, 487, L77

\item[] Strohmayer, T.~E.,  Zhang, W., Swank, J.~H., et al.\ 1996, ApJ, 469, L9 

\item[] Suleimanov, V., Poutanen, J., \& Werner, K.\ 2011, A\&A, 527, A139

\item[] Swank, J.~H., Becker, R.~H., Boldt, E.~A., et al.\ 1977, ApJL, 212, L73 

\item[] Takahashi, T.,  Mitsuda, K., Kelley, R., et al.\ 2010, Proc. SPIE, 7732, 27

\item[] Tetzlaff, N., Schmidt, J.~G., Hohle, M.~M., \& Neuh{\"a}user, R.\ 2012, 
Pubs.\ Astr.\ Soc.\ Austr., 29, 98 

\item[] Thorne, K.~S.\ 1977, ApJ, 212, 825 

\item[] Thorsett, S.~E., \& Chakrabarty, D.\ 1999, ApJ, 512, 288

\item[] Tsai, W.-Y., \& Erber, T.\ 1975, PRD, 12, 1132

\item[] Turolla, R., Zane, S., \& Drake, J.~J.\ 2004, ApJ, 603, 265 

\item[] Turon, C., O'Flaherty, K.S., \& Perryman, M.A.C. 2005, 
The Three-Dimensional Universe with Gaia, ESA SP-576

\item[] Ulmer, A.\ 1994, ApJ, 437, L111
 
\item[] van Adelsberg, M., \& Lai, D.\ 2006, MNRAS, 373, 1495 

\item[] van der Klis, M.\ 2000, ARA\&A, 38, 717 

\item[] van Kerkwijk, M.~H., \& Kaplan, D.~L.\ 2007, ApSS, 308, 191 

\item[] van Kerkwijk, M.~H., \& Kaplan, D.~L.\ 2008, ApJ, 673, L163

\item[] van Kerkwijk, M.\ H., Kaplan, D.\ L., Durant, M., Kulkarni, S.\ R., \& 
Paerels, F.\ 2004, ApJ, 608, 432

\item[] van Paradijs, J. 1978, Nature, 274, 650

\item[] van Paradijs, J. 1979, ApJ, 234, 609

\item[] van Straaten, S., Ford, E.~C., van der Klis, M., M{\'e}ndez, M., 
\& Kaaret, P.\ 2000, ApJ, 540, 1049 

\item[] Wachter, S., Hoard, D.~W., Bailyn, C.~D., Corbel, S., \& Kaaret, P.\ 2002, 
ApJ, 568, 901

\item[] Wachter, S., Patel, S.~K., Kouveliotou, C., et al.\ 2004, ApJ, 615, 887 

\item[] Walter, F.~M., Eisenbei{\ss}, T., Lattimer, J.~M., et al.\ 2010, ApJ, 724, 669

\item[] Wang, Z., Chakrabarty, D., \& Kaplan, D.~L.\ 2006, Nature, 440, 772 

\item[] Wang, Z., Kaspi, V.~M.,  \& Higdon, S.~J.~U.\ 2007, ApJ, 665, 1292 

\item[] Watts, A.~L., \& Strohmayer, T.~E.\ 2007, APSS, 308, 625 

\item[] Watts, A.~L., Strohmayer, T.~E., \& Markwardt, C.~B.\ 2005, ApJ, 634, 547 

\item[] Webb, N.~A. \& Barret, D.\ 2007, ApJ, 671, 727

\item[] Webb, N.~A., Olive, J.-F., \& Barret, D.\ 2004, A\&A, 417, 181 

\item[] Weinberg, N., Miller, M.~C., \& Lamb, D.~Q.\ 2001, ApJ, 546, 1098 

\item[] Wijnands, R., \& van der Klis, M.\ 1998, Nature, 394, 344

\item[] Woods, P.~M., Kaspi, V.~M., Thompson, C., et al.\ 2004, ApJ, 605, 378 

\item[] Woods, P.~M., Kouveliotou, C., G{\"o}{\u g}{\"u}{\c s}, E., et 
al.\ 2001, ApJ, 552, 748 

\item[] Woods, P.~M., \& Thompson, C.\ 2006, Compact stellar X-ray sources, 547

\item[] Yakovlev, D.~G.,  Kaminker, A.~D., Gnedin, O.~Y., \& Haensel, P.\ 2001, 
Phys.\ Rep., 354, 1 

\item[] Yakovlev, D.~G., Ho, W.~C.~G., Shternin, P.~S., 
Heinke, C.~O., \& Potekhin, A.~Y.\ 2011, MNRAS, 411, 1977

\item[] Yakovlev, D.~G., \& Pethick, C.~J.\ 2004, ARA\&A, 42, 169

\item[] Yunes, N., Psaltis, D., \"Ozel, F., \& Loeb, A.\ 2010, PRD, 
81, 064020 

\item[] Zamfir, M., Cumming, A., \& Galloway, D.~K.\ 2012, ApJ, 749, 69 

\item[] Zane, S., Cropper, M., Turolla, R., et al.\ 2005, ApJ, 627, 397 

\item[] Zane, S., \& Turolla, R.\ 2006, MNRAS, 366, 727 

\item[] Zavlin, V.~E.\ 2006, ApJ, 638, 951 

\item[] Zavlin, V.~E.\ 2007, ApSS, 308, 297 

\item[] Zavlin, V.~E., \& Pavlov, G.~G.\ 1998, A\&A, 329, 583 

\item[] Zavlin, V. E., Pavlov, G. G., \& Shibanov, Y. A. 1996, A\&A, 315, 141

\item[] Zhang, B., \& Harding, A.~K.\ 2000, ApJ, 532, 1150 

\item[] Zhang, C.~M., Wang, J., Zhao, Y.~H., et al.\ 2011, A\&A, 527, A83

\end{harvard}

\end{document}